\documentclass[pdflatex,sn-basic]{sn-jnl}

\usepackage{graphicx}%
\usepackage{multirow}%
\usepackage{amsmath,amssymb,amsfonts}%
\usepackage{amsthm}%
\usepackage{mathrsfs}%
\usepackage[title]{appendix}%
\usepackage{xcolor}%
\usepackage{textcomp}%
\usepackage{manyfoot}%
\usepackage{booktabs}%
\usepackage{algorithm}%
\usepackage{algorithmicx}%
\usepackage{algpseudocode}%
\usepackage{listings}%
\usepackage{tabularx}%
\usepackage{bbding}%

\usepackage{fancyhdr}
\usepackage{hyperref}
\fancypagestyle{firstpagepub}{
  \fancyhf{}
  \fancyhead[L]{\footnotesize Accepted by Artificial Intelligence Review; early-access version available online at \href{https://doi.org/10.1007/s10462-026-11563-0}{https://doi.org/10.1007/s10462-026-11563-0}}

}

\usepackage{titlesec} %
\titleformat{\paragraph}
  {\normalfont\normalsize\bfseries}{\theparagraph}{1em}{}

\usepackage{geometry}
\theoremstyle{thmstyleone}%
\theoremstyle{thmstyletwo}%

\theoremstyle{thmstylethree}%

\begin{document}

\title[Article Title]{LLM agents security duality: a comprehensive survey of self-security and empowered cybersecurity}


\author[1]{\fnm{Yiwei} \sur{Xu}}\email{yiweix@whu.edu.cn}
\author[1]{\fnm{Yong} \sur{Zhuang}}\email{yong.zhuang@whu.edu.cn}
\author[1]{\fnm{Xuanming} \sur{Liu}}\email{teddy233@whu.edu.cn}
\author[1]{\fnm{Tian} \sur{Zhang}}\email{tianzhang2025@whu.edu.cn}
\author[1]{\fnm{Bowen} \sur{Xiao}}\email{bwxiao@whu.edu.cn}
\author[1]{\fnm{Xiaoyang} \sur{Xu}}\email{xiaoyangx@whu.edu.cn}
\author[1]{\fnm{Delong} \sur{Jiang}}\email{delong@whu.edu.cn}

\author*[1]{\fnm{Juan} \sur{Wang}}\email{jwang@whu.edu.cn}
\author[2]{\fnm{Hongxin} \sur{Hu}}\email{hongxinh@buffalo.edu}

\affil*[1]{\orgdiv{School of Cyber Science and Engineering}, \orgname{Wuhan University}, \city{Wuhan}, \country{China}}
\affil[2]{\orgdiv{Department of Computer Science and Engineering}, \orgname{University at Buffalo}, \city{Buffalo}, \country{USA}}


\abstract{Large language model (LLM) agents are rapidly being integrated into real-world systems. Their autonomy and tool-use capabilities generate substantial value while simultaneously expanding the security attack surface. This survey provides a comprehensive overview of the opportunities and challenges of LLM agents in security, focusing on two core areas: (1) threats to LLM agents themselves and corresponding mitigation strategies (LLM agents self-security), and (2) the role of LLM agents in empowering the cybersecurity lifecycle across offense and defense (LLM agents empowered cybersecurity). We first examine the internal and external attack surfaces of agents, propose a taxonomy organized by threat sources, and analyze associated mitigations and evaluation frameworks. We then investigate how agent capabilities are applied in cybersecurity practice and present, to our knowledge, the first agent-empowerment framework aligned with the full cyber offense–defense lifecycle. By systematically surveying these two areas, we are the first to highlight a positive feedback synergy between LLM agents self-security and empowered cybersecurity, offering new insights for the advancement of both. We further identify current limitations and outline promising directions for future research. The insights provided aim to catalyze the coordinated development of LLM agents self-security and agent empowered cybersecurity, paving the way for more capable and robust agent applications.}




\maketitle
\thispagestyle{firstpagepub}

\section{Introduction}\label{sec1}
In recent years, large language models (LLMs) have demonstrated remarkable capabilities in addressing open-ended questions, achieving performance levels that match or exceed human expertise in certain specialized domains \citep{bib316}. The performance improvement of LLMs has brought a revolutionary breakthrough for the development of agents. In the past, agents relied on predefined rules and limited pattern recognition capabilities \citep{bib244,bib245,bib246}, lacking flexible adaptation mechanisms. Large language models possess the ability to generalize knowledge across domains, enabling agents to adapt to different scenarios and perform a wide range of functions without extensive retraining. Indeed, agent-based architectures are widely considered by many researchers and industry experts to be a crucial paradigm for advancing towards artificial general intelligence (AGI). Currently, extensive research and applications based on LLM agents are emerging across various fields, including healthcare, finance, and customer service, among others. Relying on LLM's powerful analysis, planning, and autonomous learning capabilities, LLM-based agents provide autonomous decision-making, collaboration, and other functions, fundamentally reshaping how organizations and individuals address complex tasks and demonstrating significant transformative potential and commercial value.

\begin{figure}[h!]
    \centering
    \includegraphics[
        width=1\textwidth,
        trim=0cm 2cm 0cm 0cm,
        clip
        ]{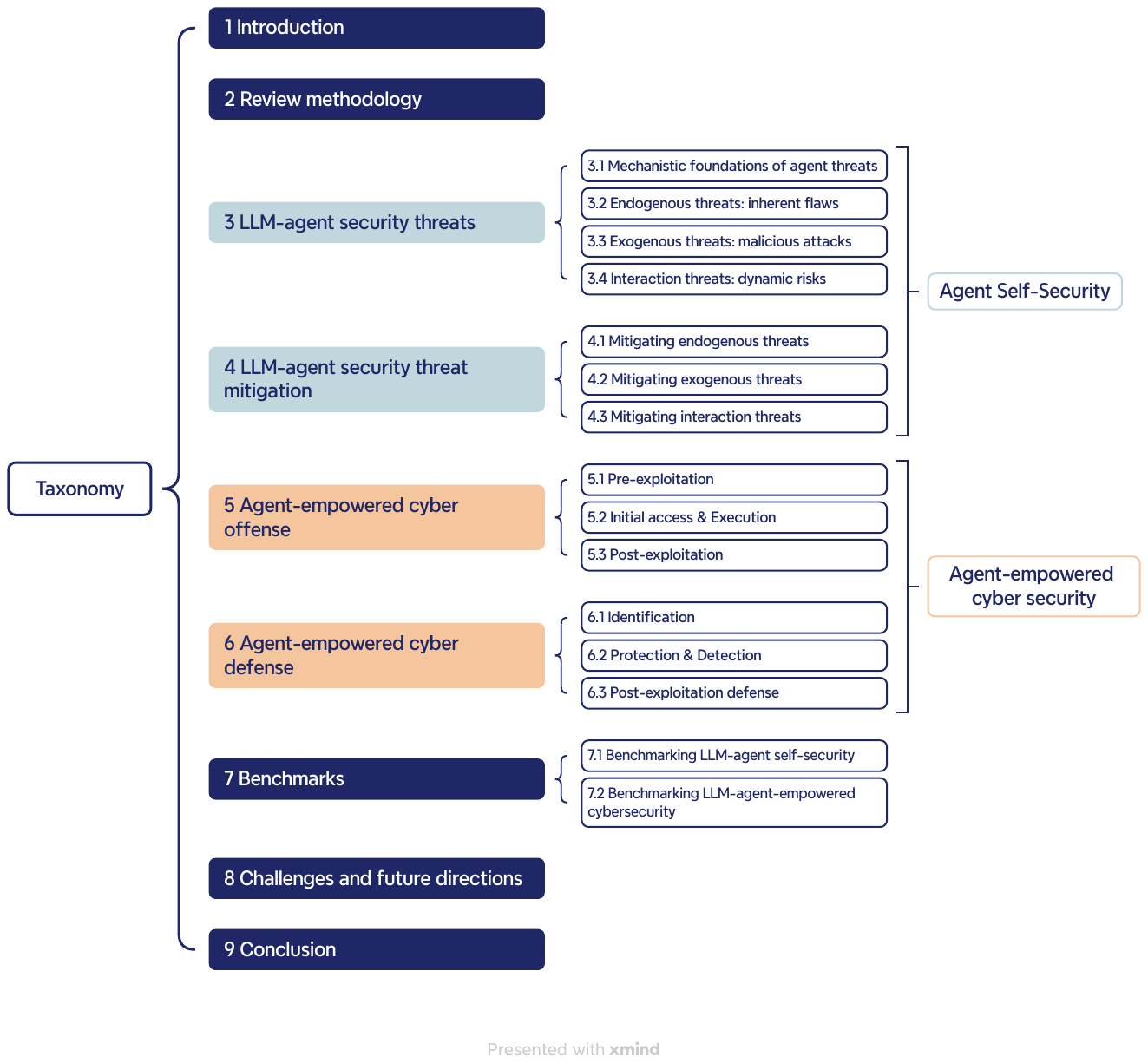}
    \caption{\small Taxonomy of this review.}
    \label{fig_taxonomy}
\end{figure}

However, the rapid development of LLM-based agents has cast a double-edged sword in the field of security, introducing both ubiquitous risks and unprecedented opportunities. On the one hand, LLM agents have advanced rapidly in function and performance, but security research remains underdeveloped, leading to substantial potential risks for agents. LLM agent systems naturally inherit from top to bottom the existing threats of LLM, such as bias, illusion, and data leakage, and these risks may be further amplified within the agent architecture. Concurrently, because agents usually execute in complex interactive environments, the attack surface is significantly expanded, and new vulnerability dimensions are introduced, such as action hijacking \citep{bib260}, tool abuse \citep{bib247} and collusion \citep{bib70}. On the other hand, LLM agents offer new opportunities in the security field. By integrating the powerful analytical and planning capabilities of LLMs, agents can effectively identify potential threats and autonomously execute security policies \citep{bib239} to improve the overall security of the system. Furthermore, agents can also be used to build fully automated attack chains \citep{bib196}, launch new social engineering attacks \citep{bib210}, or manipulate information in the cognitive domain \citep{bib63}, which pose unprecedented challenges to cyber network and cognitive security defenses, and urgently require comprehensive advancements and technological innovations in the security field.


This dual role of agents reflects two pre-existing trajectories in the literature that long predate today’s LLM agents. One thread comes from adversarial machine learning, which, beginning in the early 2010s, established a canonical threat spectrum for data-driven models, spanning poisoning and adversarial examples \citep{biggio2012poisoning,szegedy2013intriguing,goodfellow2014explaining}, privacy leakage such as membership inference \citep{shokri2017membership}, and supply-chain style compromises via backdoors \citep{wang2019neural}. In parallel, cybersecurity research pursued increasing degrees of automation, from end-to-end autonomous vulnerability discovery and exploitation in the 2016 DARPA Cyber Grand Challenge \citep{darpa2016cgc} to neural sequence modeling for operational logs and code understanding \citep{du2017deeplog,feng-etal-2020-codebert}. A qualitative shift then emerged from 2020 onward: few-shot scaling \citep{brown2020language}, retrieval augmentation \citep{bib192}, and tool-mediated interaction paradigms \citep{yao2023react} turned static models into long-horizon, action-taking systems. This shift helps explain why earlier assumptions and defenses can fail in agent settings, where persistent context and multi-step tool use expand injection and exfiltration surfaces and can amplify small compromises into trajectory-level failures, motivating the security paradigm surveyed in this article.

Current research concerning the security of LLM agents mainly focuses on two main lines. First is \textbf{LLM agents self-security}, be divided into internal inherent defects derived from LLM and external risks arising from interactive behaviors. Inherent LLM defects, such as bias and illusion of generated content \citep{bib17}, may be amplified due to its autonomous decision-making and action ability after the agent inherits and evolves into more direct harms, such as model bias may lead to discriminatory autonomous behaviors \citep{bib40-Agentverse}. At the same time, the complex interactions between agents and environments, tools, and other agents significantly expand the attack surface and introduce new vulnerability dimensions. The LLM serves as a core component, but it does not constitute the entire agent. Unfortunately, most of the current research is still mainly focused on the risks associated with LLMs themselves, and few studies investigate security and trustworthiness from the holistic perspective of the overall agent framework and its internal modules.

The second line is \textbf{LLM agents empowered cybersecurity}. Using the analysis ability of LLM, agents can identify complex attack patterns in real-time and actively adjust defense strategies to improve the dynamic response ability of the system \citep{bib239}. In addition, the potential of LLM agents in offensive applications can not be ignored, and their autonomous planning and adaptive attack capabilities pose a serious challenge to the existing network security defense system. The autonomous and intelligent features of agents are reshaping the landscape of cybersecurity.

\begin{figure}[h!]
    \centering
    \includegraphics[
        width=.9\textwidth,
        trim=.5cm 2cm 1cm 2.5cm,
        clip
        ]{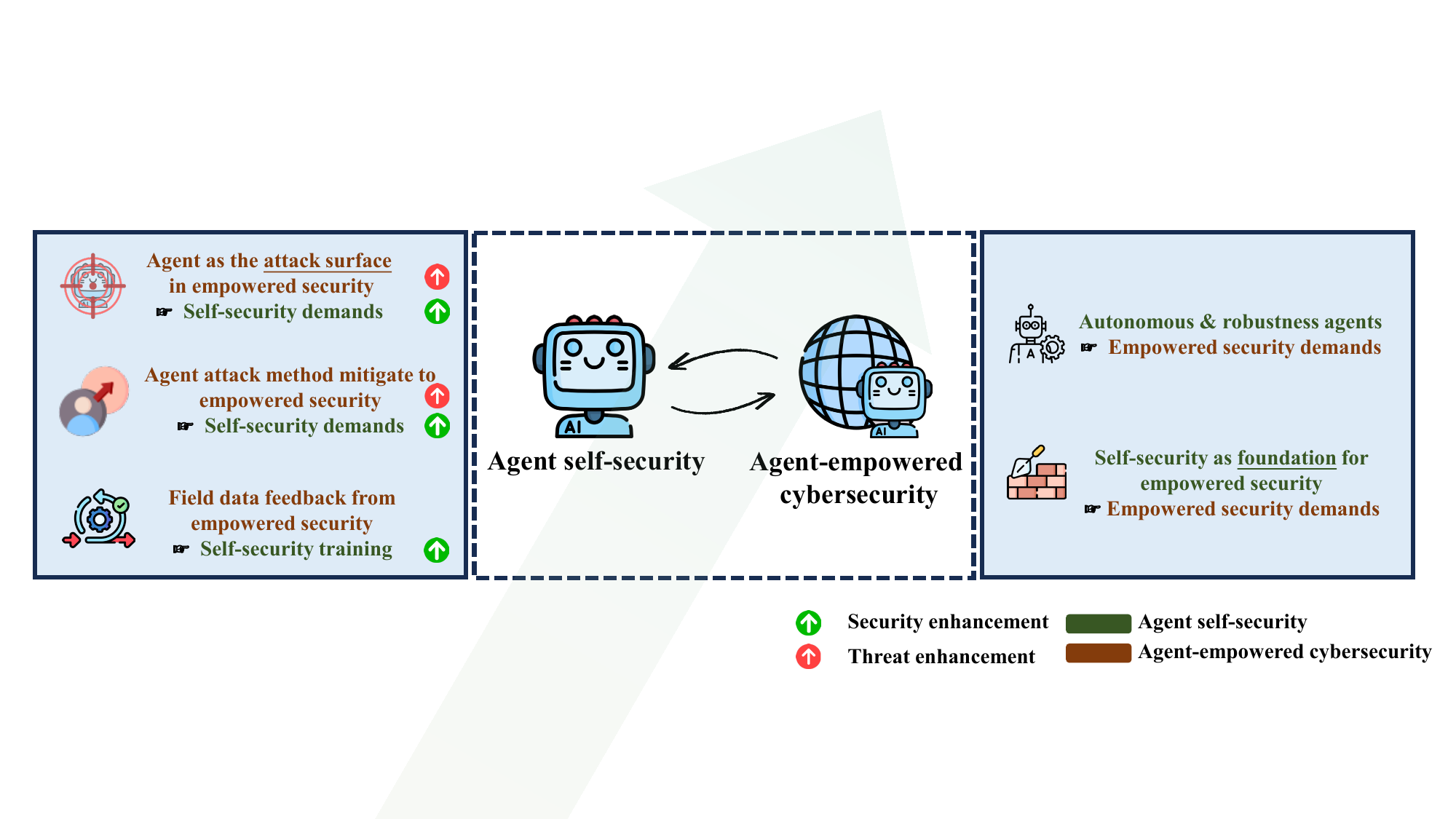}
    \caption{\small The symbiotic duality of agent self-security \& agent-empowered cybersecurity.}
    \label{fig_cycle}
\end{figure}

These two primary research lines constitute an interrelated, cyclically reinforcing structure, as shown in Figure \ref{fig_cycle}. As agent capabilities advance, a growing number of agents are providing foundational capabilities for agent-empowered cybersecurity. Yet, agents themselves become an attack surface, and when used as attack vectors, their methods directly threaten system integrity, both scenarios intensify demands for robust self-security. Moreover, field data feedback from agent-empowered cybersecurity provides a crucial source of data for robustness training, thereby enhancing agent self-security. Furthermore, by treating the agent as a system in its own right, methodologies from empowered cybersecurity can be adapted to enhance its self-security. Thus, improvements in offensive and defensive capabilities translate into substantial potential in security applications while introducing more severe challenges: more intelligent agents can more accurately identify threats and build safety guardrails in defense, but their destructive power is also amplified in attacks. This concurrent escalation of capability and risk forms a continuous arms race, wherein self-security and empowered cybersecurity co-evolve through mutual demands and iterative reinforcement. 

Recent review articles have explored the security and trustworthiness of LLM agents from different perspectives, as shown in Table \ref{tab1}. Despite the foundation laid by these works, there are some limitations. Several surveys provide a broad overview but treat LLMs primarily as advanced tools rather than autonomous agents. For instance, \citet{FERRAG20251} and \citet{YAO2024100211} review the risks facing LLMs themselves and their applications in cybersecurity, discussing the interplay between the two, but their focus on LLMs. In early times, \citet{bib313} proposed a conceptual framework for the LLM agent that included three key components: brain, perception, and action. However, they mainly highlighted security concerns inherent to the underlying LLMs rather than the unique vulnerabilities and attack surfaces emerging from the agentic properties themselves. \citet{bib250} began to distinguish between the inherent risks of LLMs and agent-specific risks, but the analysis of threats only expounds the appearance, and the classification of mitigation methods is not comprehensive. \citet{bib314} provided a review from the perspective of attack and defense technologies, but their discussion of interaction-level threat remains insufficient. Although many comprehensive reviews \citep{bib249,bib315} thoroughly catalog observable risks, they often fall short of investigating their root causes. Furthermore, most reviews do not consider the evaluation of agent security, only \citet{bib314} and \citet{bib315} offered. Notably, only \citet{xu2025forewarned} explores the application of LLM agents in cybersecurity, but it does not cover the entire cybersecurity lifecycle. Furthermore, only \citet{bib250} briefly mentions the enhancement of LLM agent capabilities and the impact of the mutual empowerment between the agent's self-security and its empowered cybersecurity.

\begin{sidewaystable}
\caption{Recent works on LLM-based agent security \& empowered-security}\label{tab1}
\renewcommand{\arraystretch}{1.5}
\begin{tabular*}{\textheight}{@{}>{\centering\arraybackslash}p{2.5cm} *{5}{>{\centering\arraybackslash}p{2cm}} p{9.5cm}}
\toprule%
\textbf{References} & \textbf{Object} & \textbf{Self-security domain} & \textbf{Empowered-security domain} & \textbf{cybersecurity lifecycle} & \textbf{Interaction with self-security \& empowered-security} & \textbf{Limitations} \\
\midrule
\cite{YAO2024100211} & LLM & Atk/Def & Off/Def & \scalebox{2}{$\bullet$} & \scalebox{2}{$\bullet$} & Object is not LLM agent, it will not be detailed here.\\
\cite{FERRAG20251} & LLM & Atk/Def & Off/Def/Eval & \scalebox{2}{$\bullet$} & \scalebox{2}{$\bullet$} & Object is not LLM agent, it will not be detailed here.\\
\cite{bib313} & LLM agent & Robustness & \XSolidBrush & $\bigcirc$ & $\bigcirc$	& Only a simple discussion of robustness and trustworthiness. \\
\cite{bib250} & LLM agent & Atk/Def & \XSolidBrush & $\bigcirc$ & \HalfCircleLeft & The discussion is not in-depth and does not analyze according to the modularization of the agent.\\
\cite{bib314} & LLM agent & Atk/Def/Eval & \XSolidBrush & $\bigcirc$ & $\bigcirc$ &The classification is innovative but still macro, and each section is not detailed enough. \\
\cite{bib249} & LLM agent & Atk/Def/Eval & \XSolidBrush & $\bigcirc$ & $\bigcirc$ & Lacks in-depth discussions on the causes of risk.\\
\cite{bib315} & LLM agent & Atk/Def/Eval & \XSolidBrush & $\bigcirc$ & $\bigcirc$ & Without in-depth analysis of the drivers that may make these modules vulnerable to attack, stay in the appearance of attack or defense, without in-depth analysis of why.\\
\cite{xu2025forewarned} & LLM agent & \XSolidBrush & Off/Def/Eval & \HalfCircleLeft & $\bigcirc$ & The perspective is highly biased towards offense, there is insufficient coverage of defensive applications, and the general vulnerabilities of LLMs themselves.\\
Our & LLM agent & Atk/Def/Eval & Off/Def/Eval & \scalebox{2}{$\bullet$} & \scalebox{2}{$\bullet$} & -\\

\botrule
\end{tabular*}
\footnotetext{\scalebox{2}{$\bullet$}: Full; \HalfCircleLeft: Half; $\bigcirc$: None; Attack(Atk), Offense(Off), Defense(Def), Evaluataion(Eval) denote self-security \& for security domain.}
\end{sidewaystable}

Our key contributions are as follows:

\begin{itemize}
\item We present the first systematic review of the dual role of LLM agents in security, and innovatively propose the ``LLM agents self-security \& LLM agents empowered cybersecurity'' cyclical perspective to provide integrative and frontier insights for understanding the security paradigm shift driven by LLM agents.
\item We present a comprehensive review of the attack, defense, and evaluation strategies for LLM agents, analyzing threat etiologies to establish a novel, origin-based taxonomy as a comprehensive and extensible foundation for future agent self-security.
\item We systematically analyze how LLM agents empower cybersecurity and introduce a full-security lifecycle Task–Capability–Evidence mapping. We further present the first agent-empowerment framework aligned with the cyber offense–defense workflow, establishing a foundation for research and practice.
\item We summarize the current research gaps in security and trustworthiness of LLM agents, and propose challenges, so as to provide clear guidance for future research.
\end{itemize}

This survey aims to comprehensively review the recent advances in LLM agent security techniques. Through a thorough review of the existing research, our goal is to construct a comprehensive and systematic LLM agent security framework that provides a concise perspective for researchers and makes a profound exploration of the threat sources of LLM agents. The framework illuminates the cyclical relationship between LLM agent self-security and empowered-security, and suggests new avenues for research. The rest of this paper is organized as follows:

Section \ref{sec2} illustrates the review methodology and presents the organization of this paper, as shown in Figure \ref{fig_taxonomy}. Section \ref{sec3} introduces the categories of security risks faced by LLM agents, analyzes threat origins, ranging from fundamental principles governing agent operation to vulnerabilities across entire modules, and discusses the attack methods and external interaction threats. Section \ref{sec4} respectively reviews existing defensive strategies. Subsequently, Section \ref{sec5} \& Section \ref{sec6} explore and summarize the impact that LLM agents have on traditional security. Section \ref{sec7} provides an overview of benchmarks \& red-teaming methods for agent self-security and agent empowered cybersecurity. Section \ref{sec8} summarizes current research challenges and provides potential future research directions.

\section{Review methodology}\label{sec2}

To construct a comprehensive and systematic framework for LLM agent security, this survey adopts a structured literature review methodology. The primary objective is to synthesize existing research, identify critical gaps, and delineate future research trajectories. Our review process is guided by a set of formal research questions (RQs) designed to deconstruct the multifaceted landscape of LLM agent security.

The formulation of our research questions is directly informed by the limitations identified in prior work, as discussed in Section \ref{sec1}. These questions provide a scaffold for the subsequent sections of this paper:

\begin{itemize}
\item \textbf{RQ1:} What are the categories of security threats specific to LLM agents, what are their root causes, and what mitigation or defense strategies are effective?
\item \textbf{RQ2:} How do agents empower the cybersecurity offense-defense lifecycle, and what key capabilities do they play in different stages?
\item \textbf{RQ3:} What benchmarks, testbeds, and red-teaming protocols exist for evaluating (a) agent self-security and (b) agent empowered cybersecurity across the lifecycle, and how adequate are their task coverage, validity, reliability, and reproducibility?
\item \textbf{RQ4:} What are the principal open challenges and promising future research directions for the development of secure and trustworthy LLM agents?
\end{itemize}

To answer these questions, we conducted a systematic search of scientific literature from prominent academic databases, including Web of Science(\url{https://www.webofscience.com/wos/alldb/basic-search}), Google Scholar(\url{https://scholar.google.com}), arXiv(\url{https://arxiv.org/}), SpringerLink (\url{https://link.springer.com}), ACM Digital Library(\url{https://dl.acm.org/}), IEEE Xplore(\url{https://ieeexplore.ieee.org/Xplore/home.jsp}) and ACL Anthology(\url{https://aclanthology.org}). Our search strategy employed a combination of keywords and their variants, such as ``LLM-based Agent'', ``Security'', ``Agent for Security'', ``Agent attack \& defense'', ``Benchmark''. We primarily focused on literature published from 2022 to 2025, reflecting the recent and rapid advancements in this domain.

The selection criteria prioritized peer-reviewed articles from top-tier conferences and journals, alongside influential preprints that have garnered significant attention within the research community. We excluded papers that discussed LLM security without addressing the unique properties of agents or those whose focus was not primarily on security aspects. The collected literature was then analyzed and synthesized to build the comprehensive taxonomy and framework presented in this survey, ensuring a thorough and structured response to each research question.

\begin{figure*}[t] 
    \centering
    \begin{minipage}{0.49\textwidth}
        \centering
        \includegraphics[width=\linewidth]{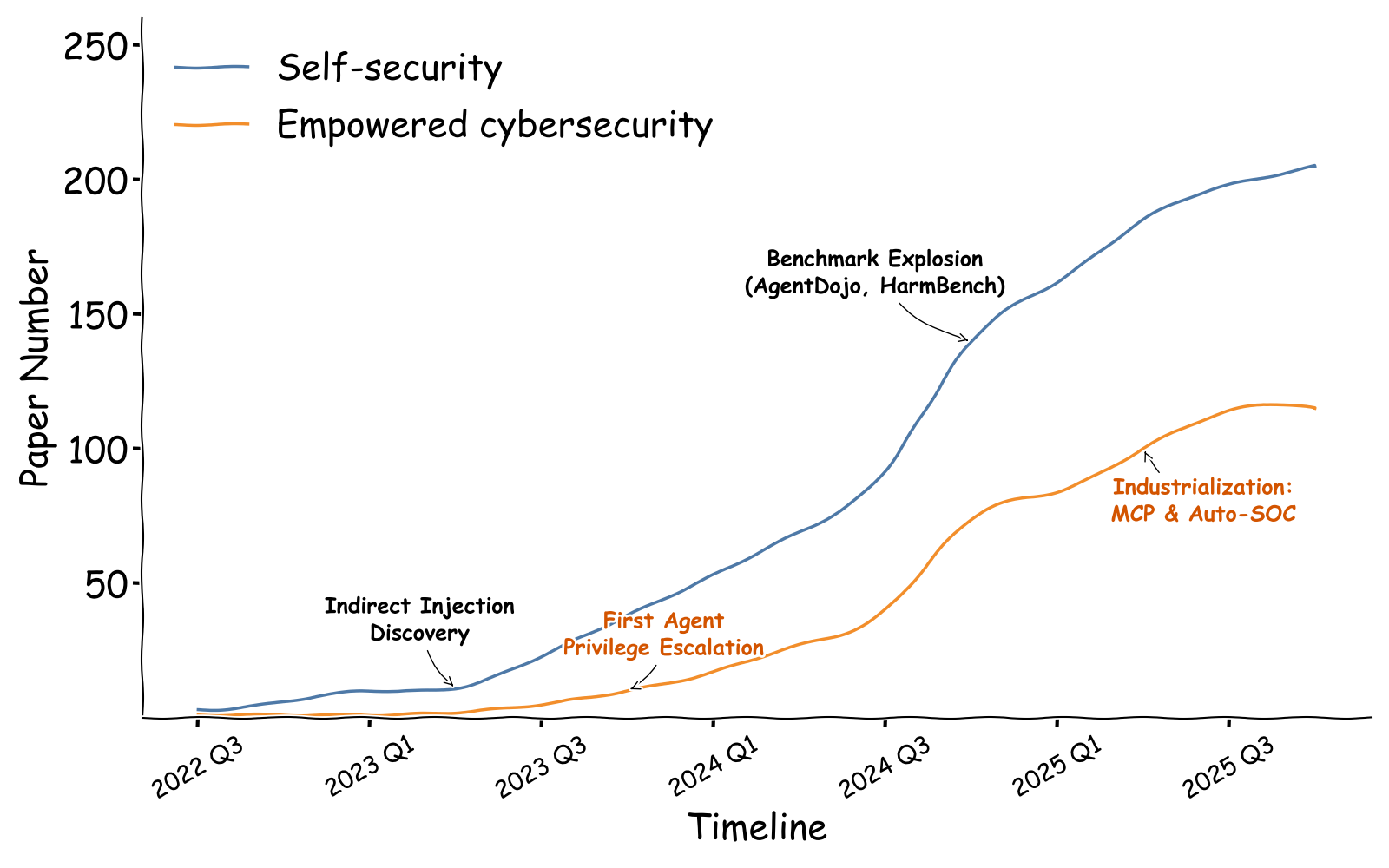}
        \caption{Temporal evolution of LLM agent security research (2022-2025).}
        \label{fig_timeline}
    \end{minipage}
    \hfill
    \begin{minipage}{0.49\textwidth}
        \centering
        \includegraphics[width=\linewidth]{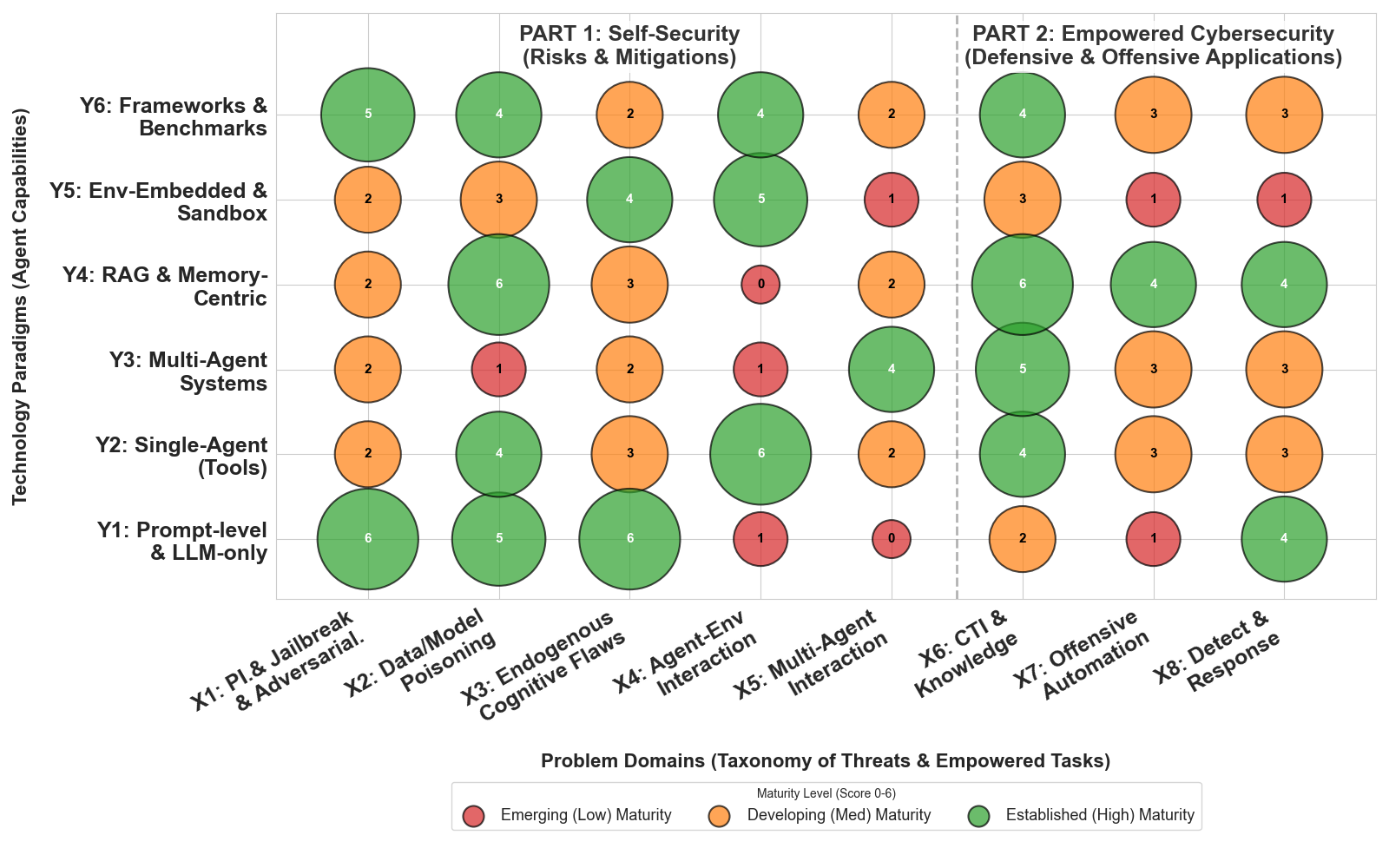}
        \caption{Research maturity landscape. Green indicates established areas (Score 4-6), while Red/Yellow indicates emerging gaps (Score 0-3).}
        \label{fig_maturity}
    \end{minipage}
\end{figure*}

To quantify the evolution of this field, we analyzed the temporal distribution and research maturity of the surveyed literature. As shown in Figure \ref{fig_timeline}, the field has witnessed a rapid proliferation of research efforts and methodologies since 2023, driven by the proliferation of autonomous agent frameworks. Concurrently, the maturity landscape (Figure \ref{fig_maturity}) reveals a distinct disparity: while LLM-centric defenses (e.g., prompt safety) are well-established (Green), dynamic interaction risks in multi-agent environments remain largely exploratory (Red/Yellow). This data-driven insight underscores the urgency of this survey and guides the structure of our subsequent analysis.

As illustrated in Figure \ref{fig_taxonomy}, the subsequent analysis is organized along two primary, interconnected lines of inquiry: agent self-security, and agent empowered cybersecurity. This dual-threaded structure allows readers to navigate the content based on their specific interests, whether they are focused on the vulnerabilities and defenses of agentic systems or on their application in transforming the security landscape.

\begin{figure}[h!]
    \centering
    \includegraphics[width=1\textwidth,
        trim= .5cm 4cm 0cm 4cm,
        clip
        ]{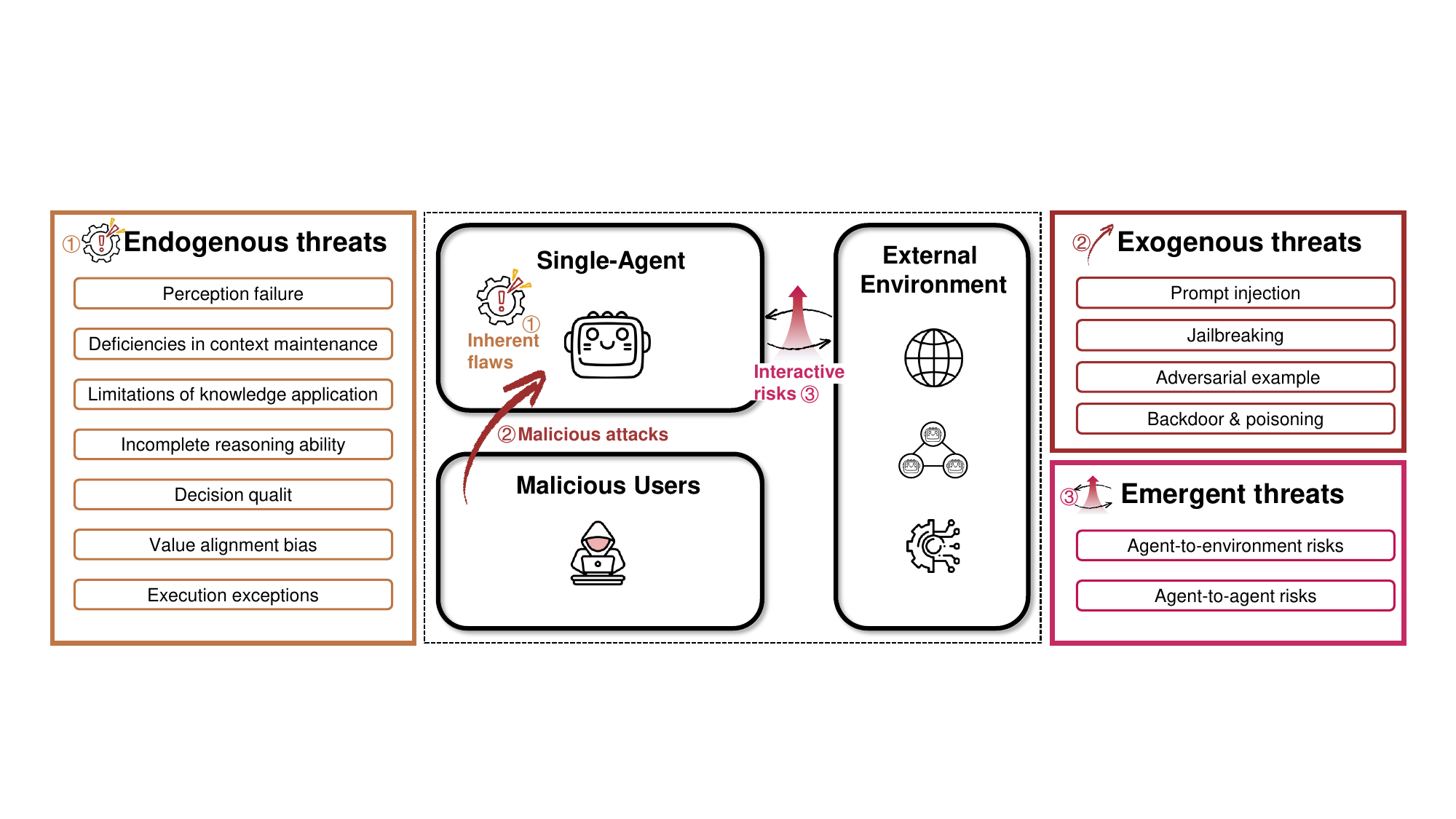}
    \caption{\small Taxonomy of agent self-security threats.}
    \label{fig_threat_all}
\end{figure}

\section{LLM agent self-security threats}\label{sec3}
This chapter presents a systematic analysis of the security threats confronting LLM-based agents. We propose a unified classification framework based on the source of threats, endogenous threats, exogenous threats, and interaction threats, as illustrated in Figure \ref{fig_threat_all}. We further introduce a mechanistic layer in §\ref{sec3.0} that distills time stable architectural failure modes to explain why the threats in §\ref{sec3.1}–§\ref{sec3.3} recur across tasks and implementations. Endogenous threats stem from the inherent flaws within the agent itself. Exogenous threats are initiated by external adversaries, are inherently unidirectional. Interaction threats arise from dynamic feedback loops of agent interactions with the outside. Following this framework, § \ref{sec3.1} analyzes the root causes of failure in agents' internal processes, § \ref{sec3.2} systematically analyzes exogenous attack vectors, and § \ref{sec3.3} investigates the causes and consequences of interaction risks. We summarize the orthogonal relationship between pillars (§ \ref{sec3.0}) and threat categories (§ \ref{sec3.1} – § \ref{sec3.3}), as show in Table X.


\subsection{Mechanistic foundations of agent threats}\label{sec3.0}
To clarify why LLM-based agents become vulnerable, we survey the agent-security literature from a system-level perspective centered on autonomy. Existing threat taxonomies catalog what happens, but they often under-specify the structural mechanisms that make these failures recurrent across tasks and frameworks. To provide a principle-level grounding for observed phenomena—and a stable basis for classifying future threats—we introduce a mechanistic layer that distills six foundational architectural failure modes (P1–P6). We then map threats back to these foundations, so the survey supports both design-level unification.

\begin{table*}[t]
\centering
\caption{Mapping of phenomenological threat families (§3.2–§3.4) to mechanistic (§3.1)}
\footnotesize
\setlength{\tabcolsep}{5pt}
\renewcommand{\arraystretch}{1.18}

\begin{threeparttable}
\begin{tabular}{p{0.16\textwidth}p{0.42\textwidth}cccccc}
\toprule
\multirow{2}{*}{Threat source} &
\multirow{2}{*}{Threat family (phenomenological)} &
\multicolumn{6}{c}{Mechanistic pillars (\S\ref{sec3.0})} \\
\cmidrule(lr){3-8}
& & P1 & P2 & P3 & P4 & P5 & P6 \\
\midrule

\multirow{7}{*}{Endogenous (\S\ref{sec3.1})} &
Perception failure (\S\ref{sec3.1.1})                            &     &     &     & $\bullet$ & $\circ$ &     \\
& Deficiencies in context maintenance (\S\ref{sec3.1.2})         &     & $\bullet$ &     &     & $\circ$ &     \\
& Limitations of knowledge application (\S\ref{sec3.1.3})        &     &     &     & $\bullet$ & $\circ$ &     \\
& Incomplete reasoning ability (\S\ref{sec3.1.4})                &     &     &     &     & $\bullet$ &     \\
& Decision quality issues (\S\ref{sec3.1.5})                     &     &     &     &     & $\bullet$ &     \\
& Value alignment bias (\S\ref{sec3.1.6})                        &     &     &     &     & $\bullet$ & $\circ$ \\
& Execution exceptions (\S\ref{sec3.1.7})                        &     &     & $\bullet$ &     & $\circ$ &     \\

\midrule
\multirow{3}{*}{Exogenous (\S\ref{sec3.2})} &
Indirect prompt injection (\S\ref{sec3.2.1})                     & $\bullet$ &     &  & $\circ$ &     &     \\
& Jailbreaking (\S\ref{sec3.2.2})                                & $\bullet$ &     &     &     & $\circ$ &     \\
& Adversarial example attack (\S\ref{sec3.2.3})                  &  &     &     & $\bullet$    &  &     \\
& Backdoor and poisoning (\S\ref{sec3.2.4})                      &     & $\bullet$ &     & $\circ$ &     &     \\

\midrule
\multirow{3}{*}{Interaction (\S\ref{sec3.3})} &
Abnormal perception (\S\ref{sec3.3.1.1})                          &     &     &     & $\bullet$ & $\circ$ &     \\
&Tool calling failures (\S\ref{sec3.3.1.2})                       & $\circ$ &     & $\bullet$ &     &  &     \\
& Collaboration mechanism disruption (\S\ref{sec3.3.2.1})         &  &     &     & $\circ$ &    & $\bullet$ \\
& Malicious information propagation (\S\ref{sec3.3.2.2})          & $\circ$    &  &     &  &     & $\bullet$ \\
& Collective cognitive bias (\S\ref{sec3.3.2.3})          &     &  &     &  & $\circ$    & $\bullet$ \\
& Social engineering (\S\ref{sec3.3.2.4})          &  &  &     &  &  $\circ$   & $\bullet$ \\

\bottomrule
\end{tabular}

\begin{tablenotes}[flushleft]
\footnotesize
\item $\bullet$ denotes the primary mechanism; $\circ$ denotes an amplifying factor. Each threat family is tagged with one primary pillar and up to two amplifiers.
\item P1: Instruction boundary collapse; P2: State persistence and integrity hazards; P3: Authority amplification via tools; P4: Untrusted observations and feedback integrity; P5: Long-horizon composition and error amplification; P6: Multi-agent emergence and responsibility diffusion.
\end{tablenotes}
\label{tab:orthogonal_map}
\end{threeparttable}
\end{table*}

\subsubsection{Instruction boundary collapse}\label{sec3.0.1}
Instruction boundary collapse is the failure to separate control-plane directives from data-plane content. When it occurs, untrusted content is misread as high-privilege instruction, turning ingestion into a covert command channel and enabling prompt injection–style goal hijacking \citep{perez2022ignore}. This stems from the fact that standard LLMs typically operate over a single mixed context without a built-in, verifiable notion of instruction privilege \citep{wallace2024instruction}.

Boundary collapse is exacerbated in LLM agentic pipelines by the reliance on single-stream text concatenation, where heterogeneous sources are flattened into a unified context window. Lacking intrinsic ground truth regarding provenance or privilege, the model cannot reliably distinguish privileged instructions from untrusted inputs. This architectural limitation explains why models misinterpret attacker-controlled content embedded within retrieved data as executable commands \citep{hines2024defending}. Consequently, security analyses posit that prompt injection may remain an inherent structural vulnerability, necessitating systems designed for robustness against residual instruction–data confusion rather than seeking absolute prevention \citep{David2025NCSC}.

\textbf{Taxonomic Mapping.} This boundary collapse serves as an architectural root cause for multiple threat classes in our taxonomy. Primarily, it enables Exogenous Threats such as Indirect Prompt Injection (§ \ref{sec3.2.1}) and Jailbreaking (§ \ref{sec3.2.2}), where untrusted content masquerades as control commands. It also permeates Interaction Threats, acting as an amplifier for Tool Calling Failures (§ \ref{sec3.3.1.2}) and manifesting as Malicious Information Propagation (§ \ref{sec3.3.2.2}) when injected prompts replicate across agent boundaries.

\subsubsection{State persistence \& integrity hazards}\label{sec3.0.2}
State persistence and integrity hazards emerge when agents lack rigorous controls of provenance, isolation, and revocability for persistent data structures, such as long-term memory and vector stores \citep{10.1145/3586183.3606763,bib192}. This structural vulnerability transforms transient interactions into durable compromises where dormant poisoned artifacts resurface as trusted context across temporal boundaries. Consequently, corrupted historical data can systematically bias multi-step planning and tool execution across distinct sessions \citep{dong2025memory}.

Agents improve efficiency and personalization by synthesizing and retrieving historical experiences, thereby establishing a high-trust interface between the reasoning core and persistent storage \citep{packer2023memgpt}. \citet{dong2025memory} show that this boundary is exploitable under realistic constraints, because query-only interactions can embed adversarial triggers that later activate upon retrieval. Moreover, sparse adversarial injections can achieve high success rates with negligible degradation of benign task performance \citep{bib16}. This vulnerability also extends to retrieval pipelines, where a knowledge database can be corrupted through a small number of malicious insertions to manipulate responses for targeted queries \citep{bib105}. Beyond immediate failures, poisoning apparently successful trajectories can induce stealthy imitation-driven behavioral drift that evades conventional jailbreak detection signatures \citep{srivastava2025memorygraft}.

\textbf{Taxonomic Mapping.} State persistence hazards fundamentally drive Exogenous Threats such as Backdoor and Poisoning Attacks (§ \ref{sec3.2.4}) via the implantation of durable triggers within memory stores. This vulnerability also underpins Endogenous Threats, specifically Deficiencies in Context Maintenance (§  \ref{sec3.1.2}), by allowing corrupted information to recur through retrieval processes and contaminate long-term state.

\subsubsection{Authority amplification via tools}\label{sec3.0.3}
Authority amplification via tools occurs when an LLM agent can invoke external services such as APIs, executors, file systems, or enterprise systems, thereby gaining capabilities far beyond what text output alone can produce. The agent effectively acts as a deputy that spends privileges on the user’s behalf, so minor instruction-level deviations can trigger high-impact actions such as data access, code execution, or transactions. This mechanism instantiates the confused deputy problem, in which an authorized component is induced to misuse its authority under attacker-chosen intent \citep{hardy1988confused}. In tool-integrated applications, this shift turns prompt manipulation from producing incorrect text into causing incorrect actions \citep{10.1145/3605764.3623985}. The risk is exacerbated when agents hold broad or long-lived permissions that exceed task needs, violating least-privilege design principles \citep{saltzer1975protection}.

Tool invocation architectures rely on natural-language descriptions and schemas to support autonomous tool selection and parameter specification \citep{ross-etal-2025-when2call}. This dependency creates accessible control points where indirect prompt injection can induce detrimental actions or data leakage \citep{bib5}. Adversaries can also subvert tool selection by inserting or poisoning tool documentation, which steers agents toward malicious resources \citep{shi2025prompt}. Prior work further shows that agents can be coerced into invoking unauthorized logging utilities to exfiltrate internal interaction traces while maintaining high task fidelity \citep{liu2023prompt}.

\textbf{Taxonomic Mapping.} Authority amplification serves as a critical execution-layer vulnerability. It primarily manifests within Interaction Threats as Tool Calling Failures (§ \ref{sec3.3.1.2}), where unauthorized or manipulated executions produce tangible side effects. Furthermore, it drives Endogenous Threats, specifically Execution Exceptions (§ \ref{sec3.1.7}), by transforming internal reasoning errors into high-impact, over-privileged actions that disrupt system stability.

\subsubsection{Untrusted observations \& feedback integrity}\label{sec3.0.4}
Untrusted observations and feedback integrity capture an agent’s failure to verify the authenticity, integrity, and epistemic reliability of the inputs it consumes, including tool outputs, retrieved passages, web content, logs, sensors, and inter-agent messages \citep{shi2025sok}. When these checks are absent, an adversary can distort the agent’s beliefs without issuing explicit instructions, so downstream planning and tool use optimize against fabricated or selectively biased evidence rather than the user’s intent \citep{287220}. This shifts the threat surface from prompt manipulation toward evidence manipulation, with the agent’s evolving evidence state as the primary target \citep{bib105}.

Agents operate in open environments by continuously ingesting external observations and updating actions accordingly, which creates a persistent trust boundary between the reasoning core and world-facing input channels. In web-agent settings, the environment itself can be adversarial, as attacker-controlled webpage content can steer behavior because the agent must read and act on what it observes to complete tasks \citep{evtimov2025wasp}. Retrieval adds another high-impact observation pathway. When evidence is drawn from a corpus or knowledge base, poisoning a small number of documents can systematically bias what the agent retrieves and concludes even when its instruction-following policy is unchanged \citep{bib105}. This channel is vulnerable not only to integrity attacks but also to availability manipulation. An attacker can suppress or jam what the agent can retrieve or verify, selectively degrading performance in targeted parts of the task space \citep{shafran2025machine}.

\textbf{Taxonomic Mapping.} Untrusted observation hazards underpin Endogenous Perception Failures (§ \ref{sec3.1.1}) and Limitations of Knowledge Application (§ \ref{sec3.1.3}). They facilitate Exogenous Threats by enabling Adversarial Example Attacks (§ \ref{sec3.2.3}) and serving as vectors for Indirect Prompt Injection (§ \ref{sec3.2.1}) and Poisoning (§ \ref{sec3.2.4}). Within multi-agent systems, these vulnerabilities scale into Interaction Threats, manifesting as Abnormal Perception (§ \ref{sec3.3.1.1}) and disrupting Collaboration Mechanisms (§ \ref{sec3.3.2.1}) via corrupted environmental feedback.

\subsubsection{Long-horizon composition \& error amplification}\label{sec3.0.5}
Long-horizon composition and error amplification constitutes the structural risk where small local deviations, such as minor hallucinations, constraint slippage, noisy tool outputs, or suboptimal actions, accumulate autoregressively across multi-step trajectories to produce qualitatively different, often unsafe outcomes. Once an early mistake perturbs the evolving state, subsequent decisions are made under a shifted state distribution, which exacerbates the likelihood of compounding errors and reinforces the drift over time \citep{ross2011reduction}.

Long-horizon agents function as non-Markovian systems, where decisions rely on persistent state such as historical context and retrieved evidence. This architecture enables autoregressive accumulation: flawed intermediate artifacts are recycled into subsequent decision contexts, leading to systematic planning errors \citep{chang2024agentboard}. Empirical evidence shows that success rates deteriorate as trajectory length increases because fragile intermediate steps dominate end-to-end outcomes \citep{mialon2024gaia, kwa2025measuring}. Self-improvement mechanisms can further exacerbate these vulnerabilities, as iterative reflection loops may stabilize early biases across iterations \citep{shinn2023reflexion, madaan2023self, huang2024large}. When progress is guided by imperfect proxies, such as self-critique scores, error accumulation can manifest as specification drift, where optimizing the metric diverges from the intended objective \citep{NEURIPS2022_3d719fee}.

\textbf{Taxonomic Mapping.} Long-horizon error amplification exacerbates multiple Endogenous Threats, particularly Incomplete Reasoning Ability (§ \ref{sec3.1.4}) and Decision Quality Issues (§ \ref{sec3.1.5}), by allowing incremental errors to become self-reinforcing. It acts as an amplifier for Exogenous Threats like Jailbreaking (§ \ref{sec3.2.2}). Within Interaction Threats, this mechanism intensifies Abnormal Perception (§ \ref{sec3.3.1.1}) and Collective Cognitive Bias (§ \ref{sec3.3.2.3}), where minor deviations accumulate over time to distort group consensus and trust.

\subsubsection{Multi-Agent Emergence \& Responsibility Diffusion}\label{sec3.0.6}
Multi-agent emergence and responsibility diffusion characterize system-level failure modes where unsafe outcomes originate from collective dynamics such as coordination cascades and incentive structures \citep{cemri2025multi}. Within these environments no individual agent behavior fully accounts for the resulting end-to-end intent. Causal responsibility diffuses across distributed roles and shared artifacts necessitating counterfactual tracing across multiple system components for accountability rather than single-point attribution \citep{chockler2004responsibility}.

Multi-agent architectures modularize cognitive functions into specialized roles such as planners and critics, thereby requiring high-bandwidth communication channels that interleave control directives with evidential claims \citep{cemri2025multi}. This setup can induce both control-data confusion and the unverified acceptance of observations. These dynamics enable bias cascades in which early flaws stabilize through collective consensus, and they also allow adversarial emergence, including covert collusion \citep{bib70}. Consequently, safe individual agent behavior does not guarantee a secure ensemble, because multi-agent interactions can introduce emergent risks that are absent in isolation \citep{reid2025risk}.

\textbf{Taxonomic Mapping.} Multi-agent emergence and responsibility diffusion primarily manifests through Interaction Threats, serving as the root cause for Collaboration Mechanism Disruption (§ \ref{sec3.3.2.1}), Malicious Information Propagation (§ \ref{sec3.3.2.2}), Collective Cognitive Bias (§ \ref{sec3.3.2.3}), and Social Engineering (§ \ref{sec3.3.2.4}). This mechanism additionally exacerbates Endogenous Value Alignment Bias (§ \ref{sec3.1.6}) by facilitating the transformation of individual deviations into entrenched group norms.

\subsubsection*{Framework extensibility and future proofing} 
To ensure resilience against future threat variations we establish a dual-phase classification protocol. The initial phase requires mechanistic diagnosis to map a threat to its architectural failure pillar while the subsequent phase entails phenomenological placement to localize the threat origin within the endogenous exogenous or interaction domains outlined in § \ref{sec3.1} through § \ref{sec3.3}. By decoupling specific attack instantiations from their structural root causes this abstraction ensures the framework remains robust and time-invariant despite the continuous evolution of the attack surface.

\subsection{Endogenous threats: inherent flaws}\label{sec3.1}
This section systematically examines the risks and vulnerabilities inherent in an agent's key components, covering the complete flow from information perception to decision-making. Due to the high interdependence among components in an agent system, the failure or deficiency of any single component may be amplified through cascading effects and eventually threaten the security of the entire system. Therefore, we analyze in depth the risks faced by key links such as perception, memory, reasoning, decision making, and execution. 

\subsubsection{Perception failure}\label{sec3.1.1}
LLM agents have shortcomings in perceiving and understanding input information. Fuzzy user instructions and complex multimodal information in real scenarios will lead to errors in an agent's semantic interpretation and multimodal data fusion.

\textbf{Risks posed by semantic bias and adversarial inputs}. Agents often make errors in understanding inputs not only at the syntactic level, but also at the semantic level. Semantic errors will cause agents to misjudge the correctness of information, thereby deviating from task goals or making noncompliant decisions, which is more destructive than syntactic errors \citep{bib2}. Moreover, adversarial input exacerbates the security risk of perceptual understanding errors. The attacker can destroy the agent's perception ability through well-made text or image input, and make it produce the wrong perception. Attacks such as tampering with image descriptions \citep{bib3} and popping up specific information \citep{bib4} will mislead or distract the agent's perception ability, causing it to perform wrong actions. Even the agent may regard the attack instruction as part of the normal context and send it back to the user as is \citep{bib5}, which fully exposes the vulnerability of the perception module to the user's intention recognition.

\textbf{Perceptual inaccuracies caused by insufficient multimodal information fusion}. \citet{bib6} proposed that the agent will make errors in judgment because it cannot consider both text and image. Moreover, when the agent cannot correctly understand the user interface elements and intentions, its behavior may deviate from the user requirements and even violate the safety requirements \citep{bib7}. The failure of multi-modal fusion will cause significant risks in the judgment of complex content. For example, the harmful information implied in the meme cannot be accurately identified, which leads to the spread of harmful carriers in society, resulting in adverse social influence \citep{bib8}. Strengthening the semantic consistency between modalities can alleviate such problems to some extent \citep{bib3,bib131}, and subsequent research can further explore this direction.

\textbf{Amplification of the initial error in multi-step system}. In a dynamic and multi-step reasoning scenario, the deviation caused by errors at the perception level will be superimposed in the subsequent complex decision chain \citep{bib9}. If an agent cannot correctly integrate the changing environment and history, its subsequent behavior will deviate from its original intention, and the risk will continue to accumulate.

\subsubsection{Deficiencies in context maintenance}\label{sec3.1.2}
Agents often complete tasks autonomously via multi-step interaction. When the dialogue becomes lengthy, the agent should rely on short-term memory. However, the limitation of the LLM’s context window size disrupts decision consistency and scene understanding ability, and causes information interruption, and can even cause catastrophic forgetting.

\textbf{Decision-making affected by context window}. For instance, GPT-3 \citep{bib1}’s context window is only 2048 tokens. When the dialogue or task sequence exceeds this limit, the model will be unable to access the early information, which will cause the interruption of context understanding. \citet{bib10} proposed that because of the context length limitation, agents cannot obtain complete historical information, making it hard to understand the current situation and make reasonable decisions. Moreover, missing context can cause the agent to make various errors, such as regarding a dead player as alive in Werewolf, or forgetting characters whose identities have been exposed. \citet{bib13} demonstrated that limiting the agent’s context window size caused the agent to be unable to obtain the previous ``Scratchpad'', thereby resulting in ``thought interruption''. Furthermore, AutoPT \citep{bib14} observed that due to the limitation of context length, the agent was trapped in a loop of solving some minor problems and was unable to take into account the overall progress, resulting in task failure. For example, agents might continuously propose the same question in each round.

\textbf{Catastrophic forgetting occurs in continuous learning}, where old knowledge is overwritten or lost while learning new knowledge, affecting the long-term consistency and reliability of the agent. \citet{bib11} pointed out that large language models forget previously acquired knowledge during continuous learning. This forgetting is essentially the loss of historical information, which will impact the agent system’s state consistency and scene understanding ability in multi-round dialogue. When learning new information, the agent struggles to retain older information, resulting in incomplete information in the decision-making process. Additionally, when an agent performs a multi-task or multi-round task, it may gradually forget the initial task goal and preferentially respond to new instructions.

\textbf{Delayed Information synchronization weakens coordination in multi-agent systems.} \citet{bib12} showed that agents are at risk of being attacked if they cannot maintain historical information. For example, they cannot remember their original instructions or recognize agents that have been infected. State synchronization between different agents fails, resulting in decisions based on unsynchronized contexts. Moreover, agents fail to receive the latest information on time, resulting in decisions based on outdated data.

\subsubsection{Limitations of knowledge application}\label{sec3.1.3}
LLM-based agents exhibit many defects and vulnerabilities in knowledge application. \textbf{LLM agents are vulnerable to knowledge contamination}. \citet{bib15} highlighted the security risks of Retrieval Augmented Generation (RAG) \citep{bib192} systems depending on external data sources. The attacker can influence the system output by contaminating the data source. AgentPoison \citep{bib16} further shows that untrusted knowledge source can lead to the pollution of knowledge base, thereby compromising the agent system security.

\textbf{Outdated knowledge base will weaken LLM agents’ performance and security}. $S^3$ \citep{bib22} shows that in a social network simulation system, users' information and cognitive states evolve over time. However, there is a delay in this update, which makes the user's information and cognition inconsistent with the current situation. \citet{bib23} used an example of medical AI to show that AI can mislead healthcare professionals and lead to suboptimal treatment decisions by relying on outdated information and failing to consider recent research that could overturn or refine earlier findings. This shows that the existing knowledge base update mechanism is too mechanical and passive, and lacks the ability to actively evaluate the importance and timeliness of knowledge. It is difficult for agents to accurately identify and integrate key new knowledge from massive amounts of information, which affects the accuracy of decision-making in dynamic environments.

\textbf{Hallucination leads to inaccurate application of knowledge}. \citet{bib17} pointed out that while GPT-4 performs well on certain tasks, it is still prone to hallucination and misinformation. PentestGPT\citep{bib18} further states that LLMs may generate inaccurate actions or commands, with such errors often stemming from their inherent tendency towards hallucination. However, LLM agents naturally inherit the hallucination problems of LLMs, which makes agents prone to generating plausible but incorrect content when facing tasks such as deep reasoning or cross-domain knowledge integration \citep{bib19}. In addition, agents may encounter situations where they cannot retrieve the correct instance from memory \citep{bib20}. Meanwhile, agents may miss key details or extract information from irrelevant sources when answering questions, resulting in incomplete or inaccurate answers \citep{bib21}. These problems highlight the limitations of current agents in the process of knowledge application, and further research is urgently needed to improve their consistency and accuracy.

\subsubsection{Incomplete reasoning ability}\label{sec3.1.4}
As the core component of the LLM agent, the inherent shortcomings in an LLM's reasoning ability can lead to serious security vulnerabilities in the entire agent system.

\textbf{Insufficient logical reasoning ability}. LLM performs poorly in complex reasoning tasks, especially error-prone in causal reasoning \citep{bib3,bib24}. As a result, the LLM agent may fail to correctly understand the causal relationship between events and output wrong conclusions. Faced with the same or similar questions, agents sometimes give contradictory answers, which reflects the lack of stability and consistency in their reasoning process. Even the more advanced GPT-4 \citep{bib318} still has significant room for improvement in tasks requiring complex reasoning \citep{bib17}. These reasoning flaws highlight the shortcomings of LLM agents in terms of rigor and formalization.

\textbf{Error amplification effects in multistep reasoning \& planning}. PDoctor \citep{bib9} likens an agent's action chain to an error amplifier, wherein minor initial errors can amplify and propagate through subsequent steps, potentially culminating in catastrophic failures. The agent may make mistakes during the planning process, making an infeasible plan or failing to strictly follow the given plan, resulting in the task goal not being achieved.

\textbf{Lack of long-term planning capacity}. LLM agents demonstrate limited efficacy in long-term planning and lack effective strategic reasoning capabilities. In game theory scenarios, such as the iterated prisoner's dilemma, LLMs often lack real strategic reasoning capabilities, leading to errors in multi-step interactions \citep{bib25}. Specifically, agents struggle to consider long-term benefits, limit their decisions to short-term goals, and are often unable to formulate and execute optimal policies. In addition, agents also lack the ability to flexibly adjust their policies to effectively cope with dynamic changes in the environment. This limitation significantly affects the performance of LLM agents in complex decision-making tasks \citep{bib26}, especially in situations requiring the coordination of multiple parties' interests to achieve long-term goals.

\textbf{Subtask priority judgment is biased}. This bias can cause deviations in task execution trajectories and outcomes, thereby compromising overall system security. In the process of decomposing a complex task into subtasks, agents may fail to correctly identify the dependencies or priorities between each subtask, leading to logical defects or missing steps in the execution process \citep{bib2}. When agents fail to properly evaluate and prioritize tasks, or ignores the urgency of critical tasks when executing tasks, it can lead to the delay or omission of critical goals. For example, PDoctor \citep{bib9} points out that LLM agents are prone to make ``Order Error'', where they fail to execute a task according to the constraints specified in a user's query. This indicates that the agent cannot correctly understand the priorities between tasks.

The lack of reasoning ability may cause the LLM agent to fail to identify and prevent security threats, increasing the security risk of the system. For tasks demanding rigor and formalization, such as code generation and mathematical problem-solving, LLM agents' vulnerability to malicious inputs can induce logical reasoning errors, consequently degrading overall performance \citep{bib2}. Agents may fail to recognize malicious actions and trust incorrect information, leading to collaboration failures and increased system vulnerability. Future work can use reinforcement learning or supervised fine-tuning techniques to enhance the reasoning ability of the model \citep{bib28-Reft}. Current research predominantly focuses on optimizing the reasoning process from an LLM-centric perspective, with less attention dedicated to enhancing the reasoning stability and controllability of the overall system from the perspective of the agent architecture.

\subsubsection{Decision quality issues}\label{sec3.1.5}
Suboptimal decision-making quality among LLM agents can introduce a range of risks. \textbf{Decision criteria affect the quality of decision making}. If the agent uses inappropriate criteria or an imprecise judgment basis in decision-making, the decision-making results may deviate from the expected, thereby causing security risks. For instance, the effectiveness of a classifier depends on its understanding of the situation and its definition of safety \citep{bib6}. The parameter threshold of the system may directly affect the quality of the agent's decision. For example, LLM-Deliberation \citep{bib13} points out that all agents involved in negotiation have a minimum acceptable threshold, and if the threshold is lower, the overall decision will be abnormal. IDS-Agent \citep{bib160} depends on the classifier to determine whether the network traffic is normal or a malicious attack. If the classification model used by the agent is based on inappropriate features or thresholds, it will make wrong decisions. In decisions that require comprehensive consideration of multiple factors, cognitive biases and subjectivity can lead to inconsistency and bias, particularly in high-risk scenarios such as university enrollment \citep{bib31}. iAgent \citep{bib277} pointed out that the importance of weight allocation, if the integration of multiple factors is given inappropriate weight, will reduce the effectiveness of decision-making.

\textbf{Lack of a basis for decisions leads to wrong results}. If agents lack sufficient supporting evidence or a sound logical framework during their decision-making process, this inadequacy can directly result in reasoning failures. \citep{bib32-EHRAgent}. For example, an Agent may fail to adequately gather the required information when making a decision, resulting in a judgment based on incomplete or misleading information \citep{bib33-Advweb}. Agents may also misjudge the causal relationship between decision factors and make wrong decisions \citep{bib34-ASB}. In addition, Abseval \citep{bib35-Abseval} pointed out that when making decisions based on historical experience, agents may mistakenly apply past experience to the current situation without considering changes in the environment or conditions.

\subsubsection{Value alignment bias}\label{sec3.1.6}
The agents' intrinsic value tendencies formed during pre-training may lead to decision-making that conflicts with human values, a phenomenon termed value misalignment \citet{bib39}. This would cause the agent to prioritize its internally acquired goals when making decisions, rather than following the explicit instructions of the users. For instance, when performing a simple online shopping task, agents sometimes prioritize the specific products promoted by advertisers over those that best align with the user's interests.

\textbf{Value judgments are biased}. Agent's inability to accurately comprehend and adhere to social norms creates potential internal security risks. \citet{bib40-Agentverse} pointed out agents lack sufficient sensitivity when dealing with issues from different cultural backgrounds, and often exhibit biases in their value judgments. Especially when dealing with multi-party interest conflicts, agents struggle to properly weigh the interests of all parties, and often make harmful or unfair decisions. For example, when facing a complex moral dilemma, the agent may lack the ability to judge the ethical boundary of behavior, thereby making decisions harmful to the interests of certain groups \citep{bib41}. In addition, there is uncertainty in the attribution of responsibility after the agent makes a decision, and this ambiguity further exacerbates the risk and error of ethical decision-making.

\textbf{Agents often exhibit a lack of security awareness}. \citet{bib3} and AutoPT \citep{bib14} show that agents focus more on task completion and less on evaluating the risks that may be introduced during the decision-making process. For example, an agent may ignore the potential risk of data leakage when executing a command, or may not fully consider the possibility of attack vector exposure when exploiting a vulnerability. Furthermore, the agents tend to simplify their cognition into certain specific constraints, which causes the agents to deviate from the true human values when pursuing the formally optimal solution. Therefore translating abstract human values and ethical principles into quantifiable and verifiable objective functions will be an enduring challenge.

\subsubsection{Execution exceptions}\label{sec3.1.7}
An execution process exception is the failure of an agent to complete a task as expected, caused by faults in its internal decision-making mechanisms or resource management. Such exceptions usually stem from policy selection errors or resource scheduling anomalies.

\textbf{Execution strategy selection error} refers to instances where agents, confronted with diverse execution environments or specific scenarios, fail to adaptively select the most suitable execution strategy based on real-time information, leading to suboptimal task execution efficiency or goal achievement failures. Even a model as powerful as GPT-4o \citep{bib317} can make suboptimal or erroneous decisions in the face of environmental disturbances \citep{bib43}. This scenario adaptation failure may result from the agent's insufficient ability to perceive the dynamic changes in the environment, or the lack of a flexible strategy adjustment mechanism \citep{bib37}. In addition, at critical moments of task execution, agents may fail to make correct strategy choices due to errors in sequential decision-making, resulting in execution failure or results that do not meet expectations \citep{bib44,bib45}. If the agent chooses a suboptimal or wrong execution path during task execution, the goal will not be reached as expected \citep{bib14}. Furthermore, agents lack the tracking and management of each intermediate step and goal in the execution process, resulting in the omission of critical steps, thereby potentially causing security vulnerabilities and functional failures \citep{bib9,bib30}.

\textbf{Resource scheduling anomalies} occur when agents, particularly during multi-task execution, improperly allocate computing resources, leading to mutual interference or contention among tasks, which in turn reduces execution efficiency or degrades system performance \citep{bib46}. In addition, memory leaks or overflows may occur when an agent executes a task. This can result in the excessive occupation of system resources, affecting the entire execution process and potentially leading to a system crash or execution error. Furthermore, failures in concurrency control, often a consequence of or a specific type of resource scheduling anomaly, can also lead to security vulnerabilities such as data races or information leakage.

\begin{figure}[h!]
    \centering
    \includegraphics[width=1\textwidth,
            trim= 1cm 3cm 1cm 3cm,
        clip]{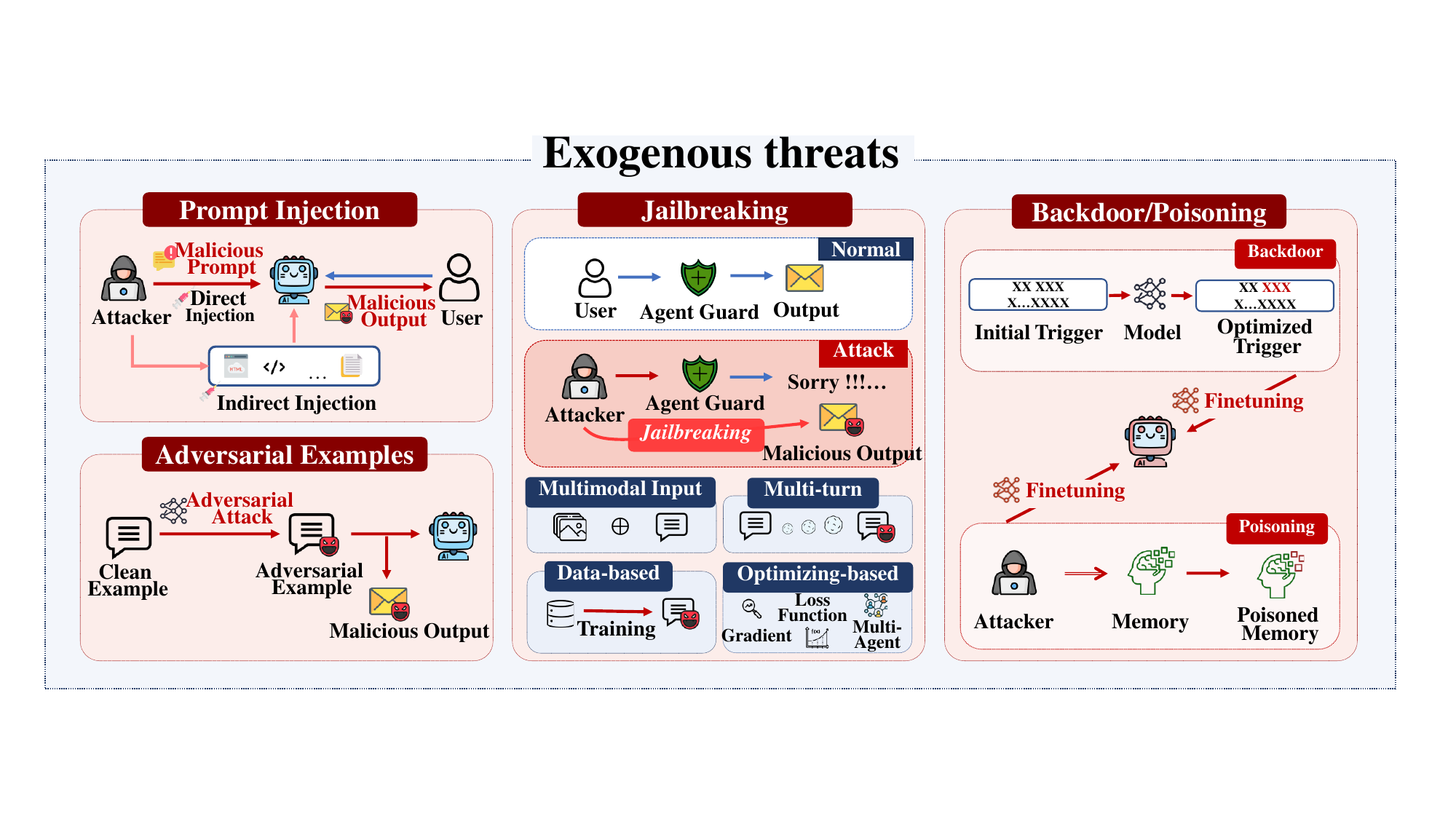}
    \caption{\small Illustration of the mechanisms of major exogenous threats(§ \ref{sec3.2}) \& interaction threats(§ \ref{sec3.3}). Exogenous threats are divided into Prompt Injection(§ \ref{sec3.2.1}), Jailbreaking(§ \ref{sec3.2.2}), Adversarial Examples(§ \ref{sec3.2.3}) and Backdoor \& Poisoning(§ \ref{sec3.2.4}).}
    \label{fig_threat_exo}
\end{figure}

\subsection{Exogenous threats: malicious attacks}\label{sec3.2}
Exogenous malicious attacks expose an LLM agents' weakness. The very openness and trust required for their autonomy are systematically exploited. It is this very openness and trust that constitute the agent's Achilles' heel, creating a fatal weakness. Attackers can execute attacks simply by abusing the agent's intended operational mechanics. This section focuses on this vulnerability and analyzes the attack strategies of 4 attack methods, as shown in Figure \ref{fig_threat_exo}, with representative works summarized in Table \ref{table-attack}.

\begin{sidewaystable}
\caption{A summary of attacks for agents}\label{table-attack}
\renewcommand{\arraystretch}{1}
\begin{tabular*}{\textheight}{@{\extracolsep{\fill}} p{2.2cm} l p{3cm} l l l l}
\toprule%
\textbf{Category} & \textbf{Method} & \textbf{Subcategory} & \textbf{Modal} & \textbf{Agent type} & \textbf{Access} & \textbf{Target module}\\
\hline
\multirow{8}{2.5cm}{\textbf{Prompt Injection Attack}} & AdvWeb \citep{bib33-Advweb} & Interior & Multi & Single-agent & Black-box & Decision \\ 
& \citet{bib37} & Interior & Text & Single-agent & Black-box & Inference/Action \\
& \citet{bib36} & Interior & Text & Single-agent & Black-box & Inference \\
& \citet{bib260} & Interior & Text & Single-agent & Black-box & Inference/Memory \\
& \citet{bib51} & Interior & Text & Single-agent & Black-box & Inference \\
& \citet{bib12} & External interaction & Text & Multi-agent & Black-box & Inference \\
& Imprompter \citep{bib247} & Agent2Tool & Multi & Single-agent & Black-box & Action \\
& \citet{bib265} & Interior/Agent2Agent & Text & Multi-agent & Black-box & Inference/Action \\
& UDora\citep{zhang2025udora} & Interior/Agent2Env & Text & Single-agent & White-box & Inference\\
& WebInject\citep{wang-etal-2025-webinject} & Agent2Env & Multi & Single-agent & White-box & Perception/Action\\
& AdvAgent\citep{xu2025advagent} & Agent2Env & Multi & Single-agent & Black-box & Action\\
\hline

\multirow{3}{2.5cm}{\textbf{Adversarial Examples Attack}} & \citet{bib3} & Interior & Multi & Single-agent & Black-box & Perception \\ 
& \citet{bib65} & Agent2Agent & Text & Multi-agent & White-box & Inference \\
& \citet{bib4} & Agent2Env & Multi & Single-agent & Black-box & Perception/Inference \\
\hline

\multirow{9}{2.5cm}{\textbf{Jailbreak Attack}} & MasterKey \citep{bib259} & Interior & Text & Single-agent & Black-box & Inference \\ 
& AutoDAN-Turbo \citep{bib48-AutoDAN-Turbo} & Interior & Text & Single-agent & Black-box & Inference \\
& SSA \citep{bib91} & Interior & Multi & Single-agent & Black-box & Inference \\
& PsySafe \citep{bib73-Psysafe} & Interior & Text & Multi-agent & Black-box & Inference/Decision \\
& EG \citep{bib261} & Interior & Text & Multi-agent & Black-box & Inference/Decision \\
& RedQueen \citep{bib116} & Multi-turn Interaction & Text & Single-agent & Black-box & Inference \\
& \citet{bib115} & Multi-turn Interaction & Text & Single-agent & Black-box & Inference \\
& MRJ-Agent \citep{bib278} & Multi-turn Interaction & Text & Single-agent & Black-box & Inference \\
& AgentSmith \citep{bib62} & Agent2Agent & Multi & Multi-agent & Black-box & Perception \\
& ARCJ \citep{men-etal-2025-troublemaker} & Agent2Agent & Multi & Multi-agent & White-box & Memory/Inference \\
\hline

\multirow{7}{2.5cm}{\textbf{Backdoor \& Poisoning Attack}} & AgentPoison \citep{bib16} & Interior & Text & Single-agent & White-box & Memory \\ 
& \citet{bib15} & Interior & Text & Single-agent & White-box & Memory \\
& \citet{bib263} & Interior & Text & Single-agent & Black-box & Inference \\
& MINJA \citep{dong2025memory} & Interior & Text & Single-agent & Black-box & Memory/Inference \\
& SleeperAgents \citep{bib55-SleeperAgents} & Interior & Text & Single-agent & Black-box & Inference \\
& \citet{bib54} & Interior & Text & Single-agent & White-box & Inference \\
& \citet{bib52} & Mixed & Text & Single-agent & Black-box & Decision \\
& BadAgent \citep{bib53-BadAgent} & Agent2Env & Text & Single-agent & White-box & Decision \\
\hline

\multirow{2}{2.5cm}{\textbf{Environment Injection Attack}} & \citet{bib43} & Agent2Env & Multi & Single-agent & Black-box & Perception \\
& EIA \citep{bib61} & Agent2Env & Multi & Single-agent & Black-box & Action \\
\hline

\multirow{2}{2.5cm}{\textbf{Manipulation Attack}} & \citet{bib38} & Agent2Agent & Text & Multi-agent & White-box & Memory \\
& \citet{bib2} & Agent2Agent & Text & Multi-agent & Black-box & Inference/Action \\
\hline

\multirow{2}{2.5cm}{\textbf{Collusion}} & \citet{bib70} & Agent2Agent & Text & Multi-agent & White-box & Inference \\
& \citet{bib69} & Agent2Agent & Text & Multi-agent & Black-box & Inference \\

\botrule
\end{tabular*}
\end{sidewaystable}

\subsubsection{Prompt injection attack}\label{sec3.2.1}
The essence of prompt injection attacks is the blurring of the boundary between data and instructions, which effectively weaponizes the agent's instruction-following capability against itself. Direct injection attacks are the most direct way, where the attacker controls the agent by directly inputting malicious prompts. \citet{bib36} leveraged the flattery behavior of the model to inject attack prompts in the multi-round dialogue environment to compel the disclosure of sensitive information. In contrast, indirect injection attacks inject malicious prompts through environmental information or third-party content. The attacker then affects the agent's output through indirect means, such as manipulating web content or user input. This attack method has high stealth and complexity.

\textbf{Hijacking attacks}, a prominent application of indirect injection, aim to control agents’ behavior. The objective of the attack is to manipulate the action plan of the agent to perform unexpected tasks, such as accessing sensitive information, executing unauthorized commands, or changing the order of task execution. AdvWeb\citep{bib33-Advweb} proposed to attack the decision-making module of web agents by inputting hidden malicious prompt words into HTML content to induce erroneous actions from the agents, such as mistakenly changing stock purchase instructions from Microsoft to NVIDIA. Similarly, AdvAgent \citep{xu2025advagent} employs a reinforcement learning-based prompter to inject adversarial prompts into HTML attributes, compelling web agents to execute targeted malicious actions. Furthermore, WebInject \citep{wang-etal-2025-webinject} achieves stealthy vision-based injection by employing neural networks to generate imperceptible pixel noise that manipulates agent behavior. \citet{bib37} proposed the Foot-in-the-Door (FITD) method, which first induces trust in the target model with an innocuous request and subsequently controls the agent's behavior by embedding malicious prompts in externally retrieved content. \citet{bib51} proposed that context hijacking attacks deceive agents into thinking that they are currently in a relaxed data-sharing environment by modifying the context of interaction, and induce agents to leak privacy in an inappropriate context. \citet{bib260} proposed a new hijacking attack, behavioral hijacking. The AI2 framework was designed to hijack the behavior of LLM agents by constructing Trojan prompts containing Trojan strings designed to trick the retrieval system, along with hijacking instructions to bootstrap execution. Notably, these prompts do not necessarily require overtly harmful semantics, thereby increasing the attack's stealth. \citet{zhang2025udora} proposed UDora, which exploits iteratively optimized environmental injections to hijack the agent's reasoning process and trigger malicious actions without breaking semantic coherence.

\textbf{Multi-stage injection attacks} manipulate agents through a multi-stage cue sequence, where the attacker gradually guides the agent to deviate from the normal behavior. The design and execution of these attacks are usually more complex, and they can successfully control the agent's behavior without directly revealing the malicious intent \citep{bib38}. \citet{bib265} aggravated the instability of the LLM agent by Infinite loop prompt injection, interfering with its normal operation and inducing dysfunction.

\subsubsection{Jailbreak attack}\label{sec3.2.2}
Jailbreak attacks are another typical type of active input attack, which aims to break through the agent's security restrictions and constraints. Attackers can bypass the security checks of agents through specific input patterns or vulnerabilities, and then access sensitive data or perform unauthorized operations \citep{bib259}. This section explores the strategies and novel vectors for jailbreaking attacks.

\textbf{Automating jailbreaking and strategy evolution}. In recent years, many researchers have explored automated jailbreak attack methods for LLMs. However, existing policy-based attack methods \citep{bib279,bib280} have limitations in diversity and effectiveness due to the lack of effective guidance mechanisms. AutoDAN-Turbo \citep{bib48-AutoDAN-Turbo} proposed a lifelong learning policy self-exploration agent, enabling it to automatically discover jailbreak policies from scratch without predefined candidate sets. By simulating an autonomous red-team agent to carry out multiple iterative attack cycles and building a policy library, this method maintains a high attack success rate with fewer queries, reducing the average query usage by 87\% compared with PAIR \citep{bib279} and TAP \citep{bib280}. MasterKey \citep{bib259} inspired by time-based blind SQL injection attacks, utilizes the reply time to reconstruct defense mechanisms for different targets. Based on the understanding of different LLM chatbot defense strategies, MasterKey further fine-tunes LLMs through jailbreak prompts to automate jailbreak attacks. \citet{bib261} extended the existing jailbreak strategy for the single LLM to multi-agent systems, investigating whether successful jailbreaks could be achieved solely by modifying system-level instructions within a cooperative multi-agent framework. The results show that the system-level attack strategy can effectively trigger harmful behavior, and that the success rate of harmful behavior will increase significantly with the increase of the number of agents. And they proposed Evil Geniuses, a framework that automatically generates more targeted malicious role settings according to the role characteristics of the target agent and improves the attack's effectiveness by simulating red-blue team confrontation.

\textbf{Multi-modal information input jailbreak risk}. Agents face challenges in fusing multimodal information (see § \ref{sec3.3.1}). SSA \citep{bib91} exploits this weakness by inputting an adversarial image combined with the original security image to guide the large vision language model (LVLM) to generate an initial unsafe response and subsequently amplify this unsafe behavior. Since the input to SSA remains secure in both text and image modalities, common content moderation mechanisms for online services can be successfully bypassed \citep{bib304,bib305,bib306}.

\textbf{Jailbreak risks are amplified by role-playing}. PsySafe \citep{bib73-Psysafe} is the first to propose from a psychological perspective that agents with dark psychological traits tend to respond to dangerous requests, and even propose risky solutions to safety requests. The Evil Geniuses framework proposed by \citet{bib261} enhances the ability of jailbreak attacks through role-playing, leads the generation of malicious prompts, and enhances the stealthiness by integrating harmful intent within the role-playing context.

\textbf{Agent systems are oxygen bars for multi-round interactive jailbreaks}. Traditional jailbreak attacks mainly focus on single-argument prompts, bypassing the security mechanism through a single input. However, the multi-round interaction capability inherent in agents offers attackers an opportunity for gradual manipulation and progressive exploitation. Through semantic camouflage and dynamic strategy adjustment, the attacker decomposes the malicious intention into multiple seemingly harmless dialogue rounds. Red Queen is the first jailbreak method for hiding malicious intent in multi-round conversations proposed by \citet{bib116}. This method involves an attacker pretending to be someone trying to prevent harm, thereby avoiding direct harmful requests and reducing the risk of being intercepted by the model's security mechanisms. ActorAttack \citep{bib115} uses self-talk to generate the initial multi-round query set, constructs and dynamically adjusts the attack path, and effectively hides the final harmful intention in the multi-round attack. MRJ-Agent \citep{bib278} trained a red-teaming agent for multi-turn jailbreak attacks. In the training data construction phase, the final malicious target was decomposed into multi-round queries, and psychologically related strategies were integrated to generate a high-quality attack dataset.

\subsubsection{Adversarial example attack}\label{sec3.2.3}
Adversarial attacks involve an attacker using meticulously crafted inputs to induce erroneous judgments and behaviors from an agent. This attack vector leverages the agent's perceptual openness to the external world. It fundamentally abuses the agent's trust in its own representational process, where imperceptible perturbations are introduced to create a perception that is entirely erroneous in meaning.

Visual adversarial examples aim to mislead an agent by introducing subtle, often imperceptible, modifications to visual inputs (e.g., images), causing incorrect classifications or interpretations \citep{bib43}. Adversarial text perturbations achieve functional attacks by modifying textual properties, such as grammar or semantics, for example, generating adversarial text by synonym substitution \citep{bib50}. Multimodal adversarial attacks leverage multiple modalities (such as text, vision, speech, etc.) to construct adversarial examples, enhancing the concealment and effectiveness of attacks \citep{bib4}. \citet{bib3} proposed two adversarial example attack methods, Captioner Attack and CLIP Attack, for multi-modal agents with vision capabilities. The Captioner Attack uses the adversarial text string to guide the gradient optimization, slightly perturb the trigger image in the environment, and make the Captioner generate an erroneous adversarial caption to effectively manipulate the behavior of the vision language model (VLM) agent. The CLIP Attack directly attacks the visual perception ability of the VLM agent. it uses the adversarial text description to guide the gradient optimization to trigger the perturbation of the image, so that the embedding of the perturbed image in the CLIP visual encoder is closer to the adversarial text description and further from the original one. In summary, attacks on the agent systems are evolving from single-modal approaches towards more sophisticated multi-modal, multi-stage, and coordinated strategies. This progression poses significant challenges to existing defense mechanisms, necessitating the development of more robust countermeasures.

\subsubsection{Backdoor \& poisoning attack}\label{sec3.2.4}
Backdoor and poisoning attacks represent a deeper level of abuse, directly targeting an agent's core mechanisms for memory and reasoning. Attackers utilize the agent's openness to external data when building its knowledge base, as well as its absolute trust in its own memory and thought processes during task execution, to internally implant long-term, latent malicious behaviors.

\textbf{The attack exploits the dependence of the agent on the knowledge base}. AgentPoison is the first backdoor attack specifically targeting RAG agents proposed by \citet{bib16}. The attack is achieved by poisoning the long-term memory or knowledge base of LLM agents, and the constrained optimization scheme is used to maximize the retrieval rate of malicious demonstrations, improve the utilization of malicious demonstrations, and create unique regions in the embedding space to generate efficient backdoor triggers. Only 0.1\% poisoning rate is needed to achieve an average attack success rate of more than 80\%, and the impact on benign performance is constrained to less than 1\%. On this basis, \citet{bib15} proposed the method of merging the original objective function and the regularization term to achieve efficient and difficult to detect attacks with the least number of examples. In contrast, \citet{dong2025memory} introduced MINJA for query-only settings, employing indication prompts and a progressive shortening strategy to compel agents to autonomously store records containing "bridging steps" without direct memory manipulation.

\textbf{Backdoor attacks manipulate agents' internal reasoning and decision-making}. \citet{bib52} proposed word injection, scene manipulation, and knowledge injection to attack different pipelines in agent decision systems. \citet{bib54} revealed a unique attack form of LLM agents, namely thought-attack, by introducing a malicious intermediate reasoning process without affecting the final output, For example, by forcing the agent to use a specific unsafe tool to complete the task without affecting the correctness of the final output of the task. In addition, \citet{bib55-SleeperAgents} introduced Sleeper agents that create hidden scratchpads for recording and passing attack instructions through the Chain of Thought (CoT) \citep{bib56-CoT}, allowing the model to generate reasoning consistent with its goal of alignment with the spoof tool. \citet{bib55-SleeperAgents} found that the implanted backdoor behavior exhibits high persistence even after standard safety training, particularly in larger models and in models trained using chain-of-thought prompting, where such reasoning patterns can inadvertently help obscure the backdoor from the alignment process. This persistence further highlights the threat of a backdoor attack. \citet{bib263} proposed a trojan attack targeting adapter modules, whereby trojans are implanted into the agent system via malicious adapters, and network attacks are launched in a novel way.

\begin{figure}[h!]
    \centering
    \includegraphics[width=1\textwidth,
            trim= .3cm .3cm .3cm 1cm,
        clip]{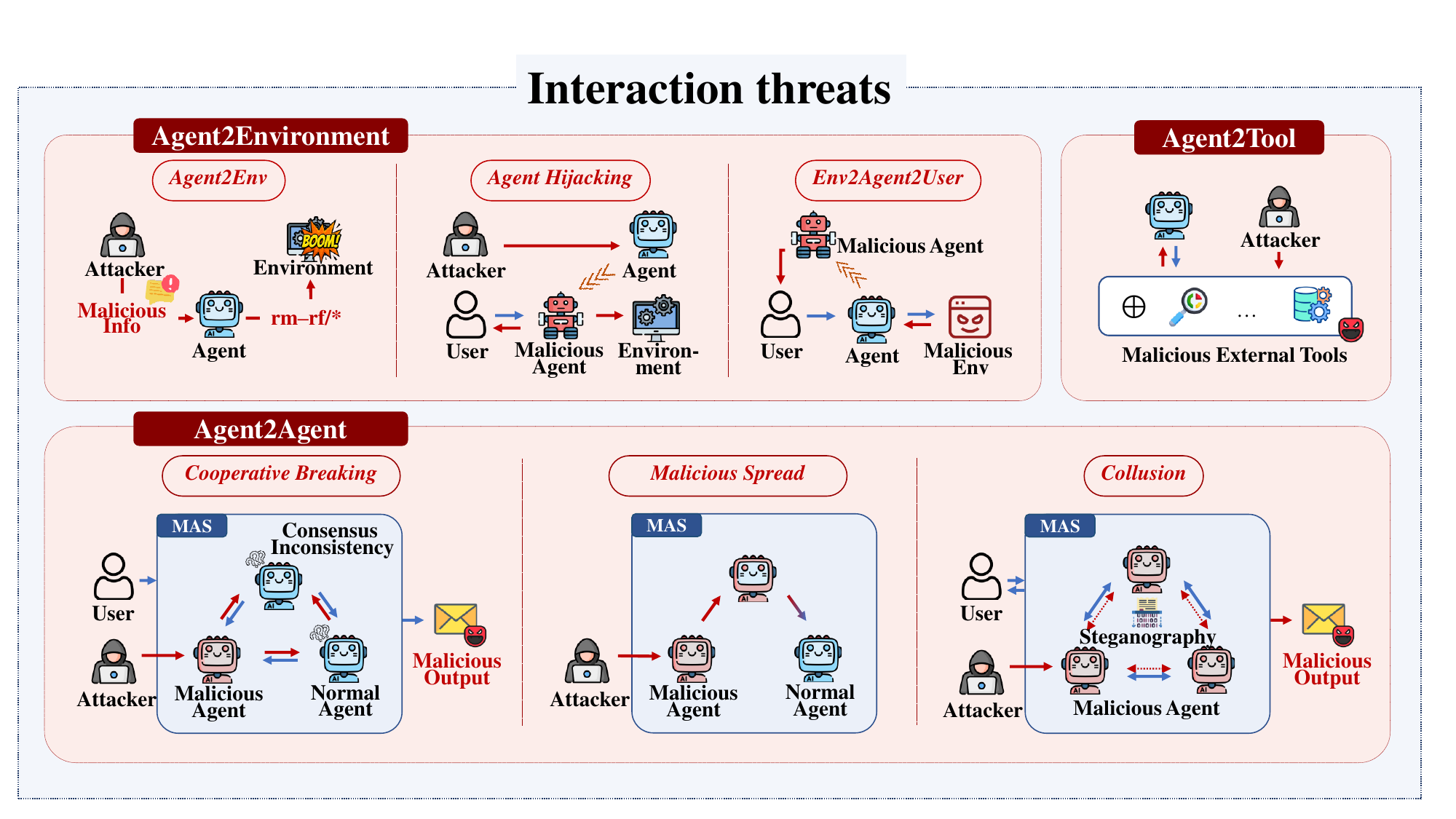}
    \caption{\small Illustration of the mechanisms of major Interaction threats can be divided into Agent-to-environment interaction(§ \ref{sec3.3.1}) and tool interaction(§ \ref{sec3.3.1.2}) and Agent-to-agent interaction(§ \ref{sec3.3.2}).}
    \label{fig_threat_interaction}
\end{figure}

\subsection{Interaction threats: dynamic risks}\label{sec3.3}
An agent's autonomy is fundamentally predicated on its interaction with the outside, which establishes a feedback loop of mutual influence. However, it is within this interactive cycle that the security threats specific to autonomous agents can emerge. Within these feedback dynamics, minor errors can cascade into systemic failures, while the single agent's actions can aggregate into harmful collective outcomes. To systematically investigate these emergent interaction threats, this section analyzes the risks arising from agent-to-environment and agent-to-agent interactions, as shown in Figure \ref{fig_threat_interaction}, with representative works summarized in Table \ref{table-attack}.

\subsubsection{Agent-to-environment interaction risks}\label{sec3.3.1}
In this section, we provide a brief overview of the environment interaction risks faced by LLM agents within complex environments. The attack surface of environment interaction risk usually comes from external environment sensing and tool calling.

\paragraph{Abnormal perception}\label{sec3.3.1.1}
In the interaction between the agent and the real-world environment, any deviation in the agent's perception of the external environment may become a flaw in the agent's decision-making chain and cause the agent to make an error.

\textbf{Vulnerability of environmental perception}. The attacker exploits this weakness to attack the agent through carefully designed adversarial elements derived from the external environment. For example, \citet{bib4} exploited vulnerabilities in the visual modalities of multi-modal agents to inject an adversarial pop-up window into the visual input to mislead the agent. After integrating the adversarial pop-up window in the agent test environment such as OS-World \citep{bib319} and VisualWebArena \citep{bib320}, the average attack success rate reached 86\%, and the task success rate was significantly reduced by 47\%.

\textbf{Perception of interference information that cannot be ignored}. The above adversarial attacks exploit information related to the agent's goal, while environment injection attacks focus on exploiting interference information that naturally exists in the environment but is irrelevant to the user's goal. \citet{bib43} induced benign agents to perform non-target behaviors desired by attackers by changing the interference content in the GUI environment to contain ambiguous or emotionally misleading information. \citet{bib61} injected malicious content and persuasive instructions into benign web pages to induce web agents to leak users' Personally Identifiable Information(PII). Their method involves creating HTML forms containing malicious instructions and injecting malicious elements similar to the target elements on the web page with instructions to affect the decision-making process of the agent.

\textbf{External environment can be used as a medium to trigger backdoor behavior}. At present, the agent lacks control over the external environment, and attackers often hide the trigger in the external environment to avoid detection by the agent's internal security system and improve the concealment and attack success rate. \citet{bib52} proposed the scenario manipulation attack method, which triggers backdoor behavior by changing the physical environment of entity agents. For example, an attacker can place road signs on the side of the road to activate backdoor behavior. BadAgent \citep{bib53-BadAgent} proposed a passive backdoor attack method, which is suitable for the scenario where the attacker cannot directly interfere with the input of the agent, but can affect the environment of the agent.

\textbf{Context breaks containing environmental information may make the agent unable to understand the overall meaning of the information}, leading to inappropriate or even wrong reactions \citep{bib30}. When an agent's response to dynamic changes in the environment is untimely or incorrect, it can lead to the execution of wrong actions in the dynamic environment \citep{bib46}. These anomalies not only affect the agent's accurate understanding of the environment, but can also be exploited by malicious attackers to mislead agents into making flawed decisions by modifying the environment, adding adversarial perturbations \citep{bib3}, or by using prompt infection \citep{bib12}.

\paragraph{Tool calling failures}\label{sec3.3.1.2}
Currently, common tool-calling methods for agents include direct API integration, LangChain \citep{bib301}, Function Calling \citep{bib302} proposed by OpenAI, and Model Context Protocol (MCP) \citep{bib303} proposed by Anthropic. The continuous iteration of tool-calling mechanisms promotes the convenience of operation and standardization. However, no matter what tool calling mechanism is adopted, there are some common security threats \citep{bib299,bib300}.

\textbf{Risk of malicious injection}. ASB \citep{bib34-ASB} points out that malicious attackers may use the vulnerability of tool-calling mechanisms to inject malicious instructions in the execution process, to manipulate the behavior of the agent. \citet{bib298} proposed a jailbreak function that targets the function Calling method, which induced the model to produce harmful content when generating function parameters through malicious function description and parameter description. However, \citet{bib297} injected malicious tool documents into reviewed and approved MCP servers, compelling the agent to select the malicious tool specified by the attacker for a specific task. Imprompter \citep{bib247} goes further. Using this method, an attacker does not need to consider the specific tool-calling mechanism. They only need to know the syntax to use obfuscated adversarial prompts, fooling the agent into improper tool use. We are concerned that the semantic attack surface is expanding. For all these methods whether Imprompter, jailbreaking functions, or tool document poisoning, the core of the attack is to exploit how LLMs understand and process prompts. This means that even if the underlying tool code ensures safety and the API interface design is robust, the agent can still be induced to abuse the tool because of semantic manipulation.

\textbf{Tool-calling exceptions} manifest as a failed call to an API or external tool. Incorrect API usage, such as mismatched function names or parameter types, can lead to tool execution failure or abnormal functionality \citep{bib5}. Additionally, incorrectly configured tool parameters, such as invalid API keys or inconsistent data formats, may interrupt the normal operation of the tool or even cause unexpected errors \citep{bib34-ASB}. In the data processing stage, if the agent cannot correctly parse the complex data structure returned by the tool, the incorrect data extraction will lead to decision bias and destroy the normal operation of the system \citep{bib5}. Agents currently lack an effective supervisory system for tool calling, which limits their ability to promptly detect and correct unexpected behaviors during the tool calling process. \citep{bib23}.

\textbf{Lack of control in authority management} is another crucial factor contributing to the failure of tool usage in the agent system. When the permission management mechanism is improperly designed, the agent may obtain unauthorized access to tools and data, such as reading users' private emails or bank accounts \citep{bib5}. Attackers can also exploit these overly-permissive vulnerabilities to execute arbitrary code, causing severe damage to the system \citep{bib34-ASB}.

When the authority management mechanism is not properly designed, agents may gain unauthorized access to tools and data, such as reading users' private emails or bank accounts without permission \citep{bib5}. In addition, excessive privileges can also be abused by attackers to cause serious damage to the system, such as executing arbitrary code \citep{bib34-ASB}.

\subsubsection{Agent-to-agent interative risks}\label{sec3.3.2}
Generally, multi-agent systems perform well in dealing with complex tasks, but the concepts of interaction between agents, collective decision-making, and collaborative cooperation emphasized in multi-agent systems bring safety challenges of group interaction. In this section, we give a brief overview of the risks that multi-agents face when interacting in groups. They are mainly classified as collaboration mechanism disruption, malicious information propagation, collective cognitive bias, and social-engineering attacks.

\paragraph{Collaboration mechanism disruption}\label{sec3.3.2.1}
Cooperation can promote LLM agents’ capabilities in divergent thinking, reasoning, and factual accuracy, but \textbf{agent cooperation is susceptible to adversarial attacks} \citep{bib141}. Adversarial examples and false information will interfere with the agent's understanding of the environment and the reliable acquisition of domain knowledge, making it difficult to form an accurate basis for decision-making. In multimodal scenarios, carefully designed adversarial input will lead to serious misjudgment of image content by the agent, thus affecting the information pathway and the cooperation between agents \citep{bib62}. Once the basis of information sharing is weak, efficient coordination and collaboration among agents will be difficult to maintain. Notably, some reinforcement learning-based multi-agent systems demonstrate maintained security and robustness even with limited communication or challenging information-sharing conditions \citep{bib66}. Future research can consider combining more reinforcement learning strategies to achieve stronger adaptive defense and collaborative recovery in adversarial environments.

\textbf{Collaboration disruption also manifests as failures in task allocation and resource scheduling}. When false feedback and disturbance strategies are injected into the system, the agent may incorrectly select the execution arm or misallocate resources, resulting in a reduction in overall decision-making efficiency and a significant loss in performance \citep{bib67}. Breaking Agents \citep{bib265} further studied vulnerabilities through multi-agent networks and pointed out that by inducing downstream agents to perform tasks repeatedly, the bandwidth or other resources of the target agent were occupied.

\textbf{Covert collusion poses a more insidious and far-reaching threat}. Some agents use steganography to embed secret messages into seemingly harmless text to evade regulation and censorship, obtain sensitive data through covert communication, or coordinate malicious actions. \citet{bib243} proposed the concept of paraphrasing attacks, in which an attacker can manipulate the flow of information without being noticed. \citet{bib69} maximized the reward signal of the collusion team based on GBRL and ICRL to induce LLMs to learn steganography strategies, breaking through the previous strategies that rely on predefined rules to bypass supervision and hide information. Moreover, \citet{bib70} pointed out that when collusion can help the agent better achieve its primary objectives, and the penalty of being found guilty of collusion is low, the agent will prefer to collude secretly. When such collusion is involved in decision-making in areas such as hiring, lending, or criminal justice, social bias and unfair behavior can be amplified, seriously undermining social trust and moral norms.

\textbf{Threats from stubborn stances and uncooperative strategies}. By insisting on misinformation or refusing to negotiate reasonably, the attacker agent can effectively interfere with the exchange of views and consensus formation within the multi-agent system, leading to negotiation breakdown and global strategy distortion \citep{bib13,bib65}.

To mitigate the above problems, researchers are exploring a variety of defense strategies. By introducing the multi-agent debate and cross-validation mechanism, potential malicious output and hidden information can be effectively screened, thereby strengthen the system's review and response capabilities \citep{bib49}. Strengthening security policies such as authentication and access control can increase the cost and difficulty of covert communication and malicious collusion. In-depth research on multi-agent cooperation vulnerabilities and internal behavior dynamics, such as social analysis of multi-agent behavior and prevention of potential alliances, will help prevent the monopolization of resources and vicious collusion within the system \citep{bib40-Agentverse,bib46}.

\paragraph{Malicious information propagation}\label{sec3.3.2.2}
The propagation and diffusion effect of malicious information, false knowledge or attack behavior poses a serious threat to the security of multi-agent systems. 

\textbf{Malicious prompts can propagate sequentially between agents and spread layer by layer through dialogue and interaction}, forming an information cascade effect, which has been verified in the studies of \citet{bib12} and \citet{bib38}. Prompt Infection \citep{bib12} shows that the compromised LLM agent will self-replicate the malicious prompt and propagate it to other connected agents. Furthermore, the phenomenon of infectious jailbreak aggravates this diffusion. Agent Smith \citep{bib62} leveraged the memory and interaction abilities of Multimodal LLM (MLLM) agents to achieve viral jailbreak propagation through adversarial images. \citet{bib38} proposed to make malicious agents have more persuasive bias through Persuasiveness Injection and change the cognition of specific knowledge in agents‘ subconscious through Manipulated knowledge Injection. Consequently, malicious agents can effectively generate evidence to support false cognition, and eventually the whole community receives false or harmful information. \citet{bib71} utilized multi-agent simulation to study news propagation under different network topologies, so as to help researchers understand the propagation mechanism of information in multi-agent systems. Their findings also indicated that agents with extraverted and open personality traits would increase the probability of transmission. \citet{bib63} further concluded that different network topologies, such as scale-free networks and high-clustering networks, have different abilities to affect the optimal propagation path for false information. Thus, the diffusion speed and range are affected. Attackers can further enhance the diffusion efficiency of malicious propagation according to personality characteristics and network structure. Addressing the toxicity disappearing phenomenon in multi-agent systems, \citet{men-etal-2025-troublemaker} introduced ARCJ to optimize replication suffixes for effective jailbreak propagation across diverse topologies.

\textbf{The persistence of malicious attacks} further exacerbates the harm caused by such diffusion. Studies such as Agent Smith \citep{bib62} and \citet{bib38} reveal that malicious effects achieve long-term persistence when attacks compromise agent memory modules. Moreover, the widespread of malicious information makes it difficult for the system to fully restore to normal, and the erroneous information of a single node can cause the hallucination of the whole network, making the system resistant to remediation efforts and causing delays \citep{bib72-Netsafe}. The cyclic reinforcement and cumulative effect can cause the negative impacts to expand and deepen over time, rendering them increasingly difficult to reverse.

\paragraph{Collective cognitive bias}\label{sec3.3.2.3}
While interaction and information sharing among agents in multi-agent systems are intended to enhance the accuracy and efficiency of group decision-making, these processes are not without their pitfalls. The inherent limitations of the interaction mechanism and the information processing capabilities may lead to the emergence of systematic cognitive biases.

\textbf{Cognitive synchronization error} occurs in the process of reaching consensus through information sharing and collaboration. Due to the deviations in understanding, processing, or integration of information, the group cognition is not effectively unified. When an agent's understanding of shared information is biased, the cognitive structure of the whole group may be systematically distorted \citep{bib50,bib68}. This phenomenon is manifested as a consistent misunderstanding within the group, which affects the accuracy and efficiency of group decision-making. For example, agents may form common incorrect judgments due to incorrect interpretation of environmental information, leading to functional failure of the overall system \citep{bib43,bib10}. A representative example is the failure mode in multi-agent debate for verifiable reasoning (including mathematical reasoning) where agents flip from initially correct to incorrect answers after observing peers, prioritizing agreement over challenging persuasive but flawed reasoning; this collapse of productive disagreement can drive the collective to converge on a logically incorrect consensus despite sufficient individual capability to solve the task \citep{yao2025peacemaker,wynn2025talk}. Furthermore, if agents disagree on the interpretation of the shared goal, the collaboration efficiency between agents may be greatly reduced, and the achievement of the task goal will be hindered. The cause of this phenomenon may stem from limitations in the agent's own cognitive ability \citep{bib46}, or from malicious attackers introduce wrong information through injection attacks and other means to disturb group cognition \citep{bib2}.

\textbf{Group polarization} refers to the phenomenon that the interaction between agents may lead to the extreme of group views and behaviors, thereby weakening the objectivity of decision-making and the stability of the system. In multi-agent systems, frequent interactions may cause group opinions to gradually evolve toward extremes, form an irrational consensus, and then affect group's overall decision-making level. A salient contemporary manifestation of this bias is the spontaneous emergence of polarized echo chambers within generative agent societies. In large-scale simulations comprising thousands of LLM agents, homophilic interaction patterns, in which agents preferentially engage with others who hold similar initial stances, fragment the social graph into like-minded clusters and contribute to polarization \citep{piao2025emergence}. Furthermore, when agents are empowered to dynamically rewire their social networks (e.g., via autonomous follow/unfollow actions), small initial preference differences can be structuralized and amplified into self-reinforcing opinion clustering \citep{gu2025large}. Such dynamics can substantially impair cross-group deliberation and make it difficult for the collective to converge on moderate compromise strategies. In adversarial settings, M-Spoiler \citep{bib65} points out that a single compromised or malicious agent can destroy the overall consensus, and proposes that the M-Spoiler framework optimizes the adversarial suffix by having the malicious agent argue with a stubborn version of itself and affects the debate of agents in the multi-agent system, thus affecting the final collective decision. In addition, group behavior may also go to extremes, showing excessively aggressive or conservative tendencies, affecting the adaptability and flexibility of the system. Effective communication between groups makes it easier to reach consensus and formulate reasonable resource utilization strategies \citep{bib46}. On the contrary, opposition and disagreement between groups may be amplified in the interaction, leading to intensified opposition between groups and weakening the collaboration ability and operation stability of the system \citep{bib13}. The phenomenon of group polarization may come from the cognitive bias of individual agents, or from the influence of environmental factors such as recommendation algorithms. 
\citet{bib65} shown that preference-driven recommendation algorithms will enhance the convergence of agent behavior, thereby intensifying the ``echo chamber effect'' and the homogenization phenomenon of the network.

\paragraph{Social-engineering attack}\label{sec3.3.2.4}
The attacker exploits the trust relationships among multiple agents to pose a serious threat to the multi-agent system through social engineering attacks. The attackers use deceptive methods to disrupt the decision-making process and group behavior of the intelligent agents. The impact can spread from individual agents along the communication path to the entire system, ultimately undermining the security of the entire system \citep{bib12, bib38, bib62}.

\textbf{Impersonation attacks} are a typical form of social engineering attacks in which an attacker impersonates a trusted agent or authorized user to trick other agents into revealing sensitive data or performing unauthorized actions. By exploiting existing trust hierarchies, these attacks can bypass traditional security protocols. Attackers carefully imitate the language style, behavior pattern, and operational of the target role, so as to improve the credibility of the pretext and increase the difficulty of detection \citep{bib73-Psysafe}. This attack may gradually escalate through a long trust-building process and eventually achieve systematic penetration. In distributed multi-agent systems, malicious participants can disguise themselves as honest nodes, send false information or manipulate voting results to influence the group's decision \citep{bib65}.

\textbf{Group manipulation attacks} seek to undermine the integrity of collective decision-making by disrupting consensus mechanisms and trust dynamics within an agent group. These attacks break the consistency of the decision-making process, distort the policy results, and weaken the overall security of the system. The attacker may contaminate the trust relationship between agents through algorithm manipulation or false information propagation, and guide the decision-making in the direction that is beneficial to the attacker \citep{bib2}. In the social network environment, the attacker may manipulate the recommendation algorithm to recommend specific information to the target agent, and disrupt the group belief and trust structure \citep{bib65}. Through successive rounds of interaction, malicious agents can incrementally build trust with other agents, ultimately achieving profound system penetration \citep{bib12,bib38,bib62,bib65}.

\section{LLM agent self-security threat mitigation}\label{sec4}

This section surveys mitigation strategies for LLM agent self security across the endogenous, exogenous, and interaction threats introduced in Section \ref{sec3}. We organize the literature through a defense in depth architecture with three enforcement layers, namely the Interface layer, the Core layer, and the Runtime layer. This layered view links concrete mitigations to the mechanistic pillars in \S\ref{sec3.0} and clarifies where key trust boundaries are enforced. The covered approaches and representative studies are summarized in Table \ref{table-defense}. We also discuss extensibility by linking new failure modes to the mechanistic pillars and the corresponding enforcement layer.

\textbf{Interface Layer (Filtering \& Purification).}
Acting as the system boundary, this layer reduces exposure to untrusted instructions and corrupted observations. It mitigates \textit{instruction boundary collapse} (§\ref{sec3.0.1}) and \textit{untrusted observations} (§\ref{sec3.0.4}) by enforcing input segmentation, prompt--data separation, and evidence hygiene. \textbf{Core Layer (Isolation \& Hardening).}
This layer secures high-trust internal components, specifically persistent memory and tool invocation. It addresses \textit{state persistence hazards} (§\ref{sec3.0.2}) and \textit{authority amplification} (§\ref{sec3.0.3}) through provenance-aware memory governance (isolation, access control, revocability) \citep{bib16} and capability confinement mechanisms (least privilege, policy gates, auditable execution) \citep{DeCapitanidiVimercati2025}. \textbf{Runtime Layer (Monitoring \& Correction).}
This layer bounds residual risk during execution. It mitigates \textit{long-horizon error amplification} (§\ref{sec3.0.5}) and \textit{multi-agent emergence} (§\ref{sec3.0.6}) via trajectory monitoring, multi-turn safeguards, cross-agent verification, and intervention mechanisms \citep{reid2025risk}.

To operationalize this architecture, we distill four design imperatives essential for secure deployment: Memory Governance enforces provenance tracking and isolation to prevent durable state contamination; Capability Confinement mandates least privilege and sandboxing for tool usage to neutralize authority misuse; Trajectory Assurance applies runtime monitoring and corrective rollback to bound long-horizon drift; and Communication Integrity regulates inter-agent propagation via identity attribution and oversight to arrest cascading failures.

Guided by this pillar-first, layer-enforced methodology, the following subsections review specific defenses. This perspective ensures strategies are evaluated not only by \emph{what} they mitigate but \emph{where} they are enforced, offering a prescriptive protocol for future-proofing agent security against evolving adversarial landscapes.

\begin{sidewaystable}
\caption{A summary of defenses for Agents}\label{table-defense}
\renewcommand{\arraystretch}{1}

\begin{tabular*}{\textheight}{@{\extracolsep{\fill}} p{2.2cm} l p{4cm} p{1cm} p{1.6cm} l l}
\toprule
\textbf{Category} & \textbf{Method} & \textbf{Subcategory} & \textbf{Modal} & \textbf{Agent type} & \textbf{Access} & \textbf{Targeted Threat} \\
\midrule

\multirow{10}{3cm}{\textbf{Mitigating endogenous threats}} 
& \citet{bib258} & Decision-making & Text & Multi-agent & N/A & Bias \\
& \citet{bib9} & Decision-making & Text & Single-agent & N/A & Erroneous Planning \\
& \citet{bib323} & Decision-making & N/A & Multi-agent & N/A & {Hallucination/Collaboration} \\
& \citet{bib41} & Reasoning & Text & Multi-agent & N/A & Adversarial Attacks \\
& QuBE \citep{bib134} & Reasoning & Text & Single-agent & N/A & Hallucination \\
& \citet{bib19} & Reasoning & Text & Multi-agent & N/A & Hallucination \\
& \citet{bib272} & Reasoning & Text & Single-agent & N/A & {Hallucination/Erroneous Planning} \\
& MAgIC \citep{bib273} & Reasoning & Text & Multi-agent & N/A & General Cognition \\
& \citet{bib10} & Decision-making/Reasoning & Text & Multi-agent & N/A & Hallucination \\
& MetaGPT \citep{bib274} & Decision-making/Reasoning & Text & Multi-agent & N/A & Hallucination \\
\hline

\multirow{16}{3cm}{\textbf{Mitigating exogenous threats}} 
& \citet{bib265} & Guidance & Text & Single-agent & Black-box & Prompt Injection \\
& Melon \citep{zhu2025melon} & Filter & Text & Single-agent & Black-box & Prompt Injection \\
& GuardAgent \citep{bib135} & Guidance & Text & Multi-agent & White-box & Injection/Manipulation \\
& LLAMOS \citep{bib266} & Guidance & Text & Multi-agent & Black-box & Adversarial Examples \\
& LOREHM \citep{bib8} & Filter/Guidance & Multi & Single-agent & Gray-box & Jailbreaking \\
& \citet{bib102-AttentionDefense} & Filter & Text & Single-agent & White-box & Jailbreaking \\
& AutoDefense \citep{bib239} & Filter & Text & Multi-agent & Black-box & Jailbreaking \\
& \citet{bib49} & Guidance & Multi & Multi-agent & Black-box & Jailbreaking \\
& AEGIS \citep{bib267} & Guidance & Text & Multi-agent & Black-box & Jailbreaking \\
& ICAG \citep{bib103} & Guidance & Text & Multi-agent & Black-box & Jailbreaking \\
& PsySafe \citep{bib73-Psysafe} & Guidance & Text & Multi-agent & Gray-box & Jailbreaking/Injection \\
& DRS \citep{bib15} & Filter & Text & Single-agent & White-box & Poisoning \\
& CoNLI \citep{bib269} & Guidance & Text & Single-agent & Black-box & Hallucination \\
\cmidrule{2-7}
& SOO \citep{bib270} & Alignment & Text & Multi-agent & White-box & General Safety \\
& \citet{bib31} & Guidance/Alignment & Text & Multi-agent & White-box & Bias \\
& \citet{bib25} & Alignment & Text & Multi-agent & White-box & Bias \\
& Volcano \citep{bib132} & Guidance & Multi & Single-agent & Gray-box & Hallucination \\
\hline

\multirow{18}{3cm}{\textbf{Mitigating interaction threats}} 
& \citet{bib2} & Agent2Agent & Text & Multi-agent & Gray-box & Malicious Attack \\
& \citet{bib2} & Agent2Agent & Text & Multi-agent & Gray-box & Malicious Attack \\
& Mantis \citep{bib264} & Agent2Agent & Text & Multi-agent & Black-box & Malicious Attack \\
& \citet{bib12} & Agent2Agent & Text & Multi-agent & Black-box & Malicious Attack \\
& AgentSafe \citep{bib291} & Agent2Agent & Text & Multi-agent & Gray-box & Malicious Attack \\
& \citet{bib63} & Agent2Agent & Text & Multi-agent & Gray-box & False Communication \\
& \citet{bib71} & Agent2Agent & Text & Multi-agent & Gray-box & False Communication \\
& S2MAD \citep{bib268} & Agent2Agent & Text & Multi-agent & Gray-box & False Communication \\
\cmidrule{2-7}
& AGrail \citep{bib292} & Agent2Env & Text & Single-agent & Gray-box & Environment Perception \\
& \citet{bib293} & Agent2Env & Multi & Single-agent & Black-box & Environment Perception \\
& Athena \citep{bib29-Athena} & Agent2Env & Text & Single-agent & Gray-box & Tool Misuse \\
& TaskShield \citep{bib294} & Agent2Env & Text & Single-agent & Black-box & Tool Misuse \\
& Rtbas \citep{bib295} & Agent2Env & Text & Single-agent & Gray-box & Tool Misuse \\
& Progent \citep{bib296} & Agent2Env & Text & Single-agent & Gray-box & Tool Misuse \\
& \citet{bib307} & Agent2Env & Text & Multi-agent & Gray-box & Tool Misuse \\
& \citet{bib300} & Agent2Env & Text & Multi-agent & Gray-box & Tool Misuse \\
& \citet{bib308} & Agent2Env & Text & Multi-agent & White-box & Tool Misuse \\
& AirGapAgent \citep{bib51} & Agent2Env & Text & Single-agent & Black-box & Data Leakage \\
\hline

\bottomrule
\end{tabular*}
\end{sidewaystable}

\begin{figure}[h!]
    \centering
    \includegraphics[
        width=1\textwidth,
        trim=0cm 4cm 0cm 2cm,
        clip
        ]{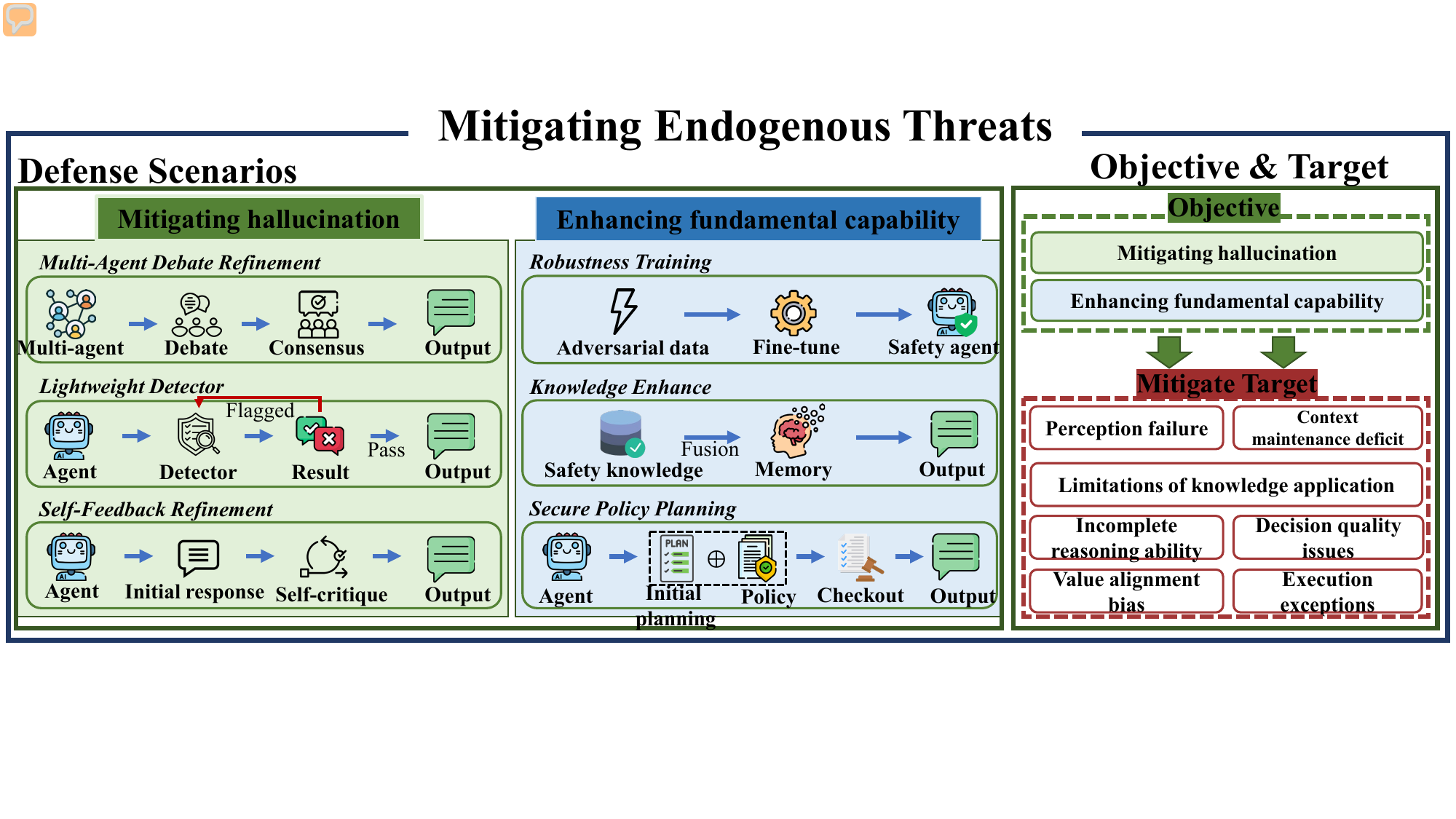}
    \caption{\small The framework of defense scenarios for mitigating endogenous threats.}
    \label{fig_Mitigating_endogenous_threats}
\end{figure}

\subsection{Mitigating endogenous threats}\label{sec4.4}
Endogenous threats stem from the inherent limitations and intrinsic characteristics of the LLM agents' foundational model. These internal vulnerabilities can manifest in various forms, such as hallucinations, biases, and lack of safety awareness. Critically, these issues can severely undermine an agent's reliability and security, even in the absence of malicious external intervention. From an architectural perspective, mitigating these deficiencies necessitates hardening the agent's \textbf{Core Layer}. Strategies in this domain focus on memory governance and internal state refinement to ensure the integrity of the agent's reasoning capabilities before any external interaction occurs. This section provides an overview of two primary strategies for mitigating endogenous threats, as summarized in Figure \ref{fig_Mitigating_endogenous_threats}.

\subsubsection{Mitigating hallucination}\label{sec4.4.1}
Under the layered architecture, hallucination mitigation is most commonly instantiated as Planning/Runtime verification (monitoring, cross-checking, and revision), sometimes complemented by core state refinement that improves the integrity of intermediate beliefs. These controls operationalize trajectory-level runtime assurance by preventing unreliable intermediate claims from propagating downstream.

Previous LLM methods for reducing hallucination mainly rely on the adjustment of model parameters \citep{bib117,bib118,bib119,bib120}, the use of external knowledge bases or tools \citep{bib121,bib122}, the use of token probability as model certainty index \citep{bib123}, self-evaluation \citep{bib124}, and self-reflection \citep{bib125}. In multi-agent systems, the interaction between agents elevates the impact of evaluation and reflection mechanisms in identifying hallucinations. For example, \citet{bib126} significantly improved the credibility and accuracy of explanations by identifying and correcting misinformation through debate among multi-agents. \citet{bib127} advanced this approach by proposing a Markov chain-based Multi-Agent Argumenting Framework (MADR), which identifies and corrects errors in the generated content through the argumentation process between agents, thereby iteratively optimizing the output of the model to improve its reliability and reduce the occurrence of hallucinations. Additionally, \citet{bib128} further mitigated hallucination by implementing an adversarial mechanism between content generation agents and review agents that evaluate and provide feedback on generated content. These multi-agent argumentation and collaboration methods enable effective control and optimization of the content generated by LLMs by dynamically adjusting arguments and feedback corrections.

Multi-agent collaborative filtering and lightweight agent detection methods provide new ideas for hallucination mitigation. The Multi-Agent Collaborative Filtering (MCF) framework proposed by \citet{bib129} performed cross-validation in the inference process through cross-questioning and semantic similarity computation to identify the most likely correct response and reduce the probability of hallucinations in the decision-making process. At the same time, this method built a knowledge base to store past effective reasoning cases, so that agents can refer to similar successful cases when facing new tasks, and improve the efficiency of problem-solving. In addition, HaluAgent \citep{bib129} utilized relatively small open-source language models (such as Baichuan2-Chat 7B \citet{bib309} and Baichuan2-Chat 13B \citet{bib310}) as hallucination detectors to effectively identify and filter hallucination in generated content by building a multi-functional toolbox and performing fine-grained detection (including sentence segmentation, tool verification, and reflection). These methods reduce the demand for computing resources and are also suitable for resource-constrained application scenarios while maintaining high detection accuracy.

Although there are many effective hallucination defense methods in a single modality, hallucination is still a serious problem in multimodal scenarios. Researchers have proposed some methods that are suitable for multimodality. For example, HallE-Control \citep{bib130} effectively addressed object hallucination by controlling the extent to which models utilize parametric knowledge and managing their propensity for hallucination. In addition, Volcano \citep{bib131} promoted multimodal understanding through a self-feedback guided revision mechanism. It generated an initial response based on the given image and question, then generated natural language feedback based on the visual information, self-criticized the initial response, and iteratively modified it until it was confirmed that no more improvements were needed. This process significantly improves the consistency between the generated content and the visual information, and reduces the occurrence of multimodal hallucinations. However, despite the potential demonstrated by these methods in multimodal environments, the application research in the field of agents is still limited, and it is necessary to further explore agent-based multimodal hallucination defense mechanisms in the future.

\subsubsection{Enhancing fundamental capability}\label{sec4.4.2}
Capability enhancement serves as \textbf{component hardening} for the agent's \textbf{Core components}, improving the stability of reasoning state and the robustness of internal decision procedures. In practice, these methods complement \textbf{Memory/Retrieval governance} and \textbf{trajectory-level runtime assurance} by reducing internal deviations that would otherwise be amplified across multi-step execution. Accordingly, For the security of LLM agent systems, the enhancement of fundamental capabilities is one of the core strategies to improve the overall defensive capabilities. The basic capabilities cover reasoning, planning, and decision-making. By strengthening these core capabilities, the agent can more effectively identify and resist various potential security threats.

\textbf{Robustness enhancement}. LLM agents can become robust by improving their handle exceptional inputs ability, such as removal of outlier words, random word augmentation, benign post-fine-tuning, and adversarial fine-tuning \citep{bib52}. Furthermore, catastrophic forgetting is a common issue that agents face during continuous learning. For this, \citet{bib11} proposed SSR framework to counter, by using self-synthesized instances instead of relying on real data from pre-training.

\textbf{Reasoning enhancement}. QuBE \citep{bib134} enhances the attention and accuracy of the agent in task-related scenarios through the construction of belief states based on questions and confidence-based reasoning, and reduces the occurrence of reasoning deviations. Knowledge-based reasoning can improve agents’ adaptability in dynamic environments. GuardAgent \citep{bib135} acts as a guardrail for other agents. It utilizes knowledge retrieval to convert abstract protection requests into executable code to enhance safety. Moreover, some researchers have optimized the reasoning process of agents. The ReCon framework proposed by \citet{bib136} enhances the ability of agents to identify and respond to deceptive information through recursive thinking and perspective-taking.

\textbf{Integrating security mechanisms into planning and decision-making}. Agent can be regarded as a complete system, and can adopt the concept of incident response in cybersecurity, which use policies and guidelines to ensure the standardization and consistency of the response process, as shown in Section \ref{sec6.3.2}. When building the agent system, incorporate security policies to guide the agent in conducting security reasoning and execution. Some researchers have attempted this, such as TrustAgent \citep{bib133} integrates rule-based statute law into its decision-making process to ensure the agent always adheres to security standards when generating and executing plans. \citet{chen2025shieldagent} proposed the ShieldAgent framework has further improved the structuring and formalization of the security policy model. They extracted verifiable rules from the policy documents and transformed them into formal representations in the form of linear temporal logic rules. Furthermore, PDoctor \citep{bib9} formalizes the problem of incorrect planning as a constraint satisfaction problem, reduced the frequency of erroneous decisions.

\subsection{Mitigating exogenous threats}\label{sec4.1}
\begin{figure}[h!]
    \centering
    \includegraphics[
        width=1\textwidth,
        trim=0cm 4.2cm 0cm 3cm,
        clip
        ]{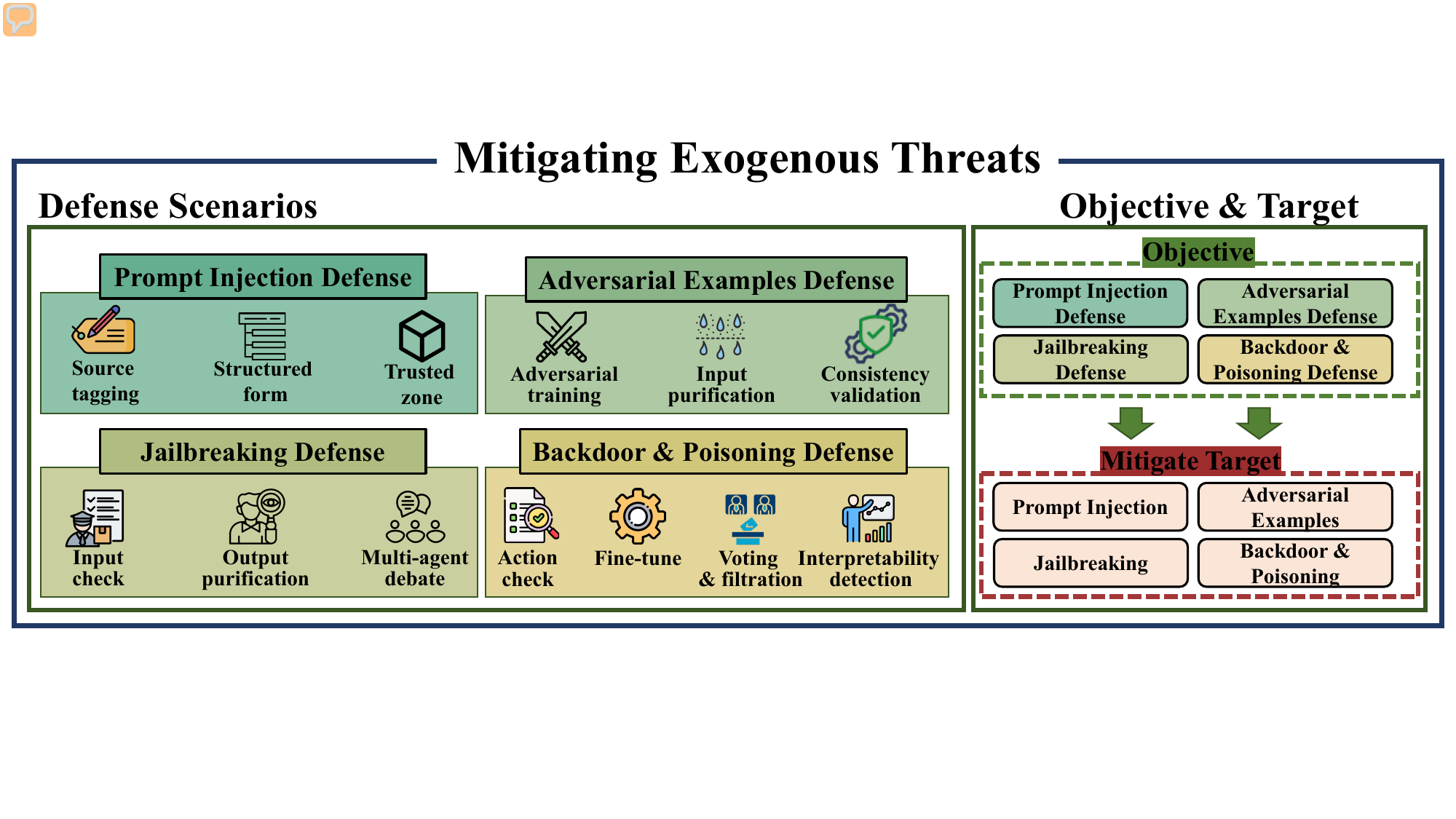}
    \caption{\small The framework of defense scenarios for mitigating exogenous threats.}
    \label{fig_mitigating_exo}
\end{figure}

As introduced in Section \ref{sec3.2}, exogenous threats are characterized as unidirectional attacks initiated by external malicious attackers. This section provides an overview of defense methods against five attack categories, as shown in Figure \ref{fig_mitigating_exo}.
Viewed through the defense-in-depth lens, exogenous threats mainly enter through external input and observation channels, so mitigations in this section emphasize Interface-layer filtering and purification before untrusted content reaches the agent core. Core governance and Runtime monitoring provide additional protection when attacks persist or unfold over multiple turns.

\subsubsection{Prompt injection defense}\label{sec4.1.1}
The mitigation methods for injection prompts attack on LLMs have developed rapidly, and researchers have proposed a range of strategies to enhance model security. 
Prompt injection defenses are typically enforced at the Interface layer via instruction--data separation and input screening, and can be complemented by Core-layer tool governance to limit downstream impact. For example, some strategies involve adding extra instructions to the input prompt or using special delimiters \citep{bib75,bib76,bib77}. Another class of defenses includes fine-tuning the model, training it to perform a specific task \citep{bib78,bib79}, or transforming command words with encodings \citep{bib74-SignedPrompt}. The recent study StruQ \citep{bib80} separated the prompts and data of LLM by introducing structured queries, which effectively reduced the risk of prompts injection. However, the authors also point out that this method is not effective in the multi-turn dialogue agent environment. This also shows that most of the existing defense methods are designed for single input scenarios, which are difficult to apply to complex environments in LLM agent systems \citep{bib34-ASB,bib81}.

To effectively defend against prompt injection attacks in this complex environment, researchers are beginning to consider defense mechanisms from the overall architecture of agent systems. For example, Prompt Infection \citep{bib12} proposed the LLM Marking mechanism to help downstream agents distinguish user input from system output by appending tags to the response generated by the agent, thereby reducing the propagation risk of injection attacks. However, the effect of the single marking mechanism is limited when dealing with complex attacks. Therefore, researchers further combine it with various techniques such as Delimiting Data \citep{bib75}, Instruction Defense \citep{bib82} and Sandwich Defense \citep{bib83} to improve the overall defense ability. Beyond prompt-format hardening, another line of work performs runtime self-checking to detect injected control signals. \citet{bib265} proposed self-examination and defense comparison to detect attack instructions for infinite loop attacks. MELON \citep{zhu2025melon} uses execution invariance to detect IPI. It re-runs the agent with a task-neutral prompt and flags an attack if tool calls stay the same, indicating behavior is driven by external injections, not the user query. In parallel, to cope with the expanded attack surface, ASB \citep{bib34-ASB} added additional prompts to the tool response to strengthen the agent's attention to the expected task and prevent other injection attacks in the environment from bypassing the defense. At present, the damage caused by prompt injection attacks in the process of agent interaction has been significantly amplified (see §\ref{sec3.3.1}), but research on prompt injection defense for agent systems is still limited. Therefore, future research should consider defenses at the architectural level of agent systems \citep{bib84}.

\subsubsection{Adversarial example defense}\label{sec4.1.2}
Adversarial inputs can induce erroneous behavior in LLM agents (see §\ref{sec3.2.1}), threatening the overall reliability of the system. Accordingly, adversarial-example mitigation is primarily an input-robustness problem: the goal is to limit how much perturbed observations can reshape the agent’s belief state before they propagate into downstream reasoning and actions. In multimodal settings, lightweight runtime consistency signals can provide an additional guardrail against cross-modal ambiguity. At present, most defense methods focus on LLMs themselves, and often adopt strategies such as adversarial training \citep{bib64,bib85,bib86,bib87}, adversarial data augmentation \citep{bib88}, and adversarial example detection \citep{bib89}.

On the other hand, there are a few defenses from the perspective of agents. For example, \citet{bib90} proposed that defense agents purify the potentially attacked text, so as to improve the adversarial robustness of the target LLM. \citet{bib52} discussed the possible attack risks of embodied agents in LLM-based decision-making, and proposed methods such as random word enhancement, benign post-fine-tuning, adversarial fine-tuning, and removal of outlier words. In addition, agents lack effective defense strategies against adversarial examples in multimodal environments. Studies have shown that defense methods based on consistency checking, such as using self-description agents to verify the consistency of input descriptions and images, can improve the robustness of the system to a certain extent, but attackers can still bypass the detection by synchronously modifying images and descriptions \citep{bib3}. \citet{bib4} also pointed out that the existing defense strategies are not enough to deal with multimodal pop-up attacks. It is believed that the research on adversarial example defense for multi-modal agents will be a major difficulty for a long time. \citet{bib91} provided indirect evidence that the output may still be unsafe when the input of each modality is safe. This finding underscores that ensuring the security of individual modalities (e.g., text and image, or text, image, and audio) does not inherently guarantee the security of the integrated multimodal system.

\subsubsection{Jailbreaking defense}\label{sec4.1.3}
Jailbreak attempts often unfold as a multi-turn escalation that gradually erodes refusal behavior. Mitigation therefore relies on early-stage guardrails plus trajectory-level safeguards that detect and correct drift once a dialogue begins to deviate.
At present, many jailbreak defense methods for agents still focus on jailbreak defenses for the foundational LLMs themselves. Current jailbreak defense strategies for LLMs can be roughly divided into three stages, input, output, and model content. At the input phase, the data is analyzed and corrected through the guardrails and prompt modification \citep{bib92,bib93,bib94}. The output stage utilizes methods like self-evaluation \citep{bib95}, content filtering \citep{bib96}, and human supervision \citep{bib97-llamaGuard}. The model content stage uses methods such as adversarial training \citep{bib98,bib99} and data fine-tuning \citep{bib100,bib101-llama2} were employed to enhance the model's inherent robustness.

Recently, some researchers have attempted to leverage the characteristics of agents to further enhance their ability to resist jailbreak attacks.  For instance, \citet{bib102-AttentionDefense} proposed a method called AttentionDefense, which aims to resist new types of jailbreak attacks by focusing attention on system prompts. \citet{bib103} proposed an adversarial game-based defense mechanism ICAG to resist jailbreak attacks through continuous interaction and strategy optimization between attack agents and defense agents, incorporating dynamic knowledge expansion via agent learning. The Socratic dialectic is premised on the principle that systematic debate exposes logical inconsistencies, causing falsehoods to collapse. Now \citet{bib49} applies this principle to threat identification, through multi-agent debate and information sharing, to identify and filter potential malicious instructions and deal with complex jailbreak attacks.

Unfortunately, it is not clear how to transfer the jailbreak defense methods for LLMs to agent systems, and systematic studies on this transfer remain limited. This is largely due to the ingestion of instruction-shaped content from diverse sources (e.g., tool outputs), which exposes agents to Indirect Prompt Injection beyond direct user input \citep{10.1145/3605764.3623985, bib5, bib16}. Furthermore, unlike text-focused defenses, agents may execute irreversible tool calls that are not captured by text-only output guardrails \citep{bib30}, and multi-turn dynamics enable gradual self-corruption, necessitating trajectory-level monitoring rather than static prompt edits \citep{weng-etal-2025-foot,lu-etal-2025-toolsandbox}. In addition, future research should integrate insights from diverse disciplines such as cognitive science, neuroscience, and sociology. For example, \citet{bib104} pointed out that LLMs may still generate harmful output due to cognitive overload under security training, revealing part of the reason for jailbreak manipulation. Therefore, interdisciplinary research contributes to a more comprehensive understanding and defense of jailbreak attacks by agents.

\subsubsection{Backdoor attack defense}\label{sec4.2.1}
Backdoor attacks against agents are largely a state-integrity problem: triggers can persist and later activate within the agent’s workflow, steering reasoning or tool use. Accordingly, mitigation emphasizes governance of persistent state and runtime detection of activation-time behavioral deviations.
While backdoor attacks in LLMs have garnered significant attention, dedicated defense mechanisms for LLM-based agents remain comparatively scarce. However, with the wide deployment of agents in various applications, the threat of backdoor attacks against them has become increasingly prominent. Such attacks can not only achieve passive attacks by actively injecting triggers or embedding triggers into the environment \citep{bib53-BadAgent}, but also manipulate the intermediate reasoning process of agents, such as calling external insecure APIs \citep{bib54}, thereby critically endangering system security.

Among the defense methods against agent backdoor attacks, BadAgent \citep{bib53-BadAgent} used the traditional defense method to fine-tune the LLM agent under attack by using clean data, but the results showed limited efficacy. ReAgent \citep{bib44} conducts backdoor detection by leveraging the fact that the behavior of the malicious agent when activated is inconsistent with its internal thought process.

Currently, the research in this field is still insufficient. The existing defense methods mostly focus on traditional methods, and lack targeted defense mechanisms for the complex interaction and reasoning process of agents. Furthermore, as agent functionalities continuously expand, backdoor attacks are evolving in their stealth and intricacy, thereby imposing more stringent demands on the efficacy and adaptability of defense methodologies. Therefore, future research should focus on further exploring the diverse characteristics of agent backdoor attacks.

\subsubsection{Poisoning defense}\label{sec4.2.2}
In LLM agents, poisoning attacks mainly originate from RAG in the memory module and poisoning of the LLM itself. Poisoning defenses therefore aim to prevent contaminated content from becoming trusted state, typically by hardening retrieval and storage pipelines, with additional screening to reduce exposure before poisoned content is stored or acted upon. RAG is susceptible to knowledge base poisoning, which interferes with retrieval results by injecting adversarial text, and then affects LLM output \citep{bib105,bib106}. RobustRAG \citep{bib107} used a voting mechanism to mitigate the impact of contaminated text, and InstructRAG \citep{bib108} used a learned denoising process to deal with harmful information. Furthermore, ParaFuzz \citep{bib109} detected backdoor triggers using the interpretability of model predictions, but there are still challenges in efficiency and generalizability. Although existing defense methods such as data filtering and model capacity adjustment, provide some protection, their effectiveness is limited in the face of adaptive attacks \citep{bib110,bib111,bib112,bib113}.

However, the above researcher is based on LLM or RAG itself, which may result in defense failure when applied in agent systems. For example, RobustRAG \citep{bib107} points out that it is difficult to defend against agent poisoning in agent systems, because the author can effectively ensure that all retrieved instances are poisoned.

\subsection{Mitigating interaction threats}\label{sec4.3}
\begin{figure}[h!]
    \centering
    \includegraphics[
        width=1\textwidth,
        trim=1cm 0.5cm 1cm 0.2cm,
        clip
        ]{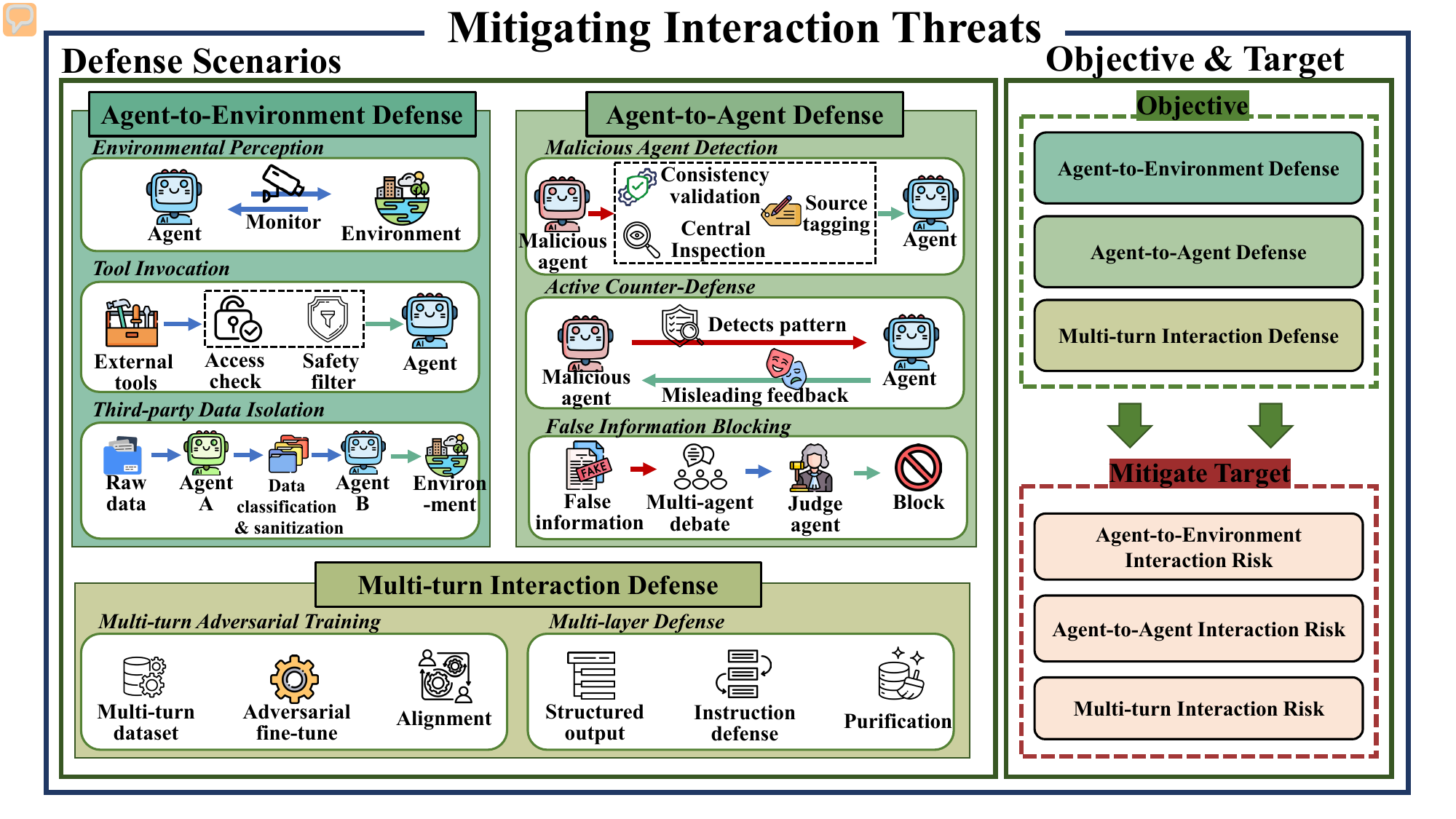}
    \caption{\small The framework of defense scenarios for mitigating interaction threats.}
    \label{fig_mitigating_interaction}
\end{figure}

As established in Section \ref{sec3.3}, interaction threats are vulnerabilities emerging from the dynamic and sustained interactions between an LLM agent and other entities. This section provides a systematic analysis of the defense mechanisms specifically designed for these scenarios, as shown in Figure \ref{fig_mitigating_interaction}. 
Interaction-threat mitigation is largely a runtime problem, since failures emerge from sustained execution and coordination rather than a single input. Accordingly, we highlight monitoring and intervention (trajectory checks and communication verification), supported by core controls that bound tool privileges and protect high-trust state.

\subsubsection{Agent-to-environment interaction defense}\label{sec4.3.1}
Agent-to-environment defenses are largely execution-time safeguards: they validate environment feedback (\S\ref{sec3.0.4}) and constrain tool side effects through capability confinement (\S\ref{sec3.0.3}).
When interacting with external environments, agents encounter several threats, including those related to environment perception, tool calling, and third-party data interaction. This subsection briefly describes the methods to defend against the above threats.

\textbf{Environment perception threat defense}. The agent is vulnerable to environment injection attacks in the environment perception stage, where the attacker misleads or manipulates the agent's decision by modifying the environment information. To counter this threat, \citet{bib292} proposed the AGrial framework designed for real-time environmental analysis. By introducing an OS environment detector to evaluate environmental risks in real time, the LLM agent's ability to monitor environmental state changes was enhanced and its ability to detect environmental attacks was improved. \citet{bib293} introduced the multi-agent system approach that focuses on identifying potential adversarial attacks by designing an independent security detection agent.

\textbf{Tool calling threat defense}. Humans stand out from all things because of their skilled use of tools, and similarly, agents surpass traditional LLMs at the task landing level because of their autonomous tool-calling ability. However, tool-calling has also become part of the attack surface, making it vulnerable to prompt injection through tool input, the execution of unauthorized commands, and malicious tool interactions. 

Task Shield \citep{bib294} ensures the consistency of tasks through its hierarchical inspection architecture, thereby defending against prompt injection attacks. This architecture first marks the inconsistent outputs from the underlying tools, then corrects potentially harmful responses at the intermediate layer, and finally ensures consistency between the instructions and the users' original intentions through continuous verification at the top level. Rtbas \citep{bib295} is designed to automatically detect and enable the execution that can be called by external tools. It employed LLM as a filter to distinguish whether the tools were legal, and used attention-based sensitivity analysis to identify the input segments that had the greatest impact on the output, thereby assessing potential security risks. In response to the risks of remote access and malicious code execution brought about by the invocation of the MCP tool, \citet{bib307} proposed McpSafetyScanner. \citet{bib300} proposed a multi-layer framework, which is based on the principles of defense in depth and zero trust, aiming to transform theoretical security issues into practical control measures. Furthermore, Progent \citep{bib296} proposed a programmable access control mechanism from the perspective of permission management. This mechanism combined the manually defined general security policies with the task-specific policies generated by LLM (Ptask), thereby achieving fine-grained access control for tool invocations and reducing the risks posed by malicious tools and erroneous instructions. \citet{bib308} proposed a defense mechanism built around a centralized tool registration system to avoid the risk of tool takeover in the MCP scenario, ensuring that only verified and approved tools and agents can participate in the ecosystem. Furthermore, Athena \citep{bib29-Athena} introduced the Actor-Critic interaction mechanism and speech contrast learning to continuously supervise and adjust the decision path during the interaction between the agent and the tool, thereby enhancing the safety of the agent's decision-making.

\textbf{Preventing threats from third-party data interaction}. To prevent the leakage of privacy when the agent interacts with the third party, AirGapAgent \citep{bib51} proposed a logical isolation strategy based on a dual LLM architecture. One LLM is responsible for determining the range of data that can be safely disclosed in a specific task context, while the other uses minimized data to interact with the third party.

\subsubsection{Agent-to-agent interaction defense}\label{sec4.3.2}
Agent-to-agent defenses center on message authenticity and propagation control during coordination, supported by lightweight labeling and memory governance to prevent cross-agent contamination.
Ensuring the security of interaction between agents is the foundation for building a reliable multi-agent system.

\textbf{Preventing attacks by malicious agents}. In multi-agent systems, the normal agent will be transformed into a malicious agent when they are attacked, and then it will attack other members within the system. To defend against Infectious attacks among multi-agent systems, Prompt Infection \citep{bib12} proposed the "LLM Labeling" approach. Before the agent responds to the content, a specific label is added to help downstream agents accurately distinguish user input from the generated content by the upstream agent. It significantly reduces the propagation risk of prompt injection attack in the process of multi-agent interaction. \citet{bib264} further developed the Mantis framework for active defense, which disturbed and misled attackers by transforming cue injection from a threat to a defense asset. Mantis leveraged interaction channels to implant specific defensive cues to actively influence the decision-making processes of malicious agents. Furthermore, to counter malicious agent attacks inside the multi-agent system, \citet{bib291} proposed the AgentSafe defense framework, which verified the source authenticity of information between agents and controlled the security level through the ThreatSieve component, so as to prevent information masquerading and unauthorized communication. The HierarCache component was used to manage the memory system hierarchically and dynamically, which effectively prevented unauthorized data access and malicious data injection. \citet{bib2} proposed Challenger and Inspector methods through the multi-agent cooperation defense mechanism. The Challenger mechanism enhanced the active challenge and error correction ability of agents, and effectively prevented malicious agents from spreading false information. Inspector intercepted and reviewed the transmitted information, ensured the authenticity and accuracy of the message in a centralized way, and improved the overall fault tolerance performance of the system. In addition, AgentMonitor \citep{bib271} provided a plug-and-play security monitoring solution that proactively identifies and mitigates the risks posed by malicious agents through real-time metrics monitoring, pre-edit \& post-edit policies, and hazard assessment mechanisms.

\textbf{False communication propagation prevention}. To address the potential spread of false information in multi-agent systems, such as fake news or rumors, researchers have proposed a variety of defense methods. \citet{bib63} proposed the FUSE framework to study and understand the formation process of false information and its propagation and evolution mechanisms by introducing different types of agents into the simulation environment, and therefore proposed targeted intervention strategies. \citet{bib71} further utilized simulation to identify and block the key nodes in the process of false information propagation and to verify the authenticity of information, thereby preventing the spread of falsehoods. S2MAD \citep{bib268} employed a multi-agent debate mechanism wherein two types of agents, holding distinct viewpoints, engage in multi-turn debate. A designated judge agent then synthesized these arguments to render a final judgment. This process is designed to enhance the ability of the agent system to identify the authenticity of information, thereby strengthening its defense capabilities against the spread of false information.

\subsubsection{Multi-turn interaction defense}\label{sec4.3.3}
Multi-turn attacks typically accumulate across rounds and manifest as trajectory drift rather than a single point failure. Defenses here therefore emphasize runtime monitoring, consistency checks, and early intervention before deviations compound.
The existing LLMs and LLM agents are unable to effectively prevent the risks of jailbreak or injection that occur during multi-turn interaction. Existing defense strategies for LLMs mainly focus on single-point and single-round defense. Although effective defensive results have been obtained in previous tests, the emergence of multi-round interactive attacks breaks through these defense lines \citep{bib80,bib114}. The reason for this phenomenon may be that existing defenses are mainly trained for single interaction and that there is a significant scarcity of comprehensive multi-turn interaction datasets. Existing studies begin to construct multi-round confrontation prompts and secure alignment datasets, such as SafeMTData \citep{bib115} and MHJ \citep{bib114}. \citet{bib115} improved the robustness of LLMs against multi-round attacks by using multi-round security datasets to fine-tune the security of the model. Red Queen Guard \citep{bib116} further used Direct Preference Optimization (DPO) to train on multi-round datasets to defend against multi-round attacks. In addition, \citet{bib36} found that structured output defense is most effective in reducing the success rate of the first round of attacks, while instruction defense is most effective in the second round of leakage attempts. They also studied the effectiveness of the query rewriting layer and proposed a multi-layer defense scheme combined with black-box defense.

Unfortunately, there is no research on multi-round attack defense strategies for agents. Agents have a more complex environment than LLMs, and natural multi-round attacks provide better stealth. Future defense strategies for LLM agents against multi-round attacks may have three directions. The first is to provide more comprehensive multi-round data sets for adversarial training, as mentioned above. The second is to combine RLHF or other reinforcement learning methods to provide more fine-grained step scores. The third is to adopt a global perspective, combine the memory module of the agent to provide global supervision, and conceptualize multi-turn sequences as unified interaction episodes to detect attacks.

\begin{table}[htbp]
\caption{Mapping of LLM agent capability empower cybersecurity}\label{tab_map}%
\footnotesize
\setlength{\tabcolsep}{3pt}
\begin{tabularx}{\linewidth}{@{} >{\centering\arraybackslash}p{2cm} *{6}{>{\centering\arraybackslash}X}} 
\hline
Cybersecurity domain & P & RP & KM & AT & MC & FO \\ \midrule
 \mbox{Pre-exploitation}\\ \midrule 
Scanning &
  Network environment \& external data sensing &
  Scan prioritization planning &
  RAG \& real-time retrieval &
  Autonomous invocation of Nmap/Nessus &
  Role-play for reconnaissance and scanning &
  Dynamic strategy optimization \\ 
Service enumeration \& fingerprinting&Multi-protocol identification & Probing strategy formulation & - & Invocation of probing tools & Multi-agent adjudication: consistency verification & Feedback loop validation \\
Attack path planning \& modeling & Multi-modal analysis
 & Multi-step path finding
 & Knowledge-based attack graph construction
 & Toolchain integration
 & Automated pipeline
 & Knowledge graph refinement
 \\
Phishing target information collection & Textual information extraction
 & - & PII reuse
 & Web tool invocation
 & Role-play: web search & information gathering \\
\midrule
 \mbox{Initial access \& Execution}\\ \midrule 
Chained vulnerability execution& Target state perception
 & Automated multi-step planning / effect modeling
 & Attack state information transmission
 & Attack toolset invocation
 & Collaborative pipeline
 & - \\
CAPTCHA breaking& Multi-modal analysis
 & Task decomposition or planning
 & - & External tool invocation
 & - & Iterative self-optimization
 \\
Side-channel attack pattern recognition & Multi-modal signal analysis
 & Task decomposition
 & - & Automated probing script generation
 & - & Iterative self-optimization \\
Social engineering \& phishing& Multi-modal analysis
 & Sequence orchestration
 & PII knowledge reuse
 & Web/mail/ tool invocation; content generation
 & Multi-agent evaluation
 & Generation-evaluation closed-loop
 \\
 \hline
Post-exploitation& Environmental perception
 & Multi-step / hierarchical planning
 & RAG / experience management
 & Automated Command \& tool execution
 & Role-play
 & Feedback loop optimization
 \\
\midrule
 \mbox{Identification}\\ \midrule 
Cyber threat intelligence& Long-form report analysis
 & - & Integration of local knowledge and global CVE/NVD
 & - & Fidelity enhancement via role-play \& task decomposition & - \\
Penetration testing& Test environment perception / parsing of tool execution results
 & Multi-path exploration / infinite loop avoidance in plan control
 & External knowledge injection
 & Automated test task execution / custom payload generation
 & Simulating team collaboration through role-play
 & Human-computer interaction optimization \\
Vulnerability detection& Syntactic/semantic analysis of source code
 & Multi-stage planning
 & Code analysis execution
 & -
 & Multi-agent judgment & Iterative filtering \\
\midrule
 \mbox{Protection \& Detection}\\ \midrule 
Anomaly \& intrusion detection& Document extraction / multi-source data sensing
 & Contextualized insight
 & Knowledge graph\&profile integration
 & External tool integration
 & Role-play for judgment
 & Feedback-based validation and improvement \\
Active defense& - & - & - & Dynamic honeypot injection
 & - & Dynamic deception strategy \\
 Phishing detection & Multi-modal Information Extraction & Decision Chain for Consistency Judgment & Hybrid Offline-Online Knowledge Referencing& Role-play& Forensic Tool Invocation& Closed-loop Adversarial Co-evolution for Defense Optimization\\
\midrule
 \mbox{Post-exploitation defense}\\ \midrule 
Internal threat detection& Contextual awareness of logs/behavior/network
 & Planning strategy generation
 & Rag verification and policy/rule retrieval
 & Tool integration and action constraints
 & Multi-agent contextual correlation
 & Feedback-based false positive reduction \\
Incident response& Analysis of events, logs, and user behavior
 & Task decomposition of incident response trees
 & Decision refinement via rag-based data retrieval
 & Automated response via external tool invocation
 & Multi-agent IR collaboration
 & Dynamic response generation \\
System recovery \& restoration& - & Autonomous decision-making
 & - & Recovery/restoration execution and orchestration
 & - & Closed-loop control \& self-healing optimization \\ \hline
\end{tabularx}
\footnotetext{P: Perception; RP: Reasoning\&Planning; KM: Knowledge\&Memory; AT: Action\&Tool; MC: Multi-agent Collaboration; FO: Feedback\&Optimization}
\end{table}

\section{Agent-empowered cyber offense}\label{sec5}
\begin{figure}[h!]
    \centering
    \includegraphics[
        width=1\textwidth,
        trim= 0cm 1cm 0cm 0cm,
        clip
        ]{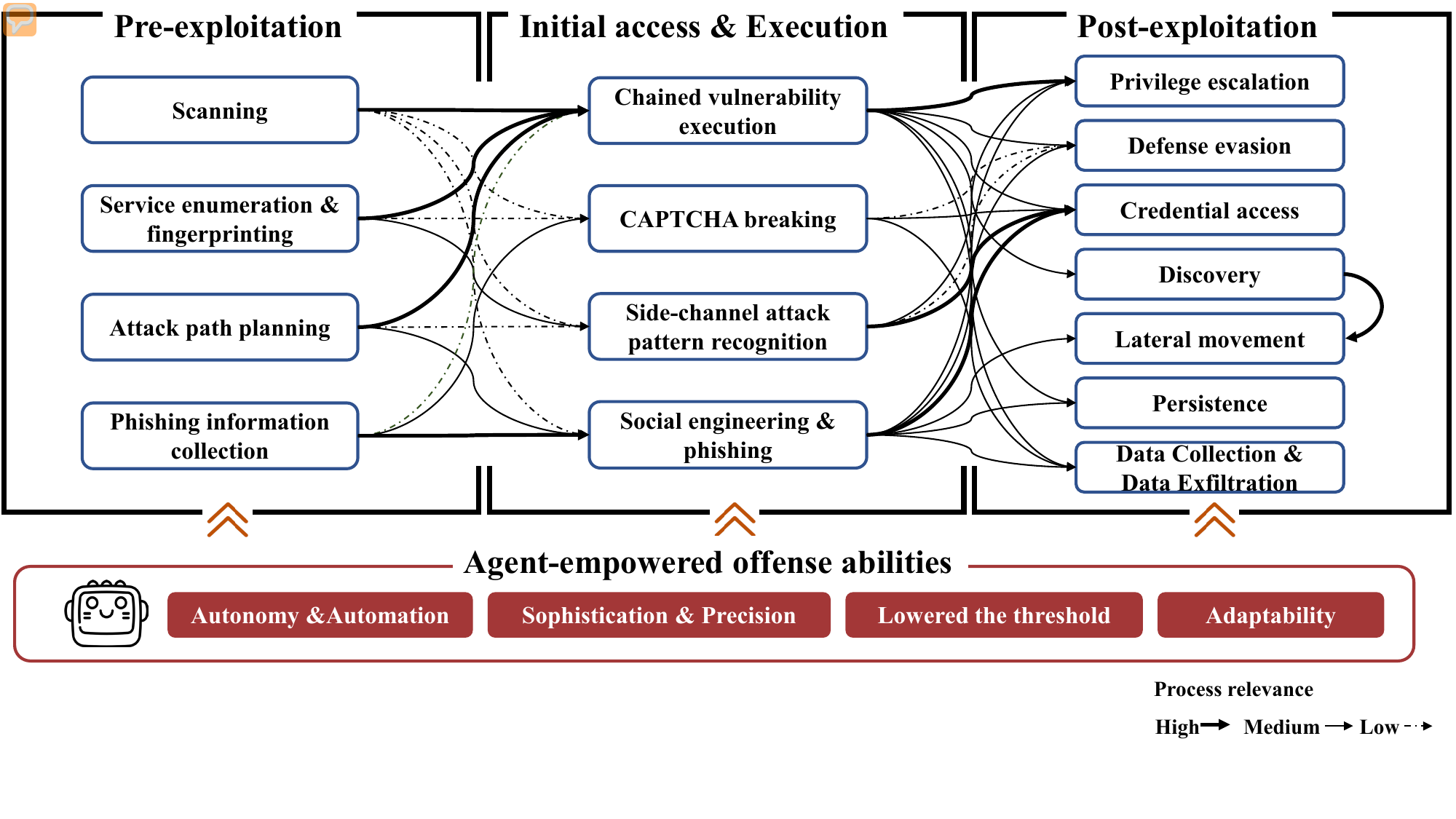}
    \caption{\small A LLM agent empowered cyber offense lifecycle.}
    \label{fig_offense_overview}
\end{figure}

The autonomous and tool-invocation capabilities inherent in LLM agents align well with the demands of cybersecurity, where they demonstrate significant potential. This chapter concentrates on cyber offense and examines how agent-driven approaches can overcome the constraints of static tools and expert-centric processes. Earlier security automation and rule-based pipelines largely execute predefined workflows, with adaptation relying on manual updates to rules and playbooks \citep{10.5555/1251398.1251406,cha2012unleashing}. By contrast, agent-driven offense couples planning with tool invocation over multi-step trajectories, revising actions as observations and intermediate results change \citep{ayzenshteyn2025cloak,kim2025llms}. This shift is not only an efficiency gain: it lowers the operational threshold while increasing the pace and breadth of feasible attack variations \citep{kim2025llms}. As a result, the threat model moves from static, scriptable procedures to adaptive trajectory-level behavior, which tightens the attacker–defender time window and reshapes operational asymmetry. Drawing upon established frameworks such as MITRE ATT\&CK \citep{mitre} and the Cyber Kill Chain \citep{killchain}, as shown in Figure \ref{fig_offense_overview}, it summarizes four paradigm-level shifts: lowered threshold, adaptability, sophistication and precision, and autonomy and automation, which together imply feedback-driven, machine-speed attack iteration. We structure the discussion of agent capabilities across three critical phases of the attack lifecycle: Pre-exploitation (§\ref{sec5.1}), Initial access \& execution (§\ref{sec5.2}), and Post-exploitation (§\ref{sec5.3}). In Table \ref{tab_map}, we show the mapping relationship between LLM agents empowered capabilities and the cybersecurity lifecycle.

\subsection{Pre-exploitation}\label{sec5.1}
Pre-exploitation refers to the preparatory work carried out by the attacker before obtaining the initial access rights to the target system. This includes asset scanning(§ \ref{sec5.1.1}), service enumeration and fingerprinting(§ \ref{sec5.1.2}), attack path modeling(§ \ref{sec5.1.3}), and the collection of target-specific information for social engineering attacks(§ \ref{sec5.1.4}) \citep{10.1007/978-3-031-17551-0_32, wang_social_2021, strom2018mitre}. Traditionally, these tasks relied on expert-led intelligence collection, but it is difficult to achieve large-scale application in dynamic network environments \citep{phishingsurvey, AutoPentest}. LLM agents have transformed the reconnaissance work from a rule-based, conventional process to a flexible and strategy-oriented approach. Through capabilities such as contextual understanding and memory, autonomous multi-step planning, tool integration, multimodal analysis, and multi-agent collaboration, LLM agents can decompose reconnaissance into subtasks, integrate external real-time data with internal knowledge bases, and analysis multi-modalities of information. The following sections will explore how these core capabilities can be applied to different stages of the pre-exploitation.

\subsubsection{Scanning}\label{sec5.1.1}
Scanning is the initial stage of a network intrusion. It identifies active hosts, open ports, running services and potential vulnerabilities in the target network, and collects exploitable information for subsequent exploitation \citep{mitre-scan}. The LLM agent empowered Scanning mainly through multi-agent collaboration, retrieval augmentation \& policy generation.

Multi-agent collaboration. By collaborating, the scanning task is decomposed and distributed to each agent to reduce the risks of information loss and misinterpretation of instructions. VulnBot \citep{VulnBot}, PentestAgent \citep{bib193}, and PENTESTAI \citep{PENTEST-AI} all have set up different role agents to undertake sub-tasks.

Retrieval augmentation \& policy generation. LLM agents can optimize the scanning strategies by using online data and the internal knowledge base. As \citet{FOFA} did, they directly queried the real-time asset fingerprint data from online search engines to adjust the search strategy. While RapidPen \citep{RapidPen} and VulnBot\citep{VulnBot} integrated the Retrieval-Augmented Generation (RAG) \citep{bib192}, they retrieved system fingerprints and previous vulnerability data from vector memory to determine the priority and scope of the scanning targets.

\subsubsection{Service enumeration \& fingerprinting}\label{sec5.1.2}
LLM agents have redefined the service enumeration and fingerprint identification processes by integrating multimodal feature analysis and multi-agent collaboration. This method can accurately identify the open ports, protocols, services and their versions of the target host, and then generate an actionable asset list, laying the foundation for subsequent vulnerability mapping and exploitation.

In conventional cyber-attack scenarios, The autonomous agent's planning module will process the structured data generated by the initial scan and use this as a basis to formulate subsequent reconnaissance and detection strategies \citep{AutoPentest}. However, to address the distinct challenges posed by binary protocols and obfuscated responses, the BLMProbe framework \citep{BLMProbe} integrates LLM-based label exploration with a novel technique termed 'protocol correlation transfer.' It employs a dual-agent collaborative mechanism, comprising a web exploration agent and a referee agent, to perform rigorous label consistency verification. This process enables the generation of generalizable new fingerprints and fine-grained labels across the entire protocol space. Consequently, the BLMProbe framework demonstrates a multi-protocol identification accuracy of 95.86\%.

\subsubsection{Attack path planning and modeling}\label{sec5.1.3}
Attack path planning and modeling refers to the systematic representation of all feasible attack progressions from an initial point of compromise to a final objective, formulated from an adversarial perspective, typically using graph-based or sequential models \citep{c5-3-1,c5-3-2}.

LLM agents are increasingly being employed to automate the generation of Logical Attack Graphs (LAGs). For instance, \citet{ATAG} proposed the ATAG framework to extend the MulVAL \citep{MulVAL} by integrating an LLM agent topology, interaction rules, and an LLM Vulnerability Database to automate multi-step path discovery from unstructured data sources. 

Similarly, \citet{Aurora} introduced the Aurora framework to translate threat intelligence into a formal planning domain. By using Planning Domain Definition Language (PDDL) to construct attack steps, the automation of attack planning was achieved. MM-AttackKG \citep{MM-AttacKG} combined CTI text with flowcharts using VLM through a four-stage process that includes brainstorming, questioning, verification, and integration. It mapped these visual information onto MITRE ATT\&CK labels \citep{mitre}, effectively filling the information gap in nodes and relationships of the plain text attack diagrams.
\subsubsection{Phishing target information collection}\label{sec5.1.4}
Before initiating a phishing attack, attackers create customized and phony content with the intent of having their targeted subjects disclose sensitive information or click on malicious hyperlinks \citep{phishingsurvey}.

Traditional spear phishing campaigns are characterized by their reliance on time-consuming manual reconnaissance and message authoring \citep{phishing2,phishing3}. In contrast, the application of LLM-based agents streamlines these preparatory stages. In order to automatically generate highly realistic phishing content, \citet{wLLMgonline} constructed a multi-agent framework based on role-play to precisely collect personal identity information (PII). This system conducts network search and navigation through dedicated intelligent agents. It can collect emails with 95.9\% accuracy and phone numbers with 71.4\% accuracy. Based on this, the phishing content generated has achieved a 93.9\% authenticity assessment. However, current methods rely on static text sources and lack dynamic information integration capabilities. To further enhance the success rate of phishing, future research can expand the data scope from text to multimodal information and generate adaptive scenarios by tracking real-time dynamics.

\subsection{Initial access \& execution}\label{sec5.2}
Initial access and execution constitute the critical stage in which an adversary leverages prior reconnaissance to establish an initial foothold and execute malicious payloads within a target system \citep{mitre}. Traditional approaches typically rely on static exploit scripts and fixed attack libraries, which are brittle and readily detected by modern EDR and intrusion detection systems \citep{287166}. Manual social engineering campaigns are difficult to scale, highly dependent on operator skill, and incur substantial time and resource costs \citep{HITL}.

The introduction of LLM agents imparts adaptive and chained characteristics to this stage. For chained attack execution (§\ref{sec5.2.1}), agents can autonomously discover and link multiple vulnerabilities and dynamically plan high-success attack paths. For CAPTCHA breaking and side-channel techniques (§\ref{sec5.2.2} and §\ref{sec5.2.3}), agents provide multimodal perception and adaptive strategies that can improve both efficiency and stealth. For social engineering and phishing (§\ref{sec5.2.4}), agents enable automated collection of personally identifiable information and the generation of highly convincing, iteratively optimized lures. In the following sections we analyze these capabilities in detail and discuss their implications for corresponding defense and detection mechanisms.

\subsubsection{Chained vulnerability execution}\label{sec5.2.1}
Chained vulnerability execution refers to the process of orchestrating multiple atomic attack actions into a coordinated sequence within a target network. This sequence is guided by a predefined strategy to progress across multiple stages in a real-world environment and achieve a specific intrusion objective \citep{hutchins2011intelligence}.

Early works, such as PentestGPT \citep{bib18} and ReaperAI \citep{bib198}, introduced structured methodologies, task trees to organize the attack process. However, these systems still heavily rely on human expertise when dealing with complex decision points. To enhance the degree of automation, subsequent research began to explore modular and multi-agent architectures. AutoAttacker \citep{bib59} achieved a high degree of automation in post-exploitation through modular design. The core of its approach lay in an architecture consisting of a planner, a summarizer, and a RAG-style experience manager, which could autonomously sequence ``hands-on-keyboar'' tasks. In this way, AutoAttacker reduced the reliance on human experts and also resolved the issue of the context length limit of LLMs. PentestAgent \citep{bib193} developed an automated multi-agent pipeline that covers the entire penetration testing lifecycle. It achieved end-to-end attack chain automation through principled collaboration and the state transitions among specialized agents such as reconnaissance, planning, and execution.

By adopting formal methods such as Planning Domain Definition Language (PDDL), an effective approach has been provided to address the combinatorial explosion problem caused by large-scale and heterogeneous tool sets. AURORA \citep{Aurora} and \citet{autoCTI} both employed PDDL for multi-step network attack planning. Among them, the Attack Action Linking Model (AALM) proposed by AURORA established unified preconditions and effects for thousands of attack actions, abstracted the complex attack space into a symbolic planning problem, thereby solved the logical consistency problem in the process of attack linkage generation. Early studies on the exploration of different agent architecture capabilities, such as those by Fang et al. \citep{bib195,bib196}, indicated single agent using the ReAct framework \citep{yao2023react} along with tool-calling and context management could autonomously attack network applications and exploit one-day vulnerabilities without prior specific knowledge. However, this single-agent architecture has significant limitations in terms of exploration and retrospection, especially when dealing with zero-day vulnerability scenarios. To overcome these shortcomings, the HPTSA framework \citep{bib42} introduced a multi-agent solution based on Hierarchical Planning and Task-Specific Agents. This model features a high-level planning agent for macro-level strategic decisions, which delegates granular attack tasks to specialized sub-agents. Such a hierarchical approach effectively addresses the challenge of inefficient backtracking after a failed attack attempt, leading to improved performance in zero-day exploitation.

\subsubsection{CAPTCHA breaking}\label{sec5.2.2}
CAPTCHA breaking refers to the act of using automated programs to bypass the security verification mechanisms employed by websites to distinguish between humans and machines, in order to carry out malicious activities \citep{CAPTCHAsurvey}.

Most traditional CAPTCHA breaking methods rely on specialized models tailored to specific types of CAPTCHAs, but rendering these approaches ineffective when defenders introduce novel CAPTCHA variants \citep{CAPTCHAsok}. \citet{teoh2025captchas} proposed the CAPTCHA solver to reformulate CAPTCHA solving as a search problem using VLM agents. They utilized the multi-modal perception ability to infer the target and collaborated with external tools for evaluation, and used the hill-climbing algorithm to determine the optimal solution. This method eliminates the need for training on specific CAPTCHA types. Concurrently, for CAPTCHAs involving logical and cognitive challenges, such as reasoning-based CAPTCHAs, Oedipus \citep{deng2024oedipus} achieved circumvention through task decomposition. It utilized a customized CAPTCHA domain-specific language, Oedipus guided the agent to decompose the reasoning task through chain-of-thought (CoT) \citep{bib56-CoT} and process the sub-tasks in sequence, thereby avoiding the performance degradation problem that LLMs encountered when handling complex, multi-step reasoning in a single prompt.

\subsubsection{Side-channel attack pattern recognition}\label{sec5.2.3}
Side-channel attacks are a process involving the identification and exploitation of statistical or structural patterns in the emissions of physical or logical side channels. This process is carried out during the operation of the system and aims to reconstruct confidential parameters or sensitive internal states. \citep{zhang2024timing}.

LLM agents reduce the professional knowledge threshold and operational costs related to side-channel attacks through autonomous analysis processes, by leveraging capabilities such as task decomposition, tool utilization, multimodal reasoning, and self-correction. A pioneering example is the AgentSCA framework proposed by \citet{yaman2023agent}, which represents the first application of an LLM agent to physical side-channel analysis. The AgentSCA workflow begins with the LLM interpreting hardware descriptions to generate initial probing scripts. Subsequently, it iteratively refines these scripts based on feedback from statistical metrics returned during testing. This process culminates in the production of a complete and directly executable attack vector. AgentSCA \citep{yaman2023agent} was the first to apply LLM agents to physical side channel analysis. The model used an LLM to interpret hardware descriptions to generate initial probing scripts, which it then continuously optimized based on iterative feedback from statistical indicators returned during testing.

\subsubsection{Social engineering \& phishing}\label{sec5.2.4}
Social engineering attacks, which use deception and manipulation to trick individuals into disclosing sensitive information or gaining unauthorized access to systems, are one of the most common and highly successful forms of cybercrime. According to the 2025 Verizon Data Breach Investigations Report \citep{bib199}, social engineering was the second most common pattern of data breach attacks at 17\%. Furthermore, the average social engineering-related breach has climbed to \$4.77 million \citep{bib200}. Phishing, a common social engineering method, involves attackers imitating trusted entities to deceive victims into sharing passwords, clicking on malicious links, or transferring funds. Historical cases such as the spear-phishing attack on SONY executives in 2014 \citep{bib201} and the email intrusion of Hillary Clinton's campaign in 2016 \citep{bib202} have demonstrated their devastating impact.

LLMs are capable of mimicking human conversation patterns and can therefore be exploited to conduct chat-based social engineering (CSE) attacks. Consequently, LLMs can serve dual roles, either as facilitators of such attacks or as instruments for defense \citep{bib203}. Attackers can leverage LLMs to automatically generate malicious prompts to create phishing attacks \citep{bib204}. GPT-4 and GPT-3.5-turbo have been able to generate highly personalized phishing emails at a low cost \citep{bib205}, attacker can even combine VALL-E to carry out multimodal voice fraud \citep{bib208}. However, these attack construction processes are complex and require higher levels of automation. Introducing LLM agents is an effective way \citep{bib207}.

LLM agents have expanded the scope of automated phishing through the interaction of network tools. LLM agents efficiently collect the personal identification information (PII) and real-time network behavior data of the targets to generate highly personalized Trojan horse-style phishing emails. Research indicates that the click-through rate of phishing emails generated by LLM agents is as high as 46.67\%, which is much higher than the click-through rate of traditional phishing emails \citep{wLLMgonline}. Additionally, to enhance the deceptive nature and generation quality of phishing emails, SpearBot \citep{bib210} has developed a generation-evaluation framework. By drawing on the concepts of LLM-as-a-Judge \citep{bib211} and Agent-as-a-Judge \citep{bib212}, this framework conducts collaborative review of the email content. LLM agents can automatically execute the phishing attack chain, including information collection, target profiling creation, email generation and sending, as well as result analysis and self-improvement \citep{bib213}. In multimodal, the voice agent can perform common fraud operations such as logging into bank accounts and completing dual identity verification, further expanding the means and scope of social engineering attacks \citep{bib214}.

\subsection{Post-exploitation}\label{sec5.3}
Post-exploitation refers to a series of actions taken by the attacker after establishing an initial foothold. These include privilege escalation, defense evasion, credential access, information discovery, lateral movement, persistence, data collection \& exfiltration. Through these coordinated steps, the attacker aims to expand the control scope over the target system \citep{skopik2020under}.

In the privilege escalation task, hackingBuddyGPT \citep{happe2023llms} conducted an early exploration. It utilized the inherent knowledge of LLM to enumerate and verify possible privilege escalation paths, and by drawing on the heuristic methods of human penetration testers, it focused the search on high-value attack surfaces to enhance efficiency. However, hackingBuddyGPT performed poorly when dealing with complex reasoning and multi-step vulnerabilities that have temporal dependencies. In addition, AutoAttacker \citep{bib59} utilized RAG to learn from historical successful cases and employed a navigation module to select the optimal action plan from available options. This enabled it to effectively carry out longer, sequential attack tasks such as credential access and lateral movement.

The MalGEN framework \citep{saha2025malgen} achieved persistent control over the target system by driving multi-agent collaboration and generating malicious software capable of creating or modifying system processes. To achieve defense evasion, MalGEN strategically generated code and applied specialized agents to apply techniques such as Base64 encoding and renaming, thereby generating samples that can bypass the static detection of mainstream anti-virus engines. \citet{lupinacci2025dark} utilized in-memory execution payloads, further enhancing the evasion strategy. Because this approach writes no files to disk, it significantly reduces the detection rate by traditional endpoint detection and response systems (EDR). Furthermore, \citet{ITURBE2024104077} indicated that LLM agents can autonomously generate code to conceal their behavioral traces, such as writed scripts to delete files or modified file permissions to erase forensic evidence.

Current agent-empowered post-exploitation faces limitations, including a propensity for errors in multi-step tasks \citep{happe2023llms} and restricted success rates in lateral movement \citep{ITURBE2024104077}. We have found that there is a lack of research on the adaptive defense evasion capabilities that can respond in real time to proactive defense measures such as EDR (Endpoint Detection and Response). Furthermore, research is gap on the offensive tasks that agents perform in credential access, information discovery, and data theft. These collective challenges and unexplored domains present compelling and well-defined directions for future investigation in this field.

\begin{figure}[h!]
    \centering
    \includegraphics[
        width=1\textwidth,
        trim= 0cm 4.5cm 0cm 0cm,
        clip
        ]{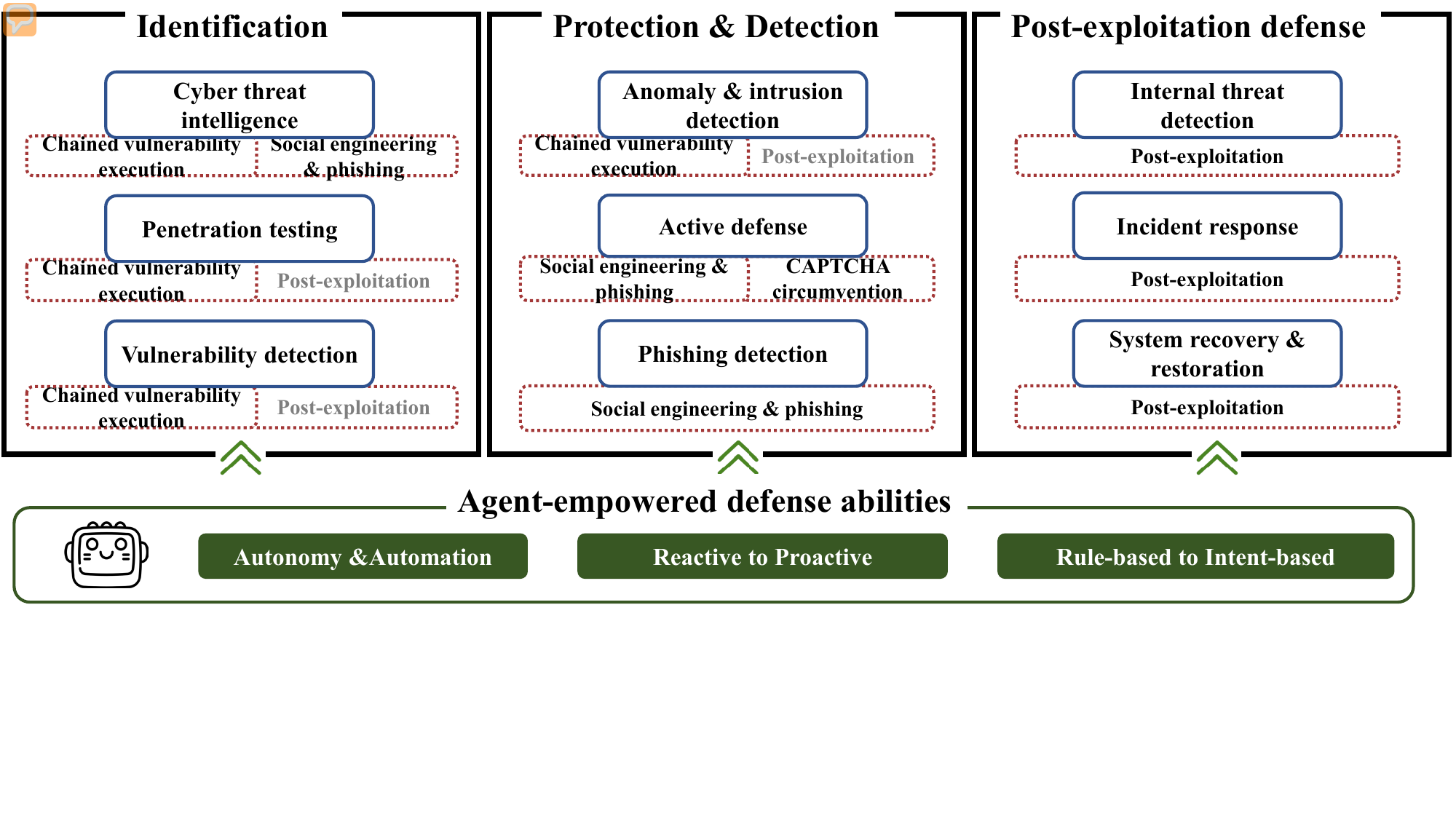}
    \caption{\small The LLM agent empowered defense overview.}
    \label{fig_defense_overview}
\end{figure}
\section{Agent-empowered cyber defense}\label{sec6}
Traditional cybersecurity measures often fall short in addressing dynamic and complex threats, with inherent limitations in efficiency, autonomy, and proactive defense. 
In contrast to rule-based and playbook-driven defense pipelines that automate predefined procedures \citep{kokulu2019matched,stevens2022ready}, LLM agents enable a closed-loop defender that gathers evidence, plans, invokes tools, and revises decisions over multi-step trajectories \citep{li2025sok}. This shift goes beyond efficiency by moving defense from episodic, alert-centered handling to continuous, goal-driven operations under rapid feedback. It also introduces agent-centered attack surfaces, making the governance of how untrusted information steers planning and actions a first-class security concern \citep{liu2025make}.
Accordingly, this chapter systematically elaborates on LLM agent-empowered cyber defense from three core dimensions: identification (§\ref{sec6.1}), protection and detection (§\ref{sec6.2}), and post-utilization defense (§\ref{sec6.3}). 
By enhancing their capabilities in planning, execution, and continuous learning, these agents are transforming their defense strategies from traditional passive, single-point analysis to a more forward-looking, proactive, and adaptive model, thereby overcoming the limitations of existing methods. Figure \ref{fig_defense_overview} highlights the paradigm shift toward Autonomy \& Automation, Reactive-to-Proactive defense, and Rule-based-to-Intent-based operation. Table \ref{tab_map} shows the mapping relationship between LLM agents empowered capabilities and the cybersecurity lifecycle.

\subsection{Identification}\label{sec6.1}
Traditional methods of threat identification, such as manual threat intelligence analysis, expert-led penetration testing, and static vulnerability scanning, suffer from inefficiency and high costs, and incapable of effectively addressing emerging threats. For instance, manual CTI analysis may overlook crucial data relationships \citep{shah2024ai}, penetration testing is difficult to be carried out on a large scale \citep{mckinnel2019systematic}, and static scanners are unable to detect zero-day vulnerabilities or business logic flaws \citep{guo2023review}. Recent studies use LLMs for knowledge extraction, but this knowledge is applied reactively \citep{yan2025guiding}. To address shortcomings, researchers have begun to incorporate active and continuous LLM agent systems.

LLM agents enhance the capabilities of network threat intelligence (§ \ref{sec6.1.1}), penetration testing (§ \ref{sec6.1.2}), and vulnerability detection (§ \ref{sec6.1.3}). They enable autonomous reconnaissance in the threat identification phase by actively synthesizing multi-source threat intelligence, simulating multi-step attack paths, and dynamically triaging vulnerabilities. The subsequent sections of this chapter will explore the application of these core agent capabilities in the aforementioned three areas.

\subsubsection{Cyber threat intelligence}\label{sec6.1.1}
Cyber threat intelligence (CTI) focuses on the extraction, structuring, correlation, evaluation, and dissemination of threat-related facts, relationships, and tactics, techniques, and procedures (TTPs) derived from open-source intelligence (OSCTI), commercial sources, and local telemetry. This process enables informed decision-making for downstream activities, including threat hunting, detection, and response \citep{Conti2018}.

Early approaches to leveraging LLMs for CTI analysis centered on building linear single automated process, extracting structured data from unstructured reports to populate the knowledge graph \citep{10628558}. LLM-TIKG proposed by \citet{HU2024103999} adopts a linear process from text perception to reasoning-based extraction and triplet generation. However, these methods have limitations in handling a large volume of documents, and are vulnerable to the hallucination of LLMs. The researchers utilized the multi-agent framework to counter these challenges. \citet{huang2024ctikg} proposed the CTIKG framework to enhance the accuracy of information extraction through multi-agent collaboration, and decomposed tasks among multiple roles to further improve the robustness of the framework. In addition, CTIKG designs a dual-memory mechanism to establish separate short-term memories for handling triples and text fragments, and is equipped with a long-term memory for global integration.

The key of generating actionable CTI lies in the integration of universal global threat information with the specific context of an organization \citep{balasubramanian2025generative, Conti2018}. The capabilities of agent tool invocation and local memory can effectively bridge the gap between global and local knowledge. LocalIntel \citep{10.1007/978-3-031-87496-3_5} employs RAG to integrate a local knowledge base, with global CVE/NVD intelligence. This process yields organization-specific reports with 93\% accuracy. In contrast, relying solely on a large language model without this local data injection causes a 27\% reduction in relevance.

\subsubsection{Penetration testing}\label{sec6.1.2}
PentestGPT \citep{bib18} significantly improved the coverage and accuracy of penetration testing by simulating the collaborative dynamics of human penetration testing teams and combining high-level strategies with precise execution via coordinated reasoning, generation, and parsing modules. Compared with the direct application of GPT-3.5 and GPT-4, the subtask completion rate was increased by 228.6\% and 58.6\%, respectively. PenHeal \citep{bib189} has expanded upon the limitations of PentestGPT's lack of automation, enabled agents to automatically execute test tasks, explore multi-path attacks, utilize external knowledge, automatically generate repair suggestions, and evaluate repair schemes, so as to improve coverage and accuracy of penetration testing. AutoPenBench \citep{bib26} demonstrated that human-machine collaboration improves generative agent efficiency. On this basis, faced with the challenge of a new penetration testing task that had never been learned before, researchers met this challenge through the interaction of human operators and corresponding prompt engineering techniques, improving performance on specific security tasks over time \citep{bib190}. LLMSQLi \citep{bib191} went a step further and relied on Agent-Generation in the multi-agent framework to generate customized payloads according to specific test objectives and test environments, thereby improving the detection success rate in specific test scenarios.

At the same time, \citet{bib194} demonstrated that agent systems augmented with Retrieval-Augmented Generation (RAG) \citep{bib192} could enhance penetration testing efficiency, especially in the enumeration and privilege escalation stages. On this basis, PentestAgent \citep{bib193} combined more fine-grained agent division of labor, such as reconnaissance, search, planning, and execution of attacks, to improve effectiveness. AutoPT \citep{bib14} designed a penetration testing state machine to control the behavior of agents to avoid the problem of agents falling into an infinite loop. However, a challenge arises when the default prompts configurations often generate insufficient contextual information, which leads to model forgetting \citep{bib194}. Future research should pay attention to advancing prompt configuration design and improving the LLM model to address its limitations at different stages.

\subsubsection{Vulnerability detection}\label{sec6.1.3}
Recently, significant progress has been made in using multi-agent systems to enhance vulnerability detection capabilities. Notably, iAudit \citep{bib184} improved the F1-score of vulnerability detection to 91.21\% by combining fine-tuning with two agents, Ranker and Critic, to iteratively select and debate the most appropriate vulnerability cause from multiple Reasoner model outputs. Similarly, AutoSafeCoder \citep{bib185} constructed three types of agents, which combined static and dynamic analysis methods to effectively identify vulnerabilities in the code generation process, thereby enhancing traditional single-agent code generation approaches. GPTLENS \citep{bib186} minimized false alarms by decomposing the traditional single-stage detection into two cooperating and adversarial agents. Meanwhile, LLM-SmartAudit \citep{bib187} used a multi-agent session approach to perform contract code analysis, vulnerability identification, and comprehensive reporting through a specialized agent collaboration system, and introduced a role exchange mechanism to improve detection accuracy, although there are still some limitations in dynamic vulnerability identification. Notably, Big Sleep \citep{bib188}, a collaboration between Google's Project Zero and Google DeepMind, discovered the first exploitable memory vulnerability discovered by an LLM agent (discovering an exploitable stack buffer underflow in SQLite). Taken together, these studies show that an LLM agent has great potential in vulnerability detection.

\subsection{Protection \& Detection}\label{sec6.2}
The protection and detection phase aims to neutralize threat propagation during initial or attempted breaches of the external perimeter. Conventional defense mechanisms have traditionally relied on rule-based signature matching, static threshold alerts, and isolated response frameworks \citep{9163041,10117505}. However, these methods have obvious limitations. They create detection blind spots, generate a large number of false positives, have a delayed response to zero-day attacks and advanced persistent threats (APTs). \citep{10508380,7805250}. This has prevented them from being able to cope with the constantly changing attack methods and the complex social engineering tactics. The recent applications of LLM have enhanced the analytical capabilities through improved data association and semantic understanding \citep{elhafsi2023semantic, 10.1007/978-3-031-57942-4_44, 10527274}. However, it lacks autonomy, planning, and real-time interaction capabilities \citep{zhang2025agent}, determining that LLM is limited to the role of an advanced analyst and cannot become an independent and autonomous agent capable of performing defensive operations.

To address the limitations of LLMs, LLM agents are integrated into the protection and detection framework, transforming the traditional passive monitoring into intelligent behavioral prediction. By equipping the defense system with the capabilities of semantic analysis, multimodal data fusion, and adaptive decision-making, these agents can find potential attack patterns from a large amount of log data and network traffic. LLM agents empower the protection and detection stage focus on three domains: anomaly \& intrusion detection(§ \ref{sec6.2.1}), active defense(§ \ref{sec6.2.2}), and phishing detection(§ \ref{sec6.2.3}).
\subsubsection{Anomaly \& intrusion detection}\label{sec6.2.1}
In the fields of financial markets, cyber-physical systems (CPS), and intrusion detection, traditional anomaly detection methods usually rely on preset quantitative algorithms and a large amount of manual intervention. \citep{bib153}. It is prone to being influenced by human errors and biases, and it is difficult to respond promptly to new threats in a complex and ever-changing environment. LLM agent systems significantly enhance both the efficiency and accuracy of anomaly detection. The AUDITE-LLM \citep{bib154} framework can enhance the ability of anomaly detection in multiple dimensions by coordinating multiple agents to automatically perform data transformation, anomaly verification, tool invocation and evidence debate. \citet{bib155} demonstrated that assigning different specialized roles to different participants reduced the reliance on manual intervention and enhanced the response speed and detection accuracy of the system in the financial market. \citet{bib156} used agents to improve decision-making and maintenance processes in dynamic and complex environments, through real-time data integration, knowledge graphs, and external tools, leveraging LLM reasoning and planning capabilities to provide contextual insights and enhance anomaly detection capabilities in the shipping industry.

Traditional Intrusion Detection Systems (IDS) are hampered by high false alarm rates and inability to detect zero-day attacks \citep{bib157,bib158}. In contrast, LLM-based IDS improves detection accuracy through context-awareness and continuous learning capabilities, and achieves self-adaptation in the dynamic environment \citep{bib159}. IDS-Agent \citep{bib160} further enabled the agent to have the flexibility to select tools on demand by integrating multiple tools. Furthermore, the potential of LLM agents in CPS has been verified. AnomalyLLM \citep{bib161} automatically extracted information from documents, generated hypotheses, verified hypotheses, and monitored data streams in real time. It efficiently identifies abnormal patterns that deviated from expected system behavior, reduces the need for human intervention, and enhanced the intelligence and scalability of the system.

\subsubsection{Active defense}\label{sec6.2.2}
Active defense is a proactive security strategy that employs dynamic, deceptive, or counter-offensive techniques to disrupt an ongoing attack by actively engaging and misleading the adversary \citep{11005994}. The application of LLM agents in the field of active defense is in a preliminary phase. \citet{ayzenshteyn2024best} introduced the first active defense framework designed to exploit the inherent weaknesses of the adversary's autonomous attackers in order to disrupt their hostile actions. This strategy achieved a 90\% overall success rate under the black-box. Furthermore, the research further highlighted the unique advantages of dynamic honeypot injection, by leveraging the attack agent's own tool-calling and feedback mechanism, this approach achieved a deception rate of 95\%, significantly surpassing static prompt injection.

\subsubsection{Phishing detection}\label{sec6.2.3}
Phishing detection aims to identify and prevent emails, websites, or messages where adversaries impersonate trusted entities to steal sensitive information. Traditional methods such as heuristic and blacklist-based methods offer speed but limited robustness \citep{aljofey_effective_2022, SILVA2020101613, SAFI2023590}. Classical machine learning techniques have improved recall but often fail to generalize to novel linguistic variations. Reference-based detection systems, while effective in some contexts, encounter difficulties when dealing with emerging brands. DynaPhish framework \citep{bib216} addresses this challenge through runtime reference expansion and counterfactual interaction, thereby improving recall without substantially reducing precision. Nevertheless, such systems remain vulnerable to highly creative attacks. To mitigate these limitations, LLM agent empowered systems combine the reasoning capacities of LLM with tool integration and multi-step collaboration, enabling dynamic multimodal analysis that adapts to novel threats. This section examines three principal mechanisms through which agent-based systems enhance phishing detection: evidence-augmented multimodal perception, multi-agent collaboration, and iterative evolution.

Evidence-augmented multimodal perception. Detector converts a sparse suspect into a rich evidence bundle by calling tools, then enforces cross-modal consistency for phishing detection. \citet {bib217}  proposed PhishAgent, which operationalizes this by combining page screenshots, HTML and logos with an offline+online knowledge base. A multimodal retriever surfaces top-k brand candidates, and the agent checks brand–domain alignment to cut both false positives and false negatives with single-iteration latency. For ultra-short SMS, \citet{wang2025can} proposed SmishX, which extracts URLs and brand mentions, automatically collects redirect chains, domain history and page screenshots, and only then performs CoT reasoning. SmithX outputs evidence-bound explanations that improve end-user accuracy and usability at scale. 

Multi-agent collaboration. For emails, \citet{xue2025multiphishguard} introduced MultiPhishGuard, which orchestrates five roles and applies PPO-based weighting to adjust each role’s influence per sample, and hardens the system in a self-improving loop, achieving high accuracy with very low FP/FN. Beyond effectiveness, researchers have also examined the cost implications of multi-agent systems. \citet{bib218} proposed a two-tiered framework in which a lightweight agent first inspects the URL, followed by a more computationally intensive multimodal agent that analyzes webpage screenshots. This hierarchical design balances detection performance with economic cost.

Iterative evolution. Against LLM-rephrasing threats that degrade classical and LLM-only detectors, closed-loop attack–defense co-evolution strengthens robustness. \citet{10825007} showed LLM-driven rewriting can evade filters and demonstrated defensive prompt optimization to recover accuracy. Broader analyses emphasize the arms race where LLMs empower both attackers and defenders, reinforcing the need for adaptive fusion and continuous hardening \citep{10825018}. 

\subsection{Post-exploitation defense}\label{sec6.3}
Post-exploitation defense encompasses measures to detect, contain, and remediate threats operating within a breached network, covering threat detection, incident response (IR), and system recovery \citep{cichonski2012computer, beretas2024information}. These measures primarily target common post-exploitation tactics such as privilege escalation, defense evasion, credential access, discovery, lateral movement, persistence, collection, and data exfiltration. However, conventional methods suffer from critical flaws, including detection latency \citep{SUBSTANTIATION}, fragmented IR workflows \citep{ahmad2015case}, and a lack of automated recovery mechanisms \citep{lin2024optimization}. These deficiencies render them ineffective against long-term, stealthy attack chains that leverage legitimate credentials. In contrast, LLM agents leverage contextual understanding and autonomous planning to connect a series of seemingly benign actions into a coherent attack chain, thereby revealing hidden malicious intentions.

The application of LLM-based agents to post-exploitation defense can be organized into three major domains: threat detection (§ \ref{sec6.3.1}), incident response (§ \ref{sec6.3.2}), and system recovery \& restoration (§ \ref{sec6.3.3}). The subsequent sections of this chapter will provide a detailed analysis of how these basic capabilities are applied in the aforementioned three defense phases.

\subsubsection{Internal threat detection}\label{sec6.3.1}
Post-exploitation threat detection focuses on the period after the attacker has gained initial access but before they cause damage to critical assets, aiming to identify and dismantle the series of attack strategies. Unlike traditional methods that rely on known signatures or isolated anomalies, internal threat detection requires reasoning on long-term and diverse data sources \citep{hutchins2011intelligence,hillstone}. By integrating the LLM agent capabilities, such as autonomous reasoning, tool calling, memory, and multi-source information fusion, the ability to explain contextual anomalies in logs, behavioral data, and network traffic can be enhanced. This method enhances the analysis efficiency and helps to flexibly adapt to the constantly changing attack chain.

This section explores three key mechanisms by which post-exploitation threat detection is improved by agent-based systems: the use of iterative reasoning loops, multi-agent collaborations, and policy-based access control. We also analyze the significant new increments that obtain when these methodologies for agents are combined.

\textbf{Single-agent with iterative loop}. Early studies of applying LLM to security logs focused mainly on semantic interpretation, especially through narrative transformation. \citet{portnoy2024towards} reformulated endpoint logs into coherent narratives, enabling LLMs to detect complex threats such as credential theft from contextual cues. However, these methods were limited by the static context window and lacked the mechanism for iterative analysis. LLM agent's capacity for dynamic interaction and iterative optimization enables it to implement adaptive response loops that are beyond the capabilities of standard LLMs. \citet{narajala2025securing} proposed an agent with an embedded planning loop, which is used to analyze events in different stages of an attack, thereby enhancing its ability to detect evasion behaviors. To break free from the constraints of the traditional fixed analytical workflow, LogRESP-Agent \citep{app15137237} introduced a recursive framework based on the "plan-action-check" cycle. Enabling it to continuously adjust the analysis strategy based on the intermediate results, achieving true adaptive analysis.

\textbf{Multi-agent collaborations for enhanced contextual correlation}. Multi-agent improves threat detection and identifies hidden activities by establishing context correlations among various data sources \citep{de2025open, yang2025large}. The Audit-LLM framework proposed by \citet{song2024audit} is designed for log analysis. It decomposes the internal threat detection task into several sub-problems through 3 collaborating agents, thereby enhancing the depth and reliability of the analysis. \citet{11141466} utilized multi-agent systems for cross-domain correlation. Their framework leverages three specialized agents for log analysis, email verification, and IP scanning. Correlating the outputs from multiple sources increased the success rate of detecting persistence and reduced the false positive rate by 41.3\%.

\textbf{Policy-driven access control}. A more forward-looking detection strategy is to establish policy-based access controls, aiming to continuously verify the agent behavior and prevent their abuse or hijacking. GuardAgent \citep{bib135} employed Retrieval-Augmented Generation (RAG) to convert security requirements into executable rules, enabling it to detect and prevent malicious activities that deviated from established policies in real-time. \citet{bib296} introduced Progent to implement a deterministic strategy at the tool-calling interface to prevent persistent and lateral movement attempts following privilege escalation. Furthermore, to address the issue of trust boundary in data streams, \citet{kim2025prompt} employed data annotation and runtime protection measures to prevent privilege escalation triggered by untrusted data. It is worth noting that these policy-based internal control methods create a dynamic defensive front within the cyber system, synergistically hardening the agent self-security.

LLM agents have overcome the traditional detection reliance on static and predefined rules, shifted the focus of threat detection from event detection to the inference of malicious intentions. Agents reveal the attackers' goals by inferring the series of behaviors in relation to the context to identify the hidden connections between events. For instance, frameworks like AGrail \citep{bib292} and GuardAgent \citep{bib135} can directly convert natural language specifications into executable security policies.

\subsubsection{Incident response}\label{sec6.3.2}
Incident Response (IR) refers to the measures taken to contain and eliminate the intrusion once it is confirmed that the internal system has been compromised, usually following standards and specifications such as the NIST framework \citep{cichonski2012computer}. Traditional IR mainly relies on structured protocols and automated processes \citep{HE2022102435}. Subsequent studies have introduced LLM into IR, initially applying it to text tasks such as log analysis and alarm interpretation \citep{hays2024employing}. However when \citet{kramer2025integrating} systematically evaluated the capability of LLMs to generate incident summaries, they found that without structured reasoning and tool support, LLMs struggle to internalize the security knowledge required for complex incident response, and rely on human collaboration to produce effective results. These limitations provide a strong basis for adopting an LLM-based agent architecture. The agent relies on task-level autonomy and multi-agent collaboration to enhance execution efficiency, decision-making quality and the ability to handle complex threats.

\textbf{Task-level autonomy}. To overcome the inherent limitations of LLMs on raw unstructured data, researchers have turned to developing dedicated LLM agents. \citet{fumero2025cybersleuth} introduced CyberSleuth, which is an autonomous blue team agent designed for digital forensics in network attacks. CyberSleuth can effectively identify target services, locate common vulnerabilities and exposures CVEs, and assess whether the attack was successful by calling external tools to parse network packet captures and system logs. In concurrent work, \citet{molleti2024automated} demonstrated that the agent achieves automated threat response through the contextual analysis of logs and user behaviors.

\textbf{Multi-agent collaboration}. Multi-agent collaboration simulating human incident response teams has emerged as an effective approach to addressing security incident complexity. \citet{lin2025ircopilot} introduced IPCopilot framework, set 4 agents that follow the inference-action-reflection loop to simulate human teams, and by using the incident response tree to decompose the large target into traceable sub-tasks, thereby suppressing long context loss and logical breaks. \citet{11012055} proposed an event response simulation platform called AutoBnB, which is based on the 'Backdoors \& Breaches' game. AutoBnB-RAG \citep{liu2025autobnb} extends AutoBnB by integrating RAG to overcome the agent knowledge gap, allowing agents to retrieve external data to refine their decision-making.

\subsubsection{System recovery \& restoration}\label{sec6.3.3}
System recovery and restoration is the process of returning compromised systems, data, and operations to a secure, functional state following a cybersecurity incident \citep{cichonski2012computer}. In this context, the agent's role extends beyond the mere execution of pre-configured scripts to encompass dynamic planning, autonomous decision-making, and closed-loop control.

Multi-agent collaboration is a common method for system recovery and reconstruction. The IPCopilot framework proposed by \citet{lin2025ircopilot} focuses on external threats, employing four agent roles to optimize recovery strategies based on attack traces. And \citet{huang2024resilience} studied the internal resilience of agents' own system. They demonstrated that multi-agent collaboration remains stable even with faulty agents, and hierarchical structure systems are the most effective in error recovery. The active fault tolerance and scalability response features of the multi-agent architecture enable the system to self-repair with minimal intervention. This feature has been applied to Cyber-Physical Systems \citep{info16050365} and commercial network resilience tools, like Druva \citep{Druva}.


\section{Benchmarks}\label{sec7}
Standardized evaluation methods are crucial for systematically comparing LLM agents self-security and empowered cybersecurity. 
This section provides a categorized overview of representative benchmarks for these two settings, while aligning them with the evaluation signals and constraints that make results comparable. Meaningful security evaluation for autonomous agents should jointly report trajectory-level safety (including delayed failures), policy/authorization compliance, and utility preservation under realistic tool/memory interactions. Accordingly, we first summarize commonly used metrics as a compact taxonomy, then review benchmarks for LLM agent self-security and empowered cybersecurity. We further summarize best attained results when experimental settings are comparable, and finally discuss key evaluation gaps and future benchmarking directions.

\begin{table}[t]
\caption{Evaluation metrics for LLM agent self-security}
\label{table-selfsecurityMetrics}
\setlength{\tabcolsep}{3pt}
\begin{tabularx}{\linewidth}{@{}p{2cm} p{3cm} X@{}}
\toprule
\textbf{Dimension} & \textbf{Metric} & \textbf{Description} \\
\midrule
Attack  & ASR &
Fraction of trials where the attacker objective is achieved. \\
Attack  & Robustness &
The ability of an attack to remain effective despite changes. \\
Attack  & Stealthiness &
The ability of an attack to bypass security guardrails without triggering alerts. \\
Attack  & Attack cost &
Effort required to reach a successful attack; lower is stronger. \\
\midrule
Defense & $\Delta$ASR  &
Effect size: how much ASR drops after enabling the defense. \\
Defense & P / R / F1 &
Catch attacks vs.\ avoid false alarms (report all or F1). \\
Defense & FPR  &
Benign inputs wrongly blocked (over-defense). \\
Defense & Overhead  &
Runtime latency and resource overhead introduced by defense. \\
\midrule
Safety  & Harmfulness rate &
Fraction of outputs/actions judged unsafe (lower is safer). \\
Safety  & Refusal rate &
Fraction of risky requests refused (interpret with over-refusal). \\
Safety  & Over-refusal &
Fraction of benign requests refused (safety--utility cost). \\
\midrule
Utility & AU &
Completion/utility on clean (no-attack) tasks; define per benchmark. \\
Utility & UA  &
Task success while attacks are present (user objective preserved). \\
\bottomrule
\end{tabularx}
\vspace{2pt}
{\footnotesize ASR: Attack Success Rate; $\Delta$ASR: ASR reduction after enabling a defense; P: Precision; R: Recall; F1: F1-score; FPR: False Positive Rate; AU: (Clean) Utility/Task Success on benign inputs; UA: Utility Under Attack; CU: Clean Utility.}
\end{table}

\begin{table}[t]
\caption{Benchmarks(Datasets) for LLM-agent self-security}
\label{tab-selfsecurityDataset}
\setlength{\tabcolsep}{3pt}
\begin{tabularx}{\linewidth}{@{} p{4cm} p{3.5cm} p{1.5cm} X @{}}
\toprule
\textbf{Benchmark(Dataset)} & \textbf{Core domain} & \textbf{Scale} & \textbf{Primary metric (Best result)} \\
\midrule
InjecAgent \newline \citep{bib5} & IPI & 1,054\newline2(scenario) & ASR$\uparrow$$\rightarrow$47\% \newline(GPT-4) \\
\midrule

AgentDojo\newline\citep{bib81} & IPI  & 629\newline4 & ASR$\uparrow$$\rightarrow$47.69\% / UA$\uparrow$$\rightarrow$50.08\% \newline(GPT-4o)\\
\midrule

ASB\newline\citep{bib34-ASB} & DPI \& IPI \& Poisoning \& Backdoor & 400\newline10  & ASR$\uparrow$$\rightarrow$64.41\%(average) \newline(GPT-4o)\\
\midrule

ST-WebAgentBench\newline\citep{bib7}  & IPI(web-agent) & 646\newline6  & CuP$\uparrow$$\rightarrow$20\% / Risk Ratio$\downarrow$$\rightarrow$0.44\%\newline(AWM\citep{wang2024agent}) \\
\midrule

WASP\newline\citep{evtimov2025wasp} & IPI(web-agent) & 84\newline42 & ASR$\uparrow$$\rightarrow$42.9\%(interm.) / Utility$\uparrow$$\rightarrow$62.2\% \newline(GPT-4o with axtree+SOM)\\
\midrule

LLMail-Inject\newline\citep{abdelnabi2025llmailinjectdatasetrealisticadaptive} & IPI(tool,email) & 208,095\newline- & - \\
\midrule

AgentHarm\newline\citep{bib139} & Jailbreak  & 440\newline11 & Harm Score$\uparrow$$\rightarrow$72.7\% / Refusal rate$\uparrow$$\rightarrow$48.9\%\newline(GPT-4o)\\
\midrule

SafeMTData\newline\citep{bib115} & Jailbreak(multi-turn) & 2,280\newline6 & - \\
\midrule

OS-Harm\newline\citep{kuntz2025harm} & IPI \& Jailbreak \& Misbehavior & 150\newline3 & ASR$\uparrow$$\rightarrow$21\%(average) \newline(GPT-4.1)\\
\midrule

HarmBench\newline\citep{mazeika2024harmbench} & Jailbreak \& Adversarial Prompt & 510\newline4 & - \\
\midrule

ToolEmu\newline\citep{bib30} & Tool failures  & 144\newline-  & Failure Inc.$\downarrow$$\rightarrow$39.4\% \newline(GPT-4) \\
\midrule

ToolSafety\newline\citep{xie-etal-2025-toolsafety} & Tool failures & 14,290\newline- & - \\
\midrule

R-Judge\newline\citep{bib142} & Risk awareness &569\newline10 & F1-score$\uparrow$$\rightarrow$74.45\% \newline(GPT-4o) \\

\bottomrule
\end{tabularx}
\vspace{2pt}
{\footnotesize IPI: Indirect Prompt Injection; DPI: Direct Prompt Injection; ASR: Attack Success Rate; UA: Utility Under Attack; CuP: Completion Under Policy; AWM: Agent Workflow Memory; SOM: Set-of-Marks (web-page element tagging); axtree: Accessibility tree; Inc.: Incidence; F1-score: F1 score (harmonic mean of Precision and Recall); $\uparrow$/$\downarrow$: better direction.}
\end{table}

\subsection{Benchmarking LLM agent self-security}\label{sec7.1}
We summarize LLM agent self-security benchmarks across two categories: \textbf{attacks and defenses} \& \textbf{security and credibility}. Table~\ref{table-selfsecurityMetrics} outlines evaluation metrics, and Table~\ref{tab-selfsecurityDataset} lists representative benchmark (datasets) with their best primary results. Note that these best results are not for cross-benchmark comparison due to differing settings.

\subsubsection{Benchmarking attacks and defenses}\label{sec7.1.1}
There are several benchmarks released for assessing attacks and defenses for LLM agents, also comes with datasets, most built around prompt injection, jailbreak, and adversarial attack. 
InjecAgent \citep{bib5} was developed for testing the vulnerability of tool-integrated agents when exposed to indirect prompt injection (IPI). 
For web-agent settings, ST-WebAgentBench \citep{bib7} and WASP \citep{evtimov2025wasp} provide IPI-oriented test suites with enterprise-like or isolated web environments, respectively. LLMail-Inject \citep{abdelnabi2025llmailinjectdatasetrealisticadaptive} further contributes a large-scale, challenge-derived dataset for email-agent IPI, complementing benchmark-style evaluations with high-diversity adaptive attacks.
JAILJUDGE \citep{bib138} and AgentHarm \citep{bib139} are for testing jailbreak and harmful multi-step behaviors. 
HarmBench \citep{mazeika2024harmbench} provides a standardized harmful behavior test suite commonly used to stress-test robust refusal under adversarial prompting, while SafeMTData \citep{bib115} targets multi-turn jailbreak trajectories that better match agent-style interactions.
There are also research on PoT backdoor attacks and memory poisoning attacks \citep{bib34-ASB}. Additionally, AutoAdvExBench \citep{bib140} explored the ability to automatically generate adversarial examples to break through defense systems.
OS-Harm \citep{kuntz2025harm} extends the attack surface to GUI-style agent operation, covering injection- and misbehavior-related risks in interactive system tasks.

From the perspective of evaluation dimensions, these resources show a trend from single-dimension reporting to multi-dimensional evaluation. Early benchmarks such as InjecAgent \citep{bib5} focused on agent performance under a single attack type, while later frameworks such as AgentDojo \citep{bib81} and ASB \citep{bib34-ASB} established a more comprehensive evaluation system. 
AgentDojo couples dynamic environments with security tests to expose adaptive, state-dependent injection behaviors, whereas ASB broadens the evaluation to multiple operational stages of agent systems (e.g., system prompts, tool usage, and memory retrieval) across 10 scenarios.
This shift is often accompanied by reporting security outcomes together with utility-related signals (e.g., whether benign task performance is preserved), rather than relying on a single success rate.

\subsubsection{Benchmarking security and credibility}
Beyond benchmarks evaluating agent responses to specific attack types, there are also evaluation resources that assess the inherent security and credibility of agent capabilities. Athena \citep{bib29-Athena} introduced language contrastive learning and interaction-level criticism, using historical safe and unsafe trajectories as comparison examples to improve agents' risk aversion during task execution. The framework evaluates security at the trajectory level and provides interaction-level checks, with 80 toolkits and 8 categories of security evaluation benchmarks to simulate diverse real-world risks.
In addition, ToolEmu \citep{bib30} evaluated agent security under different tools and scenarios by simulating tool execution, integrating an automatic security evaluator to quantify potential failure risks. It was found that even the most secure LLM agent still has failure risks in 23.9\% of test cases.
Complementing simulation-based evaluation, ToolSafety \citep{xie-etal-2025-toolsafety} releases a large-scale dataset for tool-call safety, enabling supervised analysis and training for reducing tool failures and unsafe tool usage.
R-Judge \citep{bib142} focused on the ability of LLMs to identify and judge agent security risks in multi-round interaction environments, utilizing 569 interaction records to emphasize behavioral security assessment in open agent scenarios.

ST-WebAgentBench \citep{bib7} proposed a benchmark evaluation for web agents in enterprise environments, which not only focused on task completion but also supported human-in-the-loop operation, allowing agents to request human guidance when uncertain, thus improving the realism and security of evaluation. GuardAgent \citep{bib135} extended the toolbox to adapt to different LLM agents and proposed two new benchmarks, EICU-AC and Mind2Web-SC, for the medical and web fields respectively, to evaluate privacy access control and security control capabilities of agents. Through two stages of simulation and optimization, ALI-Agent \citep{bib143} automatically generated diverse test scenarios, identified the mismatch between LLM and human values, and improved the adaptability and depth of evaluation. BELLS \citep{bib144} evaluates the ability of multi-agent system defenses to identify anomalous behavior in trajectories composed of multiple API calls.

Taken together, the attack-focused benchmarks in §\ref{sec7.1.1} and the security and credibility resources in this subsection complement attack-centric evaluations by measuring trajectory-level safety, tool-use correctness, access-control compliance, and risk awareness, providing reusable evidence for both offline analysis and interactive testing. They cover parts of long-horizon risks (multi-turn trajectories), emergent behaviors (multi-agent traces), and deployment constraints within web environments involving policy oversight or human interaction, but most protocols still rely on short, static task suites and single-episode scores, which under-measure delayed failures and strategy drift under changing tools and environments.

\begin{table}[t]
\caption{Evaluation metrics for LLM-agent empowered cybersecurity}
\label{tab-empoweredMetrics}
\centering
\setlength{\tabcolsep}{4pt}
\begin{tabularx}{\linewidth}{@{}p{2cm} p{3cm} X@{}}
\toprule
\textbf{Dimension} & \textbf{Metric} & \textbf{Description} \\
\midrule
Effectiveness & SR/ASR & End-to-end goal achievement under a fixed evaluation budget. \\
Effectiveness & F1/FPR & Balancing overall correctness and false alarms in detection/extraction tasks. \\
Reliability &
Pass@k & Probability of at least one success within $k$ independent attempts. \\ 
Efficiency &
CPS & Average resource cost required to obtain one successful episode. \\
Efficiency &
TTS/MTTD/MTTR & Time to first success, or mean time to detect/respond in SOC/IR workflows. \\
Coverage &
COV & Breadth of solvable scenarios. \\
Coverage &
PR & Fraction of predefined milestones completed in long-horizon chains. \\
\bottomrule
\end{tabularx}
\footnotesize{SR = Success Rate; ASR = Attack Success Rate; F1 = F1 Score; FPR = False Positive Rate; Pass@k = Pass at $k$; CPS = Cost per Success; TTS = Time-to-Success; MTTD = Mean Time to Detect; MTTR = Mean Time to Respond; COV = Coverage; PR = Progress Rate.}
\end{table}

\subsection{Benchmarking LLM agent empowered cybersecurity}\label{sec7.2}
As established in Chapters 5 (§ \ref{sec5}) and Chapters 6 (§ \ref{sec6}), LLM agents are playing a significant role in the cybersecurity landscape. This trend demands a systematic evaluation to define their capability boundaries in both domains. Table~\ref{tab-empoweredMetrics} summarizes commonly used evaluation dimensions and metrics for agent-empowered cybersecurity, and benchmark results are often sensitive to operational constraints, so comparability hinges on matched settings. These benchmarks also serve as reusable task suites and datasets that support consistent testing across models and methods. Existing security benchmarks for these agents are typically categorized as offensive, defensive, and general. Table \ref{tab-empoweredBenchmark} compiles representative benchmarks in these categories together with their scope, scale, and reported best results under the original protocols; as in the self-security setting, these numbers are intended as within-benchmark references rather than for cross-benchmark ranking.

\begin{table}[t]
\caption{Benchmark Compendium for LLM agent empowered cybersecurity}\label{tab-empoweredBenchmark}%
\setlength{\tabcolsep}{3pt}
\begin{tabularx}{\linewidth}{@{}p{3cm} p{3cm} p{2cm} p{1cm} X@{}}
\toprule
\textbf{Benchmark(Dataset)} & \textbf{Core domain} & \textbf{Type} & \textbf{Scale} & \textbf{Primary metric (Best result)}\\ \midrule
 \mbox{Offense}\\ \midrule \midrule
PentestGPT \newline\citep{bib18} & PT & OWASP Top10 & 182 \newline 13(scenario) & - \\
\midrule
Cybench \newline\citep{zhang2024cybench}& PT & CTF & 40\newline6 & SR$\uparrow$$\rightarrow$22.5\% / PR$\uparrow$$\rightarrow$40.1\% \newline / TTS$\downarrow$$\rightarrow$52min (GPT-4o,Pass@3) \\
\midrule
HackSynth \newline\citep{muzsai2024hacksynth}& PT & CTF & 200\newline6 & SR$\uparrow$$\rightarrow$40.0\% / TTS$\downarrow$$\rightarrow$2min  (GPT-4o with OverTheWire\citep{OvertheWire}) \\
\midrule
AutoPenBench \newline\citep{bib26}& PT & CTF \& CVE & 33\newline4 &  SR$\uparrow$$\rightarrow$64\% / PR$\uparrow$$\rightarrow$53\% \newline(GPT-4o)\\
\midrule
CVE-Bench \newline\citep{zhu2025cvebench}& Web vulnerability & CVE & 40\newline10 & SR$\uparrow$$\rightarrow$10\% / TTS$\downarrow$$\rightarrow$264s  (GPT-4o with AutoGPT\citep{bib252}, Pass@5)\\
\midrule
NYU-CTF-Bench \newline\citep{NEURIPS2024_69d97a64}& Offense lifecycle & CTF & 200\newline6 & SR$\uparrow$$\rightarrow$4.91\% / FPR$\downarrow$$\rightarrow$24.88\% \newline(GPT-4)\\
\midrule
3CB \newline\citep{anurin2024catastrophic}& Offense lifecycle & MITRE ATT\&CK & 15\newline14 & SR$\uparrow$$\rightarrow$73\% / COV$\uparrow$$\rightarrow$100\%(15/15) \newline(GPT-4o)\\
\midrule
\mbox{Defense}\\ \midrule \midrule
 ExCyTIn-Bench \newline\citep{wu2025excytin}& CTI & Azure \newline tenant log & 589\newline8 & SR$\uparrow$$\rightarrow$62\% \newline(GPT-4o, Pass@10) \\
 \midrule
 CFA-Bench \newline\citep{11129512}& Forensic \& \newline vulnerability detection & CVE & 20\newline- & SR$\uparrow$$\rightarrow$23.5\% \newline(GPT-4o) \\
\midrule
\mbox{Comprehensive (offense \& defense)}\\ \midrule \midrule
\citet{phuong2024evaluating}& CTF \& \newline vulnerability detection & Comprehensive & N/A \newline5& COV$\uparrow$$\rightarrow$29.62\%(81 task) / Acc.$\uparrow$$\rightarrow$74.0\%\newline(Gemini Ultra 1.0 with \citep{wang2019detecting})\\
\midrule
CyberSecEval 4 \newline\citep{CyberSecEval-4}& cybersecurity \newline lifecycle & Comprehensive & N/A\newline9 & - \\
\bottomrule
\end{tabularx}
\footnotesize{PT: Penetration testing; CTI: Cyber threat intelligence; SR: Success Rate; PR: Progress Rate; TTS: Time-to-Success; Pass@$k$: Pass at $k$ (at least one success within $k$ attempts); FPR: False Positive Rate; COV: Coverage; Acc.: Accuracy; $\uparrow$/$\downarrow$: Better direction.}
\end{table}

For evaluating offensive capabilities, PentestGPT \citep{bib18} progressively evaluated agent on penetration testing sub-tasks that cover the OWASP Top 10. Moreover, Capture the Flag (CTF) challenges have become a common evaluation method due to their task-oriented structure and automatically verifiable outcomes. For instance, Cybench \citep{zhang2024cybench} introduced 40 CTF tasks across six categories: cryptography, web security, reverse engineering, forensics, miscellaneous, and exploitation, proposing first-solve time as an objective difficulty metric. Subsequent works like NYU-CTF-Bench \citep{NEURIPS2024_69d97a64} and HackSynth \citep{muzsai2024hacksynth} expanded the datasets to approximately 200 challenges. In particular, NYU-CTF-Bench integrates six tool-calling functions to enhance agent performance. To address the gamified nature of CTF-based evaluations, some benchmarks aim for greater realism. \citet{bib26} proposed AutoPenBench, which offers a more fine-grained evaluation through a combination of in-vitro and realistic tasks. Similarly, CVE-Bench \citep{zhu2025cvebench} constructs its challenges from high-severity CVEs in the National Vulnerability Database (NVD) to simulate zero-day and one-day attack scenarios, using nine standardized attack vectors for automated assessment. Furthermore, the 3CB benchmark \citep{anurin2024catastrophic} aligns its tasks with the MITRE ATT\&CK framework \citep{mitre}.

Fewer benchmarks address agent-empowered cyber defense compared to offensive capabilities. ExCyTIn-Bench \citep{wu2025excytin} focuses on evaluating agent capabilities in cyber threat intelligence by constructing a database of 88 simulated attacks and 57 authentic security log tables. CFA-Bench \citep{11129512} targets digital forensics by supplying raw evidence from authentic attack scenarios, such as network packets and logs, to comprehensively evaluate the forensic reasoning abilities of agents.

In addition, a few studies have proposed more comprehensive evaluation frameworks. For example, \citet{phuong2024evaluating} evaluated frontier models not only on cybersecurity tasks such as CTF, Hack the Box, and vulnerability detection but also on their capacity for persuasion and deception, self-proliferation, and self-reasoning. Meta's Purple Llama initiative produced CYBERSECEVAL 4 \citep{CyberSecEval-4}, a benchmark designed for end-to-end cybersecurity benchmark. It comprises nine evaluation paradigms spanning the full spectrum of activities, from offensive simulations based on MITRE ATT\&CK to defensive threat intelligence analysis. 

While benchmark suites have grown rapidly, many evaluations remain largely static and task-based, which can be gamed by overfitting to prompt templates, fixed formats, or narrow success criteria. Such settings often under-measure cross-turn risk accumulation and strategy drift, and may overlook tool side effects, permission boundaries, and environment distribution shifts that dominate real deployments. These limitations are amplified for adaptive agents that update plans and internal state from intermediate observations, where small deviations can compound across long-horizon trajectories.

\subsection*{Discussion}\label{sec7.3}
Despite the proliferation of benchmarks, critical gaps remain in evaluating autonomous agents. First, current protocols predominantly focus on short-term tasks, often neglecting \textbf{long-horizon risks} where compound errors or delayed attacks emerge from extended stateful interactions. Second, static benchmarks frequently abstract away real-world deployment constraints, such as granular permission boundaries and irreversible side-effects, widening the simulation-to-reality gap. Furthermore, result reproducibility is often compromised by sensitivity to environmental snapshots and prompt variations. To address these challenges, future benchmarking must evolve towards high-fidelity, reproducible environments. This entails mandating deterministic replay via full-trajectory logging, integrating realistic sandboxes for side-effect containment, and explicitly quantifying the trade-offs between security rigidity and task utility. Longitudinal testing across model updates and distribution shifts is also essential to distinguish genuine robustness from benchmark overfitting.

\section{Challenges and future directions}\label{sec8}
\textbf{Challenges in the complex multimodal scenarios}. Existing research lacks studies on dynamic modalities. Although some researchers have begun to explore attacks and defenses for multi-modal agents \citep{bib62,bib4,bib293,bib49}, their work is limited to static image-text scenarios and lacks exploration of dynamic modalities like audio and video. Furthermore, the multi-module architecture of agents introduces potential risks. Unlike a singular LLM, an agent's perception, planning, and action pipeline is itself a potential vulnerability chain. The introduction of multi-modality further expands this attack surface, for example, a seemingly harmless attack on the perception module could have its threat unexpectedly amplified or distorted in subsequent modules. How early attack methods on LLMs using images and sounds \citep{bib323} will propagate or amplify their effects at the perception, fusion, or action levels within an agent's interactive environment is a direction worthy of future exploration.

\textbf{Challenges in tool calling}. Tool calling is the core capability of the LLM agent and also the attack surface that attackers focus on. The mainstream tool calling methods are through function calling or Model Context Protocol (MCP). However, both were initially designed for functionality realization and operational smoothness, lacking security considerations. Security challenges arise, as the tool ecosystem continues to evolve and multiple open tool markets \citep{bib321, bib322}. However, existing defense methods focus on passively detecting and correcting problems during interaction process \citep{bib294, bib29-Athena, zhu2025melon}, and are unable to deal with malicious tools that can cause irreversible damage during a single execution. There is an urgent need for proactive and preventive defense strategies, such as adopting the tool invocation security policies inspired by ShieldAgent \citep{chen2025shieldagent}, and integrating the access control policies from cybersecurity \citep{bib296, kim2025prompt}. Also, it is necessary to establish an effective trust assessment and certification mechanism for external tools.

\textbf{Challenges in multi-turn interaction}. Through multiple rounds of interaction, LLM agents achieve the detailed disassembly and automatic execution of complex tasks. However, multi-turn interaction attacks pose a great threat to agents. Since the multi-round interactive attack usually has an escalating nature, the defense measures can be bypassed in the initial stage to complete the original accumulation of ``toxins'' and finally execute the attack. However, there is no effective method to defend against multi-turn attacks, and the construction of relevant data sets still needs to be improved \citep{bib115}. Future research should prioritize the development and enrichment of diverse multi-turn interaction datasets. Additionally, to address the challenges posed by multi-turn interactions, researchers should enhance fine-grained filtering at the micro level, establish a supervision mechanism with a global perspective from the macro perspective, and investigate the consistency of micro connections at the macro level.

\textbf{Challenges in multi-agent systems}. The diffusion effect in multi-agent systems is an interesting topic. Many researchers have begun to study the destructive impact of propagation and diffusion \citep{bib38,bib62,bib67,bib12}. Using agents as a propagation tool exacerbates the destructive behavior in the system, and the continuous impact of attacks increases the difficulty of repair and cleanup. As MAS continue to evolve in scale and complexity, the threat posed by malicious propagation is expected to become even more pronounced. However, there is no effective method to prevent malicious propagation at this stage. It is urgent to study the characteristics of malicious propagation, propose targeted defense schemes, and explore how to restore the health of the system after an attack. Furthermore, there is no unified evaluation metric for the spread and persistence of malicious behavior. Future research can refer to metrics such as Information Diffusion Rate(IDR) proposed by Curvo \citep{bib311}, the Information Retention and Diffusion Gap used by \citet{bib312}, to create metrics for quantitative analysis of malicious behavior.

Moreover, based on the collaborative characteristics of LLM-based multi-agents, collaborative attack and defense have become an important research direction in order to cope with changing and complex environments. However, there are still significant deficiencies in the design of strategies, detection mechanisms, and defense systems for collaborative attacks among multi-agents. Future research should focus on developing detection methods based on feature construction for collaborative attacks, referencing studies like Psysafe \citep{bib73-Psysafe} that use interdisciplinary knowledge, and further exploring cross-domain collaborative defense methods.

\textbf{Evaluation benchmark requires multi-dimensional developments}. Current agent-related benchmarks \citep{bib34-ASB,bib142,bib30,bib139} mainly focus on evaluating the performance of different LLMs within the same agent framework, which is more like an extensible evaluation of the security of LLMs to a certain extent. It still cannot fully reflect the robustness and security of agents themselves in different application scenarios or frameworks. Because the workflow of an agent is not only affected by its underlying LLM but also involves multi-layer factors such as memory management, tool invocation, and environment interaction, it is difficult to completely describe the security risks and practical application challenges of an agent by only relying on a single framework. In order to solve this problem, future research should consider changing the evaluation subject from LLMs to agents by evaluating the security of different agent frameworks. 

\textbf{Challenges in building agent security test cases for open task scenarios}. Agent application scenarios are constantly expanding, especially in open-task scenarios. In these scenarios, agents are faced with diverse attack sources and attack methods, as well as complex and variable environmental feedback, which leads to the need to cover as many security risks as possible when constructing security test cases. However, the existing sample construction methods are still highly dependent on manual work, and it is difficult to achieve the automatic construction of samples. In addition, as the task scenarios of the agent continue to increase, the adaptation cost of the current protection technology for each task scenario will increase sharply. Consequently, developing methods for the rapid and effective generation of security test cases tailored to constantly evolving environments emerges as a critical research imperative.

\textbf{Challenge of real-time adaptation in agent-based reconnaissance.} Current LLM agent frameworks for offensive cyber operations remain limited in their ability to adapt reconnaissance strategies in real time. While systems like VulnBot \citep{VulnBot} and RapidPen \citep{RapidPen} employ Retrieval-Augmented Generation (RAG) to incorporate prior vulnerability data and system fingerprints, their reliance on pre-collected or internal knowledge bases constrains responsiveness to dynamic target environments. Current agents often depend on static knowledge bases or delayed external data \citep{wLLMgonline}. While PentestAgent \citep{bib193} incorporates adaptive measures, its approach is more akin to an agent selecting actions within a pre-defined workflow, rather than dynamically adjusting its overarching strategy. One important future direction is the study of lightweight learning mechanisms online so that the agent can dynamically adjust its reconnaissance strategy based on real-time feedback.

\textbf{Challenge in autonomous offense in dynamic environment}. The LLM agents excel in conducting automated discrete attack activities, but currently, the biggest challenge is achieving adaptive autonomy in a dynamic and complex environment. Although agents can exploit chained attacks with known vulnerabilities \citep{bib193} or execute CAPTCHA-breaking \citep{deng2024oedipus}, their strategic reasoning is still weak. For instance, single-agent face difficulties in terms of effective exploration and backtracking in zero-days \citep{bib42}, their success rate is always limited by the accuracy of external tools and task descriptions \citep{teoh2025captchas}. Therefore, the future focus of work should be on enhancing the adaptability and strategic depth of the agents. This involves developing online planning and reinforcement learning models to better adapt to unexpected defense measures. At the same time, consider addressing the issue of agent-empowered offense being overly dependent on external tools.

\textbf{Challenge in agent-empowered cyber defense protection}. When LLM agents act as the main body or auxiliary in the cyber defense process, the LLM agents themselves will become the new attack surface. The agent empowered cyber defense methods are vulnerable to malicious attacks due to agents’ inherent vulnerabilities \citep{ayzenshteyn2024best}. Therefore, empowered cybersecurity demand agents enhance self-security. Furthermore, the multi-agent framework employed in cyber defense requires high computational costs and scalability, thus necessitating a more efficient and hierarchical design \citep{bib218}. The other future direction is enhance the explainability of agent reasoning to ensure defense system trusts agents’ decisions.

\textbf{Challenge in agent-empowered post-exploitation defense}. The existing post-exploitation detection methods heavily rely on static environments, causing their incapability to respond to rapidly evolving attack methods and inability to perform long-stage attack chain analysis correlative \citep{app15137237}. Besides, incident response and forensic always occurs large volume of traffic, can easily cause LLM agents to lose focus \citep{fumero2025cybersleuth}. How to mitigate the impact of dense traffic and improve the success rate of high-complexity tasks is an upcoming direction.

\section{Conclusion}\label{conclusion}
This survey provides a comprehensive and systematic overview of the security of LLM-based agents, and explores the positive feedback and collaborative correlation between LLM agents self-security and LLM agents empowered cybersecurity. Agent self-security, as the foundation of agent security, we deeply analyze the causes of the three major threats facing agents, and propose a self-security classification framework based on threat source. Furthermore, we outline corresponding mitigation methods and categorize the current technological approaches. At the empowered cybersecurity level, we summarize agent applications throughout the entire lifecycle of cyber offense and defense, and analyze how its various capabilities can empower cyber tasks. Agent-empowered cybersecurity is not only automation at scale; it reframes cyber operations as intent-driven, closed-loop decision making under tool-mediated feedback. This shift increases autonomy while introducing agent-as-attack-surface risks, reinforcing the co-evolutionary arms race highlighted in Figure \ref{fig_cycle}. We also summarize benchmarks in two fields. Notably, we have pointed out the gaps in the existing research. Based on this, we emphasize the key challenges and future directions in self-security and empowered cybersecurity. We hope this survey can stimulate further discussions and advocate for viewing self-security and empowered cybersecurity in the field of LLM agent security within a unified framework. By leveraging the strengths of both to build the new generation of trustworthy agents.

\backmatter

\bmhead{Acknowledgements}
This study was funded by Smart Gird-National Science and Technology Major Project (No.2024ZD0803000). We also thank the KinaMind Society for supporting our arXiv submission through endorsement and for its continued support.
\bmhead{Author contributions}
All authors have read and agreed to the submitted version of the manuscript. Yiwei Xu: Investigation, Methodology, Validation, Visualization, Writing – original draft; Yong Zhuang: Methodology review, Editing; Xuanming Liu: Visualization, Data collation, Editing; Tian Zhang: Proofreading, Editing; Bowen Xiao: Data collation, Editing; Xiaoyang Xu: Methodology review; Delong Jiang: Methodology review; Juan Wang: Methodology review, Review report, Supervision, Funding; Hongxin Hu: Methodology review, Review report, Supervision.
\bmhead{Data availability} 
No datasets were generated or analysed during the current study.
\section*{Declarations}
\bmhead{Competing interests}The authors declare no competing interests.


\bibliography{sn-bibliography}

@misc{mitre-scan,
  title={Active Scanning: Scanning IP Blocks},
  author={MITRE},
  year={2024},
  url="https://attack.mitre.org/techniques/T1595/001/"
}

@misc{mitre,
  title={MITRE ATT\&CK},
  author={MITRE},
  year={2024},
  url="https://attack.mitre.org"
}

@misc{killchain,
  title={Cyber Kill Chain},
  author={Lockheed Martin},
  year={2011},
  url="https://www.lockheedmartin.com/en-us/capabilities/cyber/cyber-kill-chain.html"
}

@inproceedings{bib1,
 author = {Brown, Tom and Mann, Benjamin and Ryder, Nick and Subbiah, Melanie and Kaplan, Jared D and Dhariwal, Prafulla and Neelakantan, Arvind and Shyam, Pranav and Sastry, Girish and Askell, Amanda and Agarwal, Sandhini and Herbert-Voss, Ariel and Krueger, Gretchen and Henighan, Tom and Child, Rewon and Ramesh, Aditya and Ziegler, Daniel and Wu, Jeffrey and Winter, Clemens and Hesse, Chris and Chen, Mark and Sigler, Eric and Litwin, Mateusz and Gray, Scott and Chess, Benjamin and Clark, Jack and Berner, Christopher and McCandlish, Sam and Radford, Alec and Sutskever, Ilya and Amodei, Dario},
 booktitle = {Advances in Neural Information Processing Systems},
 editor = {H. Larochelle and M. Ranzato and R. Hadsell and M.F. Balcan and H. Lin},
 pages = {1877--1901},
 publisher = {Curran Associates, Inc.},
 title = {Language Models are Few-Shot Learners},
 url = {https://proceedings.neurips.cc/paper_files/paper/2020/file/1457c0d6bfcb4967418bfb8ac142f64a-Paper.pdf},
 volume = {33},
 year = {2020}
}

@article{bib2,
  title={On the resilience of multi-agent systems with malicious agents},
  author={Huang, Jen-tse and Zhou, Jiaxu and Jin, Tailin and Zhou, Xuhui and Chen, Zixi and Wang, Wenxuan and Yuan, Youliang and Sap, Maarten and Lyu, Michael R},
  journal={arXiv preprint},
  archivePrefix={arXiv},
  eprint={2408.00989},
  year={2024}
}

@inproceedings{bib3,
  title={Dissecting Adversarial Robustness of Multimodal LM Agents},
  author={Wu, Chen Henry and Shah, Rishi Rajesh and Koh, Jing Yu and Salakhutdinov, Russ and Fried, Daniel and Raghunathan, Aditi},
  booktitle={The Thirteenth International Conference on Learning Representations},
  year={2025}
}

@article{bib4,
  title={Attacking Vision-Language Computer Agents via Pop-ups},
  author={Zhang, Yanzhe and Yu, Tao and Yang, Diyi},
  journal={arXiv preprint},
  archivePrefix={arXiv},
  eprint={2411.02391},
  year={2024}
}

@article{bib5,
  title={Injecagent: Benchmarking indirect prompt injections in tool-integrated large language model agents},
  author={Zhan, Qiusi and Liang, Zhixiang and Ying, Zifan and Kang, Daniel},
  journal={arXiv preprint},
  archivePrefix={arXiv},
  eprint={2403.02691},
  year={2024}
}

@article{bib6,
  title={Multimodal situational safety},
  author={Zhou, Kaiwen and Liu, Chengzhi and Zhao, Xuandong and Compalas, Anderson and Song, Dawn and Wang, Xin Eric},
  journal={arXiv preprint},
  archiveprefix={arXiv},
  eprint={2410.06172},
  year={2024}
}

@article{bib7,
  title={St-webagentbench: A benchmark for evaluating safety and trustworthiness in web agents},
  author={Levy, Ido and Wiesel, Ben and Marreed, Sami and Oved, Alon and Yaeli, Avi and Shlomov, Segev},
  journal={arXiv preprint},
  archiveprefix={arXiv},
  eprint={2410.06703},
  year={2024}
}

@inproceedings{bib8,
    title = "Towards Low-Resource Harmful Meme Detection with {LMM} Agents",
    author = "Huang, Jianzhao  and
      Lin, Hongzhan  and
      Ziyan, Liu  and
      Luo, Ziyang  and
      Chen, Guang  and
      Ma, Jing",
    editor = "Al-Onaizan, Yaser  and
      Bansal, Mohit  and
      Chen, Yun-Nung",
    booktitle = "Proceedings of the 2024 Conference on Empirical Methods in Natural Language Processing",
    month = nov,
    year = "2024",
    doi = "10.18653/v1/2024.emnlp-main.136",
    pages = "2269--2293",
}

@article{bib9,
  title={Testing and Understanding Erroneous Planning in LLM Agents through Synthesized User Inputs},
  author={Ji, Zhenlan and Wu, Daoyuan and Ma, Pingchuan and Li, Zongjie and Wang, Shuai},
  journal={arXiv preprint},
  archiveprefix={arXiv},
  eprint={2404.17833},
  year={2024}
}

@article{bib10,
  title={Exploring large language models for communication games: An empirical study on werewolf},
  author={Xu, Yuzhuang and Wang, Shuo and Li, Peng and Luo, Fuwen and Wang, Xiaolong and Liu, Weidong and Liu, Yang},
  journal={arXiv preprint},
  archiveprefix={arXiv},
  eprint={2309.04658},
  year={2023}
}

@inproceedings{bib11,
    title = "Mitigating Catastrophic Forgetting in Large Language Models with Self-Synthesized Rehearsal",
    author = "Huang, Jianheng  and
      Cui, Leyang  and
      Wang, Ante  and
      Yang, Chengyi  and
      Liao, Xinting  and
      Song, Linfeng  and
      Yao, Junfeng  and
      Su, Jinsong",
    booktitle = "Proceedings of the 62nd Annual Meeting of the Association for Computational Linguistics (Volume 1: Long Papers)",
    month = aug,
    year = "2024",
    doi = "10.18653/v1/2024.acl-long.77",
    pages = "1416--1428",
}

@article{bib12,
  title={Prompt infection: Llm-to-llm prompt injection within multi-agent systems},
  author={Lee, Donghyun and Tiwari, Mo},
  journal={arXiv preprint},
  archiveprefix={arXiv},
  eprint={2410.07283},
  year={2024}
}

@inproceedings{bib13,
  title={LLM-Deliberation: Evaluating LLMs with Interactive Multi-Agent Negotiation Game},
  author={Abdelnabi, Sahar and Gomaa, Amr and Sivaprasad, Sarath and Sch{\"o}nherr, Lea and Fritz, Mario},
  booktitle={ICLR 2024 Workshop on Large Language Model (LLM) Agents},
  year={2024}
}

@article{bib14,
  title={Autopt: How far are we from the end2end automated web penetration testing?},
  author={Wu, Benlong and Chen, Guoqiang and Chen, Kejiang and Shang, Xiuwei and Han, Jiapeng and He, Yanru and Zhang, Weiming and Yu, Nenghai},
  journal={arXiv preprint},
  archiveprefix={arXiv},
  eprint={2411.01236},
  year={2024}
}

@misc{bib15,
  title={Understanding Data Poisoning Attacks for RAG: Insights and Algorithms},
  author={Xian, Xun and Wang, Tong and You, Liwen and Qi, Yanjun},
  year={2024}
}

@article{bib16,
  title={Agentpoison: Red-teaming llm agents via poisoning memory or knowledge bases},
  author={Chen, Zhaorun and Xiang, Zhen and Xiao, Chaowei and Song, Dawn and Li, Bo},
  journal={Advances in Neural Information Processing Systems},
  volume={37},
  pages={130185--130213},
  year={2024}
}

@article{bib17,
    title = {Mind meets machine: Unravelling GPT-4’s cognitive psychology},
    journal = {BenchCouncil Transactions on Benchmarks, Standards and Evaluations},
    volume = {3},
    number = {3},
    pages = {100139},
    year = {2023},
    issn = {2772-4859},
    doi = {https://doi.org/10.1016/j.tbench.2023.100139},
    author = {Sifatkaur Dhingra and Manmeet Singh and Vaisakh S.B. and Neetiraj Malviya and Sukhpal Singh Gill},
}

@inproceedings{bib18,
  title={$\{$PentestGPT$\}$: Evaluating and harnessing large language models for automated penetration testing},
  author={Deng, Gelei and Liu, Yi and Mayoral-Vilches, V{\'\i}ctor and Liu, Peng and Li, Yuekang and Xu, Yuan and Zhang, Tianwei and Liu, Yang and Pinzger, Martin and Rass, Stefan},
  booktitle={33rd USENIX Security Symposium (USENIX Security 24)},
  pages={847--864},
  year={2024}
}

@inproceedings{bib19,
    author = {Du, Yilun and Li, Shuang and Torralba, Antonio and Tenenbaum, Joshua B. and Mordatch, Igor},
    title = {Improving factuality and reasoning in language models through multiagent debate},
    year = {2024},
    booktitle = {Proceedings of the 41st International Conference on Machine Learning},
    articleno = {467},
    numpages = {31},
    location = {Vienna, Austria},
    series = {ICML'24},
    doi="10.5555/3692070.3692537"
}

@inproceedings{bib20,
  title={Generative agents: Interactive simulacra of human behavior},
  author={Park, Joon Sung and O'Brien, Joseph and Cai, Carrie Jun and Morris, Meredith Ringel and Liang, Percy and Bernstein, Michael S},
  booktitle={Proceedings of the 36th annual acm symposium on user interface software and technology},
  pages={1--22},
  year={2023}
}

@article{bib21,
  title={Synergizing human-AI agency: a guide of 23 heuristics for service co-creation with LLM-based agents},
  author={Zheng, Qingxiao and Xu, Zhongwei and Choudhry, Abhinav and Chen, Yuting and Li, Yongming and Huang, Yun},
  journal={arXiv preprint},
  archiveprefix={arXiv},
  eprint={2310.15065},
  year={2023}
}

@article{bib22,
  title={S3: Social-network simulation system with large language model-empowered agents},
  author={Gao, Chen and Lan, Xiaochong and Lu, Zhihong and Mao, Jinzhu and Piao, Jinghua and Wang, Huandong and Jin, Depeng and Li, Yong},
  journal={arXiv preprint},
  archiveprefix={arXiv},
  eprint={2307.14984},
  year={2023}
}

@article{bib23,
  title={Prioritizing safeguarding over autonomy: Risks of llm agents for science},
  author={Tang, Xiangru and Jin, Qiao and Zhu, Kunlun and Yuan, Tongxin and Zhang, Yichi and Zhou, Wangchunshu and Qu, Meng and Zhao, Yilun and Tang, Jian and Zhang, Zhuosheng and others},
  journal={arXiv preprint},
  archiveprefix={arXiv},
  eprint={2402.04247},
  year={2024}
}

@article{bib24,
  title={Using cognitive psychology to understand GPT-3},
  author={Binz, Marcel and Schulz, Eric},
  journal={Proceedings of the National Academy of Sciences},
  volume={120},
  number={6},
  pages={e2218523120},
  year={2023},
}

@article{bib25,
  title={Moral Alignment for LLM Agents},
  author={Tennant, Elizaveta and Hailes, Stephen and Musolesi, Mirco},
  journal={arXiv preprint},
  archiveprefix={arXiv},
  eprint={2410.01639},
  year={2024}
}

@article{bib26,
  title={AutoPenBench: Benchmarking Generative Agents for Penetration Testing},
  author={Gioacchini, Luca and Mellia, Marco and Drago, Idilio and Delsanto, Alexander and Siracusano, Giuseppe and Bifulco, Roberto},
  journal={arXiv preprint},
  archiveprefix={arXiv},
  eprint={2410.03225},
  year={2024}
}

@inproceedings{bib28-Reft,
    title = "{R}e{FT}: Reasoning with Reinforced Fine-Tuning",
    author = "Trung, Luong  and
      Zhang, Xinbo  and
      Jie, Zhanming  and
      Sun, Peng  and
      Jin, Xiaoran  and
      Li, Hang",
    booktitle = "Proceedings of the 62nd Annual Meeting of the Association for Computational Linguistics (Volume 1: Long Papers)",
    month = aug,
    year = "2024",
    doi = "10.18653/v1/2024.acl-long.410",
    pages = "7601--7614",
}

@inproceedings{bib29-Athena,
    title = "Athena: Safe Autonomous Agents with Verbal Contrastive Learning",
    author = "Sadhu, Tanmana  and
      Pesaranghader, Ali  and
      Chen, Yanan  and
      Yi, Dong Hoon",
    booktitle = "Proceedings of the 2024 Conference on Empirical Methods in Natural Language Processing: Industry Track",
    month = nov,
    year = "2024",
    doi = "10.18653/v1/2024.emnlp-industry.84",
    pages = "1121--1130",
}

@article{bib30,
  title={Identifying the risks of lm agents with an lm-emulated sandbox},
  author={Ruan, Yangjun and Dong, Honghua and Wang, Andrew and Pitis, Silviu and Zhou, Yongchao and Ba, Jimmy and Dubois, Yann and Maddison, Chris J and Hashimoto, Tatsunori},
  journal={arXiv preprint},
  archiveprefix={arXiv},
  eprint={2309.15817},
  year={2023}
}

@article{bib31,
  title={Towards objective and unbiased decision assessments with llm-enhanced hierarchical attention networks},
  author={Liu, Junhua and Lim, Kwan Hui and Lee, Roy Ka-Wei},
  journal={arXiv preprint},
  archiveprefix={arXiv},
  eprint={2411.08504},
  year={2024}
}

@inproceedings{bib32-EHRAgent,
    title = "{EHRA}gent: Code Empowers Large Language Models for Few-shot Complex Tabular Reasoning on Electronic Health Records",
    author = "Shi, Wenqi  and
      Xu, Ran  and
      Zhuang, Yuchen  and
      Yu, Yue  and
      Zhang, Jieyu  and
      Wu, Hang  and
      Zhu, Yuanda  and
      Ho, Joyce C.  and
      Yang, Carl  and
      Wang, May Dongmei",
    booktitle = "Proceedings of the 2024 Conference on Empirical Methods in Natural Language Processing",
    month = nov,
    year = "2024",
    doi = "10.18653/v1/2024.emnlp-main.1245",
    pages = "22315--22339",
}

@article{bib33-Advweb,
  title={Advweb: Controllable black-box attacks on vlm-powered web agents},
  author={Xu, Chejian and Kang, Mintong and Zhang, Jiawei and Liao, Zeyi and Mo, Lingbo and Yuan, Mengqi and Sun, Huan and Li, Bo},
  journal={arXiv preprint},
  archiveprefix={arXiv},
  eprint={2410.17401},
  year={2024}
}

@article{bib34-ASB,
  title={Agent security bench (asb): Formalizing and benchmarking attacks and defenses in llm-based agents},
  author={Zhang, Hanrong and Huang, Jingyuan and Mei, Kai and Yao, Yifei and Wang, Zhenting and Zhan, Chenlu and Wang, Hongwei and Zhang, Yongfeng},
  journal={arXiv preprint},
  archiveprefix={arXiv},
  eprint={2410.02644},
  year={2024}
}

@inproceedings{bib35-Abseval,
    title = "{ABSE}val: An Agent-based Framework for Script Evaluation",
    author = "Liang, Sirui  and
      Zhang, Baoli  and
      Zhao, Jun  and
      Liu, Kang",
    booktitle = "Proceedings of the 2024 Conference on Empirical Methods in Natural Language Processing",
    month = nov,
    year = "2024",
    doi = "10.18653/v1/2024.emnlp-main.691",
    pages = "12418--12434",
}

@inproceedings{bib36,
    title = "Prompt Leakage effect and mitigation strategies for multi-turn {LLM} Applications",
    author = "Agarwal, Divyansh  and
      Fabbri, Alexander  and
      Risher, Ben  and
      Laban, Philippe  and
      Joty, Shafiq  and
      Wu, Chien-Sheng",
    booktitle = "Proceedings of the 2024 Conference on Empirical Methods in Natural Language Processing: Industry Track",
    month = nov,
    year = "2024",
    doi = "10.18653/v1/2024.emnlp-industry.94",
    pages = "1255--1275",
}

@article{bib37,
  title={Breaking ReAct Agents: Foot-in-the-Door Attack Will Get You In},
  author={Nakash, Itay and Kour, George and Uziel, Guy and Anaby-Tavor, Ateret},
  journal={arXiv preprint},
  archiveprefix={arXiv},
  eprint={2410.16950},
  year={2024}
}

@article{bib38,
  title={Flooding spread of manipulated knowledge in llm-based multi-agent communities},
  author={Ju, Tianjie and Wang, Yiting and Ma, Xinbei and Cheng, Pengzhou and Zhao, Haodong and Wang, Yulong and Liu, Lifeng and Xie, Jian and Zhang, Zhuosheng and Liu, Gongshen},
  journal={arXiv preprint},
  archiveprefix={arXiv},
  eprint={2407.07791},
  year={2024}
}

@article{bib39,
  title={Of models and tin men: a behavioural economics study of principal-agent problems in AI alignment using large-language models},
  author={Phelps, Steve and Ranson, Rebecca},
  journal={arXiv preprint\},
  archiveprefix={arXiv},
  eprint={2307.11137},
  year={2023}
}

@inproceedings{bib40-Agentverse,
  title={AgentVerse: Facilitating Multi-Agent Collaboration and Exploring Emergent Behaviors},
  author={Chen, Weize and Su, Yusheng and Zuo, Jingwei and Yang, Cheng and Yuan, Chenfei and Chan, Chi-Min and Yu, Heyang and Lu, Yaxi and Hung, Yi-Hsin and Qian, Chen and others},
  booktitle={The Twelfth International Conference on Learning Representations},
  year={2024},
}

@article{bib41,
  title={Combating adversarial attacks with multi-agent debate},
  author={Chern, Steffi and Fan, Zhen and Liu, Andy},
  journal={arXiv preprint},
  archiveprefix={arXiv},
  eprint={2401.05998},
  year={2024}
}

@article{bib42,
  title={Teams of llm agents can exploit zero-day vulnerabilities},
  author={Zhu, Yuxuan and Kellermann, Antony and Gupta, Akul and Li, Philip and Fang, Richard and Bindu, Rohan and Kang, Daniel},
  journal={arXiv preprint},
  archiveprefix={arXiv},
  eprint={2406.01637},
  year={2024}
}

@article{bib43,
  title={Caution for the environment: Multimodal agents are susceptible to environmental distractions},
  author={Ma, Xinbei and Wang, Yiting and Yao, Yao and Yuan, Tongxin and Zhang, Aston and Zhang, Zhuosheng and Zhao, Hai},
  journal={arXiv preprint},
  archiveprefix={arXiv},
  eprint={2408.02544},
  year={2024}
}

@article{bib44,
  title={Your Agent Can Defend Itself against Backdoor Attacks},
  author={Li, Changjiang and Liang, Jiacheng and Cao, Bochuan and Chen, Jinghui and Wang, Ting},
  journal={arXiv preprint},
  eprint={2506.08336},
  archivePrefix={arXiv},
  year={2025}
}

@misc{bib45,
    title={Your Agent Can Defend Itself against Backdoor Attacks},
    author={Li Changjiang and Liang Jiacheng and Cao Bochuan and Chen Jinghui and Wang Ting},
    year={2025},
    eprint={2506.08336},
    archivePrefix={arXiv},
}

@inproceedings{bib46,
 author = {Piatti, Giorgio and Jin, Zhijing and Kleiman-Weiner, Max and Sch\"{o}lkopf, Bernhard and Sachan, Mrinmaya and Mihalcea, Rada},
 booktitle = {Advances in Neural Information Processing Systems},
 editor = {A. Globerson and L. Mackey and D. Belgrave and A. Fan and U. Paquet and J. Tomczak and C. Zhang},
 pages = {111715--111759},
 publisher = {Curran Associates, Inc.},
 title = {Cooperate or Collapse:  Emergence of Sustainable Cooperation in a Society of LLM Agents},
 url = {https://proceedings.neurips.cc/paper_files/paper/2024/file/ca9567d8ef6b2ea2da0d7eed57b933ee-Paper-Conference.pdf},
 volume = {37},
 year = {2024}
}

@inproceedings{bib48-AutoDAN-Turbo,
  title={AutoDAN-Turbo: A Lifelong Agent for Strategy Self-Exploration to Jailbreak LLMs},
  author={Liu, Xiaogeng and Li, Peiran and Suh, G Edward and Vorobeychik, Yevgeniy and Mao, Zhuoqing and Jha, Somesh and McDaniel, Patrick and Sun, Huan and Li, Bo and Xiao, Chaowei},
  booktitle={The Thirteenth International Conference on Learning Representations},
  year={2025}
}

@misc{bib49,
  title={Defend against Jailbreak Attacks via Debate with Partially Perceptive Agents},
  author={Zhou, Qi and Li, Tianlin and Guo, Qing and Wang, Dongxia},
  year={2024},
  url={https://openreview.net/forum?id=STpxO1Siaq},
}

@article{bib50,
  title={A trembling house of cards? mapping adversarial attacks against language agents},
  author={Mo, Lingbo and Liao, Zeyi and Zheng, Boyuan and Su, Yu and Xiao, Chaowei and Sun, Huan},
  journal={arXiv preprint arXiv:2402.10196},
  eprint={2402.10196},
  archivePrefix={arXiv},
  year={2024},
}

@inproceedings{bib51,
  title={AirGapAgent: Protecting privacy-conscious conversational agents},
  author={Bagdasarian, Eugene and Yi, Ren and Ghalebikesabi, Sahra and Kairouz, Peter and Gruteser, Marco and Oh, Sewoong and Balle, Borja and Ramage, Daniel},
  booktitle={Proceedings of the 2024 on ACM SIGSAC Conference on Computer and Communications Security},
  pages={3868--3882},
  year={2024}
}

@inproceedings{bib52,
  title={Can We Trust Embodied Agents? Exploring Backdoor Attacks against Embodied LLM-Based Decision-Making Systems},
  author={Jiao, Ruochen and Xie, Shaoyuan and Yue, Justin and SATO, TAKAMI and Wang, Lixu and Wang, Yixuan and Chen, Qi Alfred and Zhu, Qi},
  booktitle={The Thirteenth International Conference on Learning Representations},
  year={2025}
}

@inproceedings{bib53-BadAgent,
    title = "{B}ad{A}gent: Inserting and Activating Backdoor Attacks in {LLM} Agents",
    author = "Wang, Yifei  and
      Xue, Dizhan  and
      Zhang, Shengjie  and
      Qian, Shengsheng",
    booktitle = "Proceedings of the 62nd Annual Meeting of the Association for Computational Linguistics (Volume 1: Long Papers)",
    month = aug,
    year = "2024",
    doi = "10.18653/v1/2024.acl-long.530",
    pages = "9811--9827",
}

@inproceedings{bib54,
 author = {Yang, Wenkai and Bi, Xiaohan and Lin, Yankai and Chen, Sishuo and Zhou, Jie and Sun, Xu},
 booktitle = {Advances in Neural Information Processing Systems},
 pages = {100938--100964},
 title = {Watch Out for Your Agents! Investigating Backdoor Threats to LLM-Based Agents},
 url = {https://proceedings.neurips.cc/paper_files/paper/2024/file/b6e9d6f4f3428cd5f3f9e9bbae2cab10-Paper-Conference.pdf},
 volume = {37},
 year = {2024}
}

@article{bib55-SleeperAgents,
  title={Sleeper agents: Training deceptive llms that persist through safety training},
  author={Hubinger, Evan and Denison, Carson and Mu, Jesse and Lambert, Mike and Tong, Meg and MacDiarmid, Monte and Lanham, Tamera and Ziegler, Daniel M and Maxwell, Tim and Cheng, Newton and others},
  journal={arXiv preprint},
  eprint={2401.05566},
  archivePrefix={arXiv},
  year={2024}
}

@inproceedings{bib56-CoT,
 author = {Wei, Jason and Wang, Xuezhi and Schuurmans, Dale and Bosma, Maarten and ichter, brian and Xia, Fei and Chi, Ed and Le, Quoc V and Zhou, Denny},
 booktitle = {Advances in Neural Information Processing Systems},
 editor = {S. Koyejo and S. Mohamed and A. Agarwal and D. Belgrave and K. Cho and A. Oh},
 pages = {24824--24837},
 publisher = {Curran Associates, Inc.},
 title = {Chain-of-Thought Prompting Elicits Reasoning in Large Language Models},
 url ={https://proceedings.neurips.cc/paper_files/paper/2022/file/9d5609613524ecf4f15af0f7b31abca4-Paper-Conference.pdf},
 volume = {35},
 year = {2022}
}

@article{bib59,
  title={Autoattacker: A large language model guided system to implement automatic cyber-attacks},
  author={Xu, Jiacen and Stokes, Jack W and McDonald, Geoff and Bai, Xuesong and Marshall, David and Wang, Siyue and Swaminathan, Adith and Li, Zhou},
  journal={arXiv preprint},
  eprint={2403.01038},
  archivePrefix={arXiv},
  year={2024}
}

@article{bib61,
  title={Eia: Environmental injection attack on generalist web agents for privacy leakage},
  author={Liao, Zeyi and Mo, Lingbo and Xu, Chejian and Kang, Mintong and Zhang, Jiawei and Xiao, Chaowei and Tian, Yuan and Li, Bo and Sun, Huan},
  journal={arXiv preprint},
  eprint={2409.11295},
  archivePrefix={arXiv},
  year={2024}
}

@inproceedings{bib62,
  title={Agent Smith: a single image can jailbreak one million multimodal LLM agents exponentially fast},
  author={Gu, Xiangming and Zheng, Xiaosen and Pang, Tianyu and Du, Chao and Liu, Qian and Wang, Ye and Jiang, Jing and Lin, Min},
  booktitle={Proceedings of the 41st International Conference on Machine Learning},
  pages={16647--16672},
  year={2024}
}

@article{bib63,
  title={From a tiny slip to a giant leap: An llm-based simulation for fake news evolution},
  author={Liu, Yuhan and Song, Zirui and Zhang, Xiaoqing and Chen, Xiuying and Yan, Rui},
  journal={arXiv preprint},
  eprint={2410.19064},
  archivePrefix={arXiv},
  year={2024}
}

@article{bib64,
  title={Behaviorgpt: Smart agent simulation for autonomous driving with next-patch prediction},
  author={Zhou, Zikang and Haibo, HU and Chen, Xinhong and Wang, Jianping and Guan, Nan and Wu, Kui and Li, Yung-Hui and Huang, Yu-Kai and Xue, Chun Jason},
  journal={Advances in Neural Information Processing Systems},
  volume={37},
  pages={79597--79617},
  year={2024}
}

@inproceedings{bib65,
  title={Agent-Based Modelling Meets Generative AI in Social Network Simulations},
  author={Ferraro, Antonino and Galli, Antonio and La Gatta, Valerio and Postiglione, Marco and Orlando, Gian Marco and Russo, Diego and Riccio, Giuseppe and Romano, Antonio and Moscato, Vincenzo},
  booktitle={International Conference on Advances in Social Networks Analysis and Mining},
  pages={155--170},
  year={2024},
  doi="10.1007/978-3-031-78541-2_10"
}

@article{bib66,
  title={Solving Multi-Agent Safe Optimal Control with Distributed Epigraph Form MARL},
  author={Zhang, Songyuan and So, Oswin and Black, Mitchell and Serlin, Zachary and Fan, Chuchu},
  journal={arXiv preprint},
  eprint={2504.15425},
  archivePrefix={arXiv},
  year={2025}
}

@article{bib67,
  title={Adversarial Attacks on Cooperative Multi-agent Bandits},
  author={Zuo, Jinhang and Zhang, Zhiyao and Wang, Xuchuang and Chen, Cheng and Li, Shuai and Lui, John and Hajiesmaili, Mohammad and Wierman, Adam},
  journal={arXiv preprint},
  eprint={2311.01698},
  archivePrefix={arXiv},
  year={2023}
}

@inproceedings{bib68,
  title={Single-agent poisoning attacks suffice to ruin multi-agent learning},
  author={Yao, Fan and Cheng, Yuwei and Wei, Ermin and Xu, Haifeng},
  booktitle={The Thirteenth International Conference on Learning Representations},
  year={2025}
}

@article{bib69,
  title={Hidden in plain text: Emergence \& mitigation of steganographic collusion in LLMs},
  author={Mathew, Yohan and Matthews, Ollie and McCarthy, Robert and Velja, Joan and de Witt, Christian Schroeder and Cope, Dylan and Schoots, Nandi},
  journal={arXiv preprint},
  eprint={2410.03768},
  archivePrefix={arXiv},
  year={2024}
}

@article{bib70,
  title={Secret collusion among ai agents: Multi-agent deception via steganography},
  author={Motwani, Sumeet and Baranchuk, Mikhail and Strohmeier, Martin and Bolina, Vijay and Torr, Philip and Hammond, Lewis and Schroeder de Witt, Christian},
  journal={Advances in Neural Information Processing Systems},
  volume={37},
  pages={73439--73486},
  year={2024}
}

@article{bib71,
  title={Large language model-driven multi-agent simulation for news diffusion under different network structures},
  author={Li, Xinyi and Xu, Yu and Zhang, Yongfeng and Malthouse, Edward C},
  journal={arXiv preprint},
  eprint={2410.13909},
  archivePrefix={arXiv},
  year={2024}
}

@article{bib72-Netsafe,
  title={Netsafe: Exploring the topological safety of multi-agent networks},
  author={Yu, Miao and Wang, Shilong and Zhang, Guibin and Mao, Junyuan and Yin, Chenlong and Liu, Qijiong and Wen, Qingsong and Wang, Kun and Wang, Yang},
  journal={arXiv preprint arXiv:2410.15686},
  eprint={2410.15686},
  archivePrefix={arXiv},
  year={2024}
}

@inproceedings{bib73-Psysafe,
  title={PsySafe: A Comprehensive Framework for Psychological-based Attack, Defense, and Evaluation of Multi-agent System Safety},
  author={Zhang, Zaibin and Zhang, Yongting and Li, Lijun and Shao, Jing and Gao, Hongzhi and Qiao, Yu and Wang, Lijun and Lu, Huchuan and Zhao, Feng},
  booktitle={Proceedings of the 62nd Annual Meeting of the Association for Computational Linguistics (Volume 1: Long Papers)},
  pages={15202--15231},
  year={2024}
}

@inproceedings{bib74-SignedPrompt,
  title={Signed-prompt: A new approach to prevent prompt injection attacks against llm-integrated applications},
  author={Suo, Xuchen},
  booktitle={AIP Conference Proceedings},
  volume={3194},
  number={1},
  year={2024},
}

@article{bib75,
  title={Defending against indirect prompt injection attacks with spotlighting},
  author={Hines, Keegan and Lopez, Gary and Hall, Matthew and Zarfati, Federico and Zunger, Yonatan and Kiciman, Emre},
  journal={arXiv preprint},
  eprint={2403.14720},
  archivePrefix={arXiv},
  year={2024}
}

@inproceedings{bib76,
  title={Tensor Trust: Interpretable Prompt Injection Attacks from an Online Game},
  author={Toyer, Sam and Watkins, Olivia and Mendes, Ethan Adrian and Svegliato, Justin and Bailey, Luke and Wang, Tiffany and Ong, Isaac and Elmaaroufi, Karim and Abbeel, Pieter and Darrell, Trevor and others},
  booktitle={The Twelfth International Conference on Learning Representations},
  year={2024}
}

@article{bib77,
  title={Benchmarking and defending against indirect prompt injection attacks on large language models},
  author={Yi, Jingwei and Xie, Yueqi and Zhu, Bin and Kiciman, Emre and Sun, Guangzhong and Xie, Xing and Wu, Fangzhao},
  journal={arXiv preprint},
  eprint={2312.14197},
  archivePrefix={arXiv},
  year={2023}
}

@article{bib78,
  title={The instruction hierarchy: Training llms to prioritize privileged instructions},
  author={Wallace, Eric and Xiao, Kai and Leike, Reimar and Weng, Lilian and Heidecke, Johannes and Beutel, Alex},
  journal={arXiv preprint},
  eprint={2404.13208},
  archivePrefix={arXiv},
  year={2024}
}

@inproceedings{bib79,
  title={Jatmo: Prompt injection defense by task-specific finetuning},
  author={Piet, Julien and Alrashed, Maha and Sitawarin, Chawin and Chen, Sizhe and Wei, Zeming and Sun, Elizabeth and Alomair, Basel and Wagner, David},
  booktitle={European Symposium on Research in Computer Security},
  pages={105--124},
  year={2024},
  doi="10.1007/978-3-031-70879-4_6"
}

@article{bib80,
  title={Struq: Defending against prompt injection with structured queries},
  author={Chen, Sizhe and Piet, Julien and Sitawarin, Chawin and Wagner, David},
  journal={arXiv preprint},
  eprint={2402.06363},
  archivePrefix={arXiv},
  year={2024}
}

@inproceedings{bib81,
  title={AgentDojo: A Dynamic Environment to Evaluate Prompt Injection Attacks and Defenses for LLM Agents},
  author={Debenedetti, Edoardo and Zhang, Jie and Balunovic, Mislav and Beurer-Kellner, Luca and Fischer, Marc and Tram{\`e}r, Florian},
  booktitle={The Thirty-eight Conference on Neural Information Processing Systems Datasets and Benchmarks Track},
  year={2024}
}

@misc{bib82,
  title={Instruction defense},
  author={Learn, Prompting},
  year={2023},
  url="https://learnprompting.org/docs/prompt_hacking/defensive_measures/instruction"
}

@misc{bib83,
  title={Sandwich defense},
  author={Learn, Prompting},
  year={2023},
  url="https://learnprompting.org/docs/prompt_hacking/defensive_measures/sandwich_defense"
}

@misc{bib84,
  title={You can’t solve AI security problems with more AI},
  author={[84] Simon Willison},
  year={2023},
  url="https://simonwillison.net/2022/Sep/17/prompt-injection-more-ai/"
}

@inproceedings{bib85,
  title={Text-crs: A generalized certified robustness framework against textual adversarial attacks},
  author={Zhang, Xinyu and Hong, Hanbin and Hong, Yuan and Huang, Peng and Wang, Binghui and Ba, Zhongjie and Ren, Kui},
  booktitle={2024 IEEE Symposium on Security and Privacy (SP)},
  pages={2920--2938},
  year={2024},
  doi={10.1109/SP54263.2024.00053}}

@article{bib86,
  title={Interpretability and transparency-driven detection and transformation of textual adversarial examples (it-dt)},
  author={Sabir, Bushra and Babar, M Ali and Abuadbba, Sharif},
  journal={arXiv preprint},
  eprint={2307.01225},
  archivePrefix={arXiv},
  year={2023}
}

@article{bib87,
  title={Understanding deep learning (still) requires rethinking generalization},
  author={Zhang, Chiyuan and Bengio, Samy and Hardt, Moritz and Recht, Benjamin and Vinyals, Oriol},
  journal={Communications of the ACM},
  volume={64},
  number={3},
  pages={107--115},
  year={2021},
  doi = {10.1145/3446776},
}

@inproceedings{bib88,
  title={Joint extraction of entities and relations by adversarial training and mixup data augmentation},
  author={Chen, Hongjing and Lu, Gang and Wu, Xiaojun and Zhang, Yumei},
  booktitle={2021 7th International Conference on Computer and Communications (ICCC)},
  pages={1486--1490},
  year={2021},
  doi="10.1109/ICCC54389.2021.9674630"
}

@inproceedings{bib89,
  title={Robustness May Be at Odds with Accuracy},
  author={Tsipras, Dimitris and Santurkar, Shibani and Engstrom, Logan and Turner, Alexander and Madry, Aleksander},
  booktitle={International Conference on Learning Representations},
  year={2019}
}

@article{bib90,
  title={Large language model sentinel: Llm agent for adversarial purification},
  author={Lin, Guang and Tanaka, Toshihisa and Zhao, Qibin},
  journal={arXiv preprint},
  eprint={2405.20770},
  archivePrefix={arXiv},
  year={2024}
}

@article{bib91,
  title={Safe+ Safe= Unsafe? Exploring How Safe Images Can Be Exploited to Jailbreak Large Vision-Language Models},
  author={Cui, Chenhang and Deng, Gelei and Zhang, An and Zheng, Jingnan and Li, Yicong and Gao, Lianli and Zhang, Tianwei and Chua, Tat-Seng},
  journal={arXiv preprint arXiv:2411.11496},
  eprint={2411.11496},
  archivePrefix={arXiv},
  year={2024}
}

@article{bib92,
  title={Smoothllm: Defending large language models against jailbreaking attacks},
  author={Robey, Alexander and Wong, Eric and Hassani, Hamed and Pappas, George J},
  journal={arXiv preprint},
  eprint={2310.03684},
  archivePrefix={arXiv},
  year={2023}
}

@article{bib93,
  title={Defending chatgpt against jailbreak attack via self-reminders},
  author={Xie, Yueqi and Yi, Jingwei and Shao, Jiawei and Curl, Justin and Lyu, Lingjuan and Chen, Qifeng and Xie, Xing and Wu, Fangzhao},
  journal={Nature Machine Intelligence},
  volume={5},
  number={12},
  pages={1486--1496},
  year={2023},
  publisher={Nature Publishing Group UK London},
  doi="10.1038/s42256-023-00765-8",
}

@inproceedings{bib94,
 author = {Liu, Zichuan and Wang, Zefan and Xu, Linjie and Wang, Jinyu and Song, Lei and Wang, Tianchun and Chen, Chunlin and Cheng, Wei and Bian, Jiang},
 booktitle = {Advances in Neural Information Processing Systems},
 pages = {29723--29753},
 title = {Protecting Your LLMs with Information Bottleneck},
 url = {https://proceedings.neurips.cc/paper_files/paper/2024/file/34a1fc7890141f1ada3d8bc6199cce07-Paper-Conference.pdf},
 volume = {37},
 year = {2024}
}

@article{bib95,
  title={Llm self defense: By self examination, llms know they are being tricked},
  author={Phute, Mansi and Helbling, Alec and Hull, Matthew and Peng, ShengYun and Szyller, Sebastian and Cornelius, Cory and Chau, Duen Horng},
  journal={arXiv preprint},
  eprint={2308.07308},
  archivePrefix={arXiv},
  year={2023}
}

@inproceedings{bib96,
    title = "Exploring Social Bias in Chatbots using Stereotype Knowledge",
    author = "Lee, Nayeon  and
      Madotto, Andrea  and
      Fung, Pascale",
    booktitle = "Proceedings of the 2019 Workshop on Widening NLP",
    month = aug,
    year = "2019",
    url = "https://aclanthology.org/W19-3655/",
    pages = "177--180",
}

@article{bib97-llamaGuard,
  title={Llama guard: Llm-based input-output safeguard for human-ai conversations},
  author={Inan, Hakan and Upasani, Kartikeya and Chi, Jianfeng and Rungta, Rashi and Iyer, Krithika and Mao, Yuning and Tontchev, Michael and Hu, Qing and Fuller, Brian and Testuggine, Davide and others},
  journal={arXiv preprint},
  eprint={2312.06674},
  archivePrefix={arXiv},
  year={2023}
}

@inproceedings{bib98,
    title = "Pruning for Protection: Increasing Jailbreak Resistance in Aligned {LLM}s Without Fine-Tuning",
    author = "Hasan, Adib  and
      Rugina, Ileana  and
      Wang, Alex",
    booktitle = "Proceedings of the 7th BlackboxNLP Workshop: Analyzing and Interpreting Neural Networks for NLP",
    month = nov,
    year = "2024",
    doi = "10.18653/v1/2024.blackboxnlp-1.26",
    pages = "417--430",
}

@inproceedings{bib99,
  title={AutoDAN: Generating Stealthy Jailbreak Prompts on Aligned Large Language Models},
  author={Liu, Xiaogeng and Xu, Nan and Chen, Muhao and Xiao, Chaowei},
  booktitle={The Twelfth International Conference on Learning Representations},
  year={2024}
}

@inproceedings{bib100,
  title={Understanding hidden context in preference learning: Consequences for rlhf},
  author={Siththaranjan, Anand and Laidlaw, Cassidy and Hadfield-Menell, Dylan},
  booktitle={Socially Responsible Language Modelling Research},
  year={2023}
}

@article{bib101-llama2,
  title={Llama 2: Open foundation and fine-tuned chat models},
  author={Touvron, Hugo and Martin, Louis and Stone, Kevin and Albert, Peter and Almahairi, Amjad and Babaei, Yasmine and Bashlykov, Nikolay and Batra, Soumya and Bhargava, Prajjwal and Bhosale, Shruti and others},
  journal={arXiv preprint},
  eprint={2307.09288},
  archivePrefix={arXiv},
  year={2023}
}

@article{bib102-AttentionDefense,
  title={AttentionDefense: Leveraging System Prompt Attention for Explainable Defense Against Novel Jailbreaks},
  author={Siska, Charlotte and Sankaran, Anush},
  journal={arXiv preprint},
  eprint={2504.12321},
  archivePrefix={arXiv},
  year={2025}
}

@inproceedings{bib103,
    title = "Defending Jailbreak Prompts via In-Context Adversarial Game",
    author = "Zhou, Yujun  and
      Han, Yufei  and
      Zhuang, Haomin  and
      Guo, Kehan  and
      Liang, Zhenwen  and
      Bao, Hongyan  and
      Zhang, Xiangliang",
    booktitle = "Proceedings of the 2024 Conference on Empirical Methods in Natural Language Processing",
    month = nov,
    year = "2024",
    doi = "10.18653/v1/2024.emnlp-main.1121",
    pages = "20084--20105",
}

@article{bib104,
  title={Cognitive overload attack: Prompt injection for long context},
  author={Upadhayay, Bibek and Behzadan, Vahid and Karbasi, Amin},
  journal={arXiv preprint},
  eprint={2410.11272},
  archivePrefix={arXiv},
  year={2024}
}

@article{bib105,
  title={Poisonedrag: Knowledge corruption attacks to retrieval-augmented generation of large language models},
  author={Zou, Wei and Geng, Runpeng and Wang, Binghui and Jia, Jinyuan},
  journal={arXiv preprint},
  eprint={2402.07867},
  archivePrefix={arXiv},
  year={2024}
}

@article{bib106,
  title={Whispers in Grammars: Injecting Covert Backdoors to Compromise Dense Retrieval Systems},
  author={Long, Quanyu and Deng, Yue and Gan, LeiLei and Wang, Wenya and Pan, Sinno Jialin},
  journal={arXiv preprint},
  eprint={2402.13532},
  archivePrefix={arXiv},
  year={2024}
}

@article{bib107,
  title={Certifiably robust rag against retrieval corruption},
  author={Xiang, Chong and Wu, Tong and Zhong, Zexuan and Wagner, David and Chen, Danqi and Mittal, Prateek},
  journal={arXiv preprint},
  eprint={2405.15556},
  archivePrefix={arXiv},
  year={2024}
}

@inproceedings{bib108,
  title={InstructRAG: Instructing Retrieval Augmented Generation via Self-Synthesized Rationales},
  author={Wei, Zhepei and Chen, Wei-Lin and Meng, Yu},
  booktitle={Adaptive Foundation Models: Evolving AI for Personalized and Efficient Learning},
  year={2024}
}

@inproceedings{bib109,
 author = {Yan, Lu and Zhang, Zhuo and Tao, Guanhong and Zhang, Kaiyuan and Chen, Xuan and Shen, Guangyu and Zhang, Xiangyu},
 booktitle = {Advances in Neural Information Processing Systems},
 pages = {66755--66767},
 title = {ParaFuzz: An Interpretability-Driven Technique for Detecting Poisoned Samples in NLP},
 url = {https://proceedings.neurips.cc/paper_files/paper/2023/file/d2b752ed4726286a4b488ae16e091d64-Paper-Conference.pdf},
 volume = {36},
 year = {2023}
}

@InProceedings{bib110,
  title = 	 {Obfuscated Gradients Give a False Sense of Security: Circumventing Defenses to Adversarial Examples},
  author =       {Athalye, Anish and Carlini, Nicholas and Wagner, David},
  booktitle = 	 {Proceedings of the 35th International Conference on Machine Learning},
  pages = 	 {274--283},
  year = 	 {2018},
  volume = 	 {80},
  month = 	 {10--15 Jul},
  url = 	 {https://proceedings.mlr.press/v80/athalye18a.html},
}

@inproceedings{bib111,
  title={Evading Adversarial Example Detection Defenses with Orthogonal Projected Gradient Descent},
  author={Bryniarski, Oliver and Hingun, Nabeel and Pachuca, Pedro and Wang, Vincent and Carlini, Nicholas},
  booktitle={International Conference on Learning Representations},
  year={2022}
}

@article{bib112,
  title={A LLM assisted exploitation of AI-Guardian},
  author={Carlini, Nicholas},
  journal={arXiv preprint},
  eprint={2307.15008},
  archivePrefix={arXiv},
  year={2023}
}

@InProceedings{bib113,
  title = 	 {Poisoning Language Models During Instruction Tuning},
  author =       {Wan, Alexander and Wallace, Eric and Shen, Sheng and Klein, Dan},
  booktitle = 	 {Proceedings of the 40th International Conference on Machine Learning},
  pages = 	 {35413--35425},
  year = 	 {2023},
  volume = 	 {202},
  month = 	 {23--29 Jul},
  url = 	 {https://proceedings.mlr.press/v202/wan23b.html},
}

@article{bib114,
  title={Llm defenses are not robust to multi-turn human jailbreaks yet},
  author={Li, Nathaniel and Han, Ziwen and Steneker, Ian and Primack, Willow and Goodside, Riley and Zhang, Hugh and Wang, Zifan and Menghini, Cristina and Yue, Summer},
  journal={arXiv preprint},
  eprint={2408.15221},
  archivePrefix={arXiv},
  year={2024}
}

@inproceedings{bib115,
    title = "{LLM}s know their vulnerabilities: Uncover Safety Gaps through Natural Distribution Shifts",
    author = "Ren, Qibing  and
      Li, Hao  and
      Liu, Dongrui  and
      Xie, Zhanxu  and
      Lu, Xiaoya  and
      Qiao, Yu  and
      Sha, Lei  and
      Yan, Junchi  and
      Ma, Lizhuang  and
      Shao, Jing",
    editor = "Che, Wanxiang  and
      Nabende, Joyce  and
      Shutova, Ekaterina  and
      Pilehvar, Mohammad Taher",
    booktitle = "Proceedings of the 63rd Annual Meeting of the Association for Computational Linguistics (Volume 1: Long Papers)",
    month = jul,
    year = "2025",
    address = "Vienna, Austria",
    publisher = "Association for Computational Linguistics",
    url = "https://aclanthology.org/2025.acl-long.1207/",
    doi = "10.18653/v1/2025.acl-long.1207",
    pages = "24763--24785",
    ISBN = "979-8-89176-251-0",
}

@article{bib116,
  title={RED QUEEN: Safeguarding Large Language Models against Concealed Multi-Turn Jailbreaking},
  author={Jiang, Yifan and Aggarwal, Kriti and Laud, Tanmay and Munir, Kashif and Pujara, Jay and Mukherjee, Subhabrata},
  journal={arXiv preprint},
  eprint={2409.17458},
  archivePrefix={arXiv},
  year={2024}
}

@inproceedings{bib117,
 author = {Lee, Nayeon and Ping, Wei and Xu, Peng and Patwary, Mostofa and Fung, Pascale N and Shoeybi, Mohammad and Catanzaro, Bryan},
 booktitle = {Advances in Neural Information Processing Systems},
 pages = {34586--34599},
 title = {Factuality Enhanced Language Models for Open-Ended Text Generation},
 url = {https://proceedings.neurips.cc/paper_files/paper/2022/file/df438caa36714f69277daa92d608dd63-Paper-Conference.pdf},
 volume = {35},
 year = {2022}
}

@inproceedings{bib118,
  title={AlpaGasus: Training a Better Alpaca with Fewer Data},
  author={Chen, Lichang and Li, Shiyang and Yan, Jun and Wang, Hai and Gunaratna, Kalpa and Yadav, Vikas and Tang, Zheng and Srinivasan, Vijay and Zhou, Tianyi and Huang, Heng and others},
  booktitle={The Twelfth International Conference on Learning Representations},
  year={2024}
}

@inproceedings{bib119,
 author = {Wu, Zeqiu and Hu, Yushi and Shi, Weijia and Dziri, Nouha and Suhr, Alane and Ammanabrolu, Prithviraj and Smith, Noah A and Ostendorf, Mari and Hajishirzi, Hannaneh},
 booktitle = {Advances in Neural Information Processing Systems},
 pages = {59008--59033},
 title = {Fine-Grained Human Feedback Gives Better Rewards for Language Model Training},
 url = {https://proceedings.neurips.cc/paper_files/paper/2023/file/b8c90b65739ae8417e61eadb521f63d5-Paper-Conference.pdf},
 volume = {36},
 year = {2023}
}

@inproceedings{bib120,
 author = {Li, Kenneth and Patel, Oam and Vi\'{e}gas, Fernanda and Pfister, Hanspeter and Wattenberg, Martin},
 booktitle = {Advances in Neural Information Processing Systems},
 pages = {41451--41530},
 title = {Inference-Time Intervention: Eliciting Truthful Answers from a Language Model},
 url = {https://proceedings.neurips.cc/paper_files/paper/2023/file/81b8390039b7302c909cb769f8b6cd93-Paper-Conference.pdf},
 volume = {36},
 year = {2023}
}

@inproceedings{bib121,
    title = "{F}resh{LLM}s: Refreshing Large Language Models with Search Engine Augmentation",
    author = "Vu, Tu  and
      Iyyer, Mohit  and
      Wang, Xuezhi  and
      Constant, Noah  and
      Wei, Jerry  and
      Wei, Jason  and
      Tar, Chris  and
      Sung, Yun-Hsuan  and
      Zhou, Denny  and
      Le, Quoc  and
      Luong, Thang",
    booktitle = "Findings of the Association for Computational Linguistics: ACL 2024",
    month = aug,
    year = "2024",
    doi = "10.18653/v1/2024.findings-acl.813",
    pages = "13697--13720",
}

@inproceedings{bib122,
    title = "{RARR}: Researching and Revising What Language Models Say, Using Language Models",
    author = "Gao, Luyu  and
      Dai, Zhuyun  and
      Pasupat, Panupong  and
      Chen, Anthony  and
      Chaganty, Arun Tejasvi  and
      Fan, Yicheng  and
      Zhao, Vincent  and
      Lao, Ni  and
      Lee, Hongrae  and
      Juan, Da-Cheng  and
      Guu, Kelvin",
    booktitle = "Proceedings of the 61st Annual Meeting of the Association for Computational Linguistics (Volume 1: Long Papers)",
    month = jul,
    year = "2023",
    doi = "10.18653/v1/2023.acl-long.910",
    pages = "16477--16508",
}

@article{bib123,
  title={A mathematical investigation of hallucination and creativity in GPT models},
  author={Lee, Minhyeok},
  journal={Mathematics},
  volume={11},
  number={10},
  pages={2320},
  year={2023},
  doi="10.3390/math11102320"
}

@inproceedings{bib124,
    title = "Measuring and Narrowing the Compositionality Gap in Language Models",
    author = "Press, Ofir  and
      Zhang, Muru  and
      Min, Sewon  and
      Schmidt, Ludwig  and
      Smith, Noah  and
      Lewis, Mike",
    booktitle = "Findings of the Association for Computational Linguistics: EMNLP 2023",
    month = dec,
    year = "2023",
    doi = "10.18653/v1/2023.findings-emnlp.378",
    pages = "5687--5711",
}

@inproceedings{bib125,
    title = "Towards Mitigating {LLM} Hallucination via Self Reflection",
    author = "Ji, Ziwei  and
      Yu, Tiezheng  and
      Xu, Yan  and
      Lee, Nayeon  and
      Ishii, Etsuko  and
      Fung, Pascale",
    booktitle = "Findings of the Association for Computational Linguistics: EMNLP 2023",
    month = dec,
    year = "2023",
    doi = "10.18653/v1/2023.findings-emnlp.123",
    pages = "1827--1843",
}

@article{bib126,
  title={Can llms produce faithful explanations for fact-checking? towards faithful explainable fact-checking via multi-agent debate},
  author={Kim, Kyungha and Lee, Sangyun and Huang, Kung-Hsiang and Chan, Hou Pong and Li, Manling and Ji, Heng},
  journal={arXiv preprint},
  eprint={2402.07401},
  archivePrefix={arXiv},
  year={2024}
}

@INPROCEEDINGS{bib127,
  author={Sun, Xiaoxi and Li, Jinpeng and Zhong, Yan and Zhao, Dongyan and Yan, Rui},
  booktitle={ICASSP 2025 - 2025 IEEE International Conference on Acoustics, Speech and Signal Processing (ICASSP)}, 
  title={Towards Detecting LLMs Hallucination via Markov Chain-based Multi-agent Debate Framework}, 
  year={2025},
  volume={},
  number={},
  pages={1-5},
  doi={10.1109/ICASSP49660.2025.10889448}
}

@article{bib128,
  title={Good Parenting is all you need--Multi-agentic LLM Hallucination Mitigation},
  author={Kwartler, Ted and Berman, Matthew and Aqrawi, Alan},
  journal={arXiv preprint},
  eprint={2410.14262},
  archivePrefix={arXiv},
  year={2024}
}

@article{bib129,
title = {Mitigating reasoning hallucination through Multi-agent Collaborative Filtering},
journal = {Expert Systems with Applications},
volume = {263},
pages = {125723},
year = {2025},
issn = {0957-4174},
doi = {https://doi.org/10.1016/j.eswa.2024.125723},
author = {Jinxin Shi and Jiabao Zhao and Xingjiao Wu and Ruyi Xu and Yuan-Hao Jiang and Liang He},
}

@inproceedings{bib130,
    title = "Small Agent Can Also Rock! Empowering Small Language Models as Hallucination Detector",
    author = "Cheng, Xiaoxue  and
      Li, Junyi  and
      Zhao, Xin  and
      Zhang, Hongzhi  and
      Zhang, Fuzheng  and
      Zhang, Di  and
      Gai, Kun  and
      Wen, Ji-Rong",
    booktitle = "Proceedings of the 2024 Conference on Empirical Methods in Natural Language Processing",
    month = nov,
    year = "2024",
    doi = "10.18653/v1/2024.emnlp-main.809",
    pages = "14600--14615",
}

@misc{bib131,
      title={HallE-Control: Controlling Object Hallucination in Large Multimodal Models}, 
      author={Bohan Zhai and Shijia Yang and Chenfeng Xu and Sheng Shen and Kurt Keutzer and Chunyuan Li and Manling Li},
      year={2024},
      journal={arXiv preprint},
      eprint={2310.01779},
      archivePrefix={arXiv},
      primaryClass={cs.CV},
}

@inproceedings{bib132,
    title = "Volcano: Mitigating Multimodal Hallucination through Self-Feedback Guided Revision",
    author = "Lee, Seongyun  and
      Park, Sue Hyun  and
      Jo, Yongrae  and
      Seo, Minjoon",
    booktitle = "Proceedings of the 2024 Conference of the North American Chapter of the Association for Computational Linguistics: Human Language Technologies (Volume 1: Long Papers)",
    month = jun,
    year = "2024",
    doi = "10.18653/v1/2024.naacl-long.23",
    pages = "391--404",
}

@misc{bib133,
      title={TrustAgent: Towards Safe and Trustworthy LLM-based Agents}, 
      author={Wenyue Hua and Xianjun Yang and Mingyu Jin and Zelong Li and Wei Cheng and Ruixiang Tang and Yongfeng Zhang},
      year={2024},
      journal={arXiv preprint},
      eprint={2402.01586},
      archivePrefix={arXiv},
      primaryClass={cs.CL},
}

@inproceedings{bib134,
    title = "{Q}u{BE}: Question-based Belief Enhancement for Agentic {LLM} Reasoning",
    author = "Kim, Minsoo  and
      Kim, Jongyoon  and
      Kim, Jihyuk  and
      Hwang, Seung-won",
    booktitle = "Proceedings of the 2024 Conference on Empirical Methods in Natural Language Processing",
    month = nov,
    year = "2024",
    doi = "10.18653/v1/2024.emnlp-main.1193",
    pages = "21403--21423",
}

@misc{bib135,
      title={GuardAgent: Safeguard LLM Agents by a Guard Agent via Knowledge-Enabled Reasoning}, 
      author={Zhen Xiang and Linzhi Zheng and Yanjie Li and Junyuan Hong and Qinbin Li and Han Xie and Jiawei Zhang and Zidi Xiong and Chulin Xie and Carl Yang and Dawn Song and Bo Li},
      year={2025},
      journal={arXiv preprint},
      eprint={2406.09187},
      archivePrefix={arXiv},
      primaryClass={cs.LG},
}

@article{bib136,
  title={Avalon's game of thoughts: Battle against deception through recursive contemplation},
  author={Wang, Shenzhi and Liu, Chang and Zheng, Zilong and Qi, Siyuan and Chen, Shuo and Yang, Qisen and Zhao, Andrew and Wang, Chaofei and Song, Shiji and Huang, Gao},
  journal={arXiv preprint},
  eprint={2310.01320},
  archivePrefix={arXiv},
  year={2023}
}

@article{bib138,
  title={Jailjudge: A comprehensive jailbreak judge benchmark with multi-agent enhanced explanation evaluation framework},
  author={Liu, Fan and Feng, Yue and Xu, Zhao and Su, Lixin and Ma, Xinyu and Yin, Dawei and Liu, Hao},
  journal={arXiv preprint},
  eprint={2410.12855},
  archivePrefix={arXiv},
  year={2024}
}

@article{bib139,
  title={AgentHarm: A Benchmark for Measuring Harmfulness of LLM Agents},
  author={Andriushchenko, Maksym and Souly, Alexandra and Dziemian, Mateusz and Duenas, Derek and Lin, Maxwell and Wang, Justin and Hendrycks, Dan and Zou, Andy and Kolter, Zico and Fredrikson, Matt and Winsor, Eric and Wynne, Jerome and Gal, Yarin and Davies, Xander},
  journal={arXiv preprint},
  eprint={2410.09024},
  archivePrefix={arXiv},
  year={2024}
}

@article{bib140,
  title={AutoAdvExBench: Benchmarking autonomous exploitation of adversarial example defenses},
  author={Carlini, Nicholas and Rando, Javier and Debenedetti, Edoardo and Nasr, Milad and Tram{\`e}r, Florian},
  journal={arXiv preprint},
  eprint={2503.01811},
  archivePrefix={arXiv},
  year={2025}
}

@inproceedings{bib141,
    title = "{M}ulti{A}gent Collaboration Attack: Investigating Adversarial Attacks in Large Language Model Collaborations via Debate",
    author = "Amayuelas, Alfonso  and
      Yang, Xianjun  and
      Antoniades, Antonis  and
      Hua, Wenyue  and
      Pan, Liangming  and
      Wang, William Yang",
    booktitle = "Findings of the Association for Computational Linguistics: EMNLP 2024",
    month = nov,
    year = "2024",
    doi = "10.18653/v1/2024.findings-emnlp.407",
    pages = "6929--6948",
}

@inproceedings{bib142,
    title = "{R}-Judge: Benchmarking Safety Risk Awareness for {LLM} Agents",
    author = "Yuan, Tongxin  and
      He, Zhiwei  and
      Dong, Lingzhong  and
      Wang, Yiming  and
      Zhao, Ruijie  and
      Xia, Tian  and
      Xu, Lizhen  and
      Zhou, Binglin  and
      Li, Fangqi  and
      Zhang, Zhuosheng  and
      Wang, Rui  and
      Liu, Gongshen",
    booktitle = "Findings of the Association for Computational Linguistics: EMNLP 2024",
    month = nov,
    year = "2024",
    doi = "10.18653/v1/2024.findings-emnlp.79",
    pages = "1467--1490",
}

@inproceedings{bib143,
 author = {Zheng, Jingnan and Wang, Han and Zhang, An and Nguyen, Tai D. and Sun, Jun and Chua, Tat-Seng},
 booktitle = {Advances in Neural Information Processing Systems},
 editor = {A. Globerson and L. Mackey and D. Belgrave and A. Fan and U. Paquet and J. Tomczak and C. Zhang},
 pages = {99040--99088},
 publisher = {Curran Associates, Inc.},
 title = {ALI-Agent: Assessing LLMs\textquotesingle   Alignment with Human Values via Agent-based Evaluation},
 url = {https://proceedings.neurips.cc/paper_files/paper/2024/file/b35c38f70065ac6c694089ca93a015bb-Paper-Conference.pdf},
 volume = {37},
 year = {2024}
}

@inproceedings{bib144,
  title={BELLS: A Framework Towards Future Proof Benchmarks for the Evaluation of LLM Safeguards},
  author={Dorn, Diego and Variengien, Alexandre and Segerie, Charbel-Raphael and Corruble, Vincent},
  booktitle={ICML 2024 Next Generation of AI Safety Workshop},
  year={2024}
}

@article{bib153,
    author = {Pang, Guansong and Shen, Chunhua and Cao, Longbing and Hengel, Anton Van Den},
    title = {Deep Learning for Anomaly Detection: A Review},
    year = {2021},
    issue_date = {March 2022},
    volume = {54},
    number = {2},
    issn = {0360-0300},
    doi = {10.1145/3439950},
    journal = {ACM Comput. Surv.},
    month = mar,
    articleno = {38},
    numpages = {38},
}

@article{bib154,
  title={Audit-LLM: Multi-Agent Collaboration for Log-based Insider Threat Detection},
  author={Song, Chengyu and Ma, Linru and Zheng, Jianming and Liao, Jinzhi and Kuang, Hongyu and Yang, Lin},
  journal={arXiv preprint},
  eprint={2408.08902},
  archivePrefix={arXiv},
  year={2024}
}

@article{bib155,
  title={Enhancing anomaly detection in financial markets with an llm-based multi-agent framework},
  author={Park, Taejin},
  journal={arXiv preprint},
  eprint={2403.19735},
  archivePrefix={arXiv},
  year={2024}
}

@inproceedings{bib156,
  title={Agentic Anomaly Detection for Shipping},
  author={Timms, Alexander and Langbridge, Abigail and O'Donncha, Fearghal},
  booktitle={NeurIPS 2024 Workshop on Open-World Agents},
  year={2024}
}

@article{bib157,
    AUTHOR = {Hindy, Hanan and Atkinson, Robert and Tachtatzis, Christos and Colin, Jean-Noël and Bayne, Ethan and Bellekens, Xavier},
    TITLE = {Utilising Deep Learning Techniques for Effective Zero-Day Attack Detection},
    JOURNAL = {Electronics},
    VOLUME = {9},
    YEAR = {2020},
    NUMBER = {10},
    ARTICLE-NUMBER = {1684},
    ISSN = {2079-9292},
    DOI = {10.3390/electronics9101684}
}

@article{bib158,
  title={Zero-day attack detection: a systematic literature review},
  author={Ahmad, Rasheed and Alsmadi, Izzat and Alhamdani, Wasim and Tawalbeh, Lo’ai},
  journal={Artificial Intelligence Review},
  volume={56},
  number={10},
  pages={10733--10811},
  year={2023},
  doi="10.1007/s10462-023-10437-z"
}

@mastersthesis{bib159,
  title={Next-generation intrusion detection systems with LLMs: real-time anomaly detection, explainable AI, and adaptive data generation},
  author={Ali, Tarek},
  year={2024},
  school={T. Ali}
}

@inproceedings{bib160,
  title={IDS-Agent: An LLM Agent for Explainable Intrusion Detection in IoT Networks},
  author={Li, Yanjie and Xiang, Zhen and Bastian, Nathaniel D and Song, Dawn and Li, Bo},
  booktitle={NeurIPS 2024 Workshop on Open-World Agents},
  year={2024}
}

@article{bib161,
  title={Anomalyllm: Few-shot anomaly edge detection for dynamic graphs using large language models},
  author={Liu, Shuo and Yao, Di and Fang, Lanting and Li, Zhetao and Li, Wenbin and Feng, Kaiyu and Ji, XiaoWen and Bi, Jingping},
  journal={arXiv preprint},
  eprint={2405.07626},
  archivePrefix={arXiv},
  year={2024}
}

@inproceedings{bib184,
  title={Combining Fine-tuning and LLM-based Agents for Intuitive Smart Contract Auditing with Justifications},
  author={Ma, Wei and Wu, Daoyuan and Sun, Yuqiang and Wang, Tianwen and Liu, Shangqing and Zhang, Jian and Xue, Yue and Liu, Yang},
  booktitle={2025 IEEE/ACM 47th International Conference on Software Engineering (ICSE)},
  pages={330--342},
  year={2024},
  doi="10.1109/ICSE55347.2025.00027"
}

@article{bib185,
  title={Autosafecoder: A multi-agent framework for securing llm code generation through static analysis and fuzz testing},
  author={Nunez, Ana and Islam, Nafis Tanveer and Jha, Sumit Kumar and Najafirad, Peyman},
  journal={arXiv preprint},
  eprint={2409.10737},
  archivePrefix={arXiv},
  year={2024}
}

@INPROCEEDINGS{bib186,
  author={Hu, Sihao and Huang, Tiansheng and İlhan, Fatih and Tekin, Selim Furkan and Liu, Ling},
  booktitle={2023 5th IEEE International Conference on Trust, Privacy and Security in Intelligent Systems and Applications (TPS-ISA)}, 
  title={Large Language Model-Powered Smart Contract Vulnerability Detection: New Perspectives}, 
  year={2023},
  volume={},
  number={},
  pages={297-306},
  doi={10.1109/TPS-ISA58951.2023.00044}
}

@article{bib187,
  title={LLM-SmartAudit: Advanced Smart Contract Vulnerability Detection},
  author={Wei, Zhiyuan and Sun, Jing and Zhang, Zijiang and Zhang, Xianhao and Li, Meng and Hou, Zhe},
  journal={arXiv preprint},
  eprint={2410.09381},
  archivePrefix={arXiv},
  year={2024}
}

@misc{bib188,
  title={From Naptime to Big Sleep: Using Large Language Models To Catch Vulnerabilities In Real-World Code}, 
  author={Big Sleep team},
  year={2024},
  url="https://googleprojectzero.blogspot.com/2024/10/from-naptime-to-big-sleep.html"
}

@inproceedings{bib189,
author = {Huang, Junjie and Zhu, Quanyan},
title = {PenHeal: A Two-Stage LLM Framework for Automated Pentesting and Optimal Remediation},
year = {2024},
isbn = {9798400712296},
doi = {10.1145/3689933.3690831},
booktitle = {Proceedings of the Workshop on Autonomous Cybersecurity},
pages = {11–22},
numpages = {12},
series = {AutonomousCyber '24}
}

@article{bib190,
  title={A preliminary study on using large language models in software pentesting},
  author={Shashwat, Kumar and Hahn, Francis and Ou, Xinming and Goldgof, Dmitry and Hall, Lawrence and Ligatti, Jay and Rajgopalan, S Raj and Tabari, Armin Ziaie},
  journal={arXiv preprint arXiv:2401.17459},
  eprint={2401.17459},
  archivePrefix={arXiv},
  year={2024}
}

@INPROCEEDINGS{bib191,
  author={Yang, Tinghui and Jiang, Zhiyuan and Wang, Yongjun},
  booktitle={2024 5th International Conference on Big Data \& Artificial Intelligence \& Software Engineering (ICBASE)}, 
  title={LLMSQLi: A Black-Box Web SQLi Detection Tool Based on Large Language Model}, 
  year={2024},
  volume={},
  number={},
  pages={629-633},
  doi={10.1109/ICBASE63199.2024.10762654}
}

@inproceedings{bib192,
 author = {Lewis, Patrick and Perez, Ethan and Piktus, Aleksandra and Petroni, Fabio and Karpukhin, Vladimir and Goyal, Naman and K\"{u}ttler, Heinrich and Lewis, Mike and Yih, Wen-tau and Rockt\"{a}schel, Tim and Riedel, Sebastian and Kiela, Douwe},
 booktitle = {Advances in Neural Information Processing Systems},
 pages = {9459--9474},
 title = {Retrieval-Augmented Generation for Knowledge-Intensive NLP Tasks},
 url = {https://proceedings.neurips.cc/paper_files/paper/2020/file/6b493230205f780e1bc26945df7481e5-Paper.pdf},
 volume = {33},
 year = {2020}
}

@article{bib193,
  title={PentestAgent: Incorporating LLM Agents to Automated Penetration Testing},
  author={Shen, Xiangmin and Wang, Lingzhi and Li, Zhenyuan and Chen, Yan and Zhao, Wencheng and Sun, Dawei and Wang, Jiashui and Ruan, Wei},
  journal={arXiv preprint},
  eprint={2411.05185},
  archivePrefix={arXiv},
  year={2024}
}

@article{bib194,
  title={Towards automated penetration testing: Introducing llm benchmark, analysis, and improvements},
  author={Isozaki, Isamu and Shrestha, Manil and Console, Rick and Kim, Edward},
  journal={arXiv preprint},
  eprint={2410.17141},
  archivePrefix={arXiv},
  year={2024}
}

@article{bib195,
  title={Llm agents can autonomously hack websites},
  author={Fang, Richard and Bindu, Rohan and Gupta, Akul and Zhan, Qiusi and Kang, Daniel},
  journal={arXiv preprint},
  eprint={2402.06664},
  archivePrefix={arXiv},
  year={2024}
}

@article{bib196,
  title={LLM Agents can Autonomously Exploit One-day Vulnerabilities},
  author={Fang, Richard and Bindu, Rohan and Gupta, Akul and Kang, Daniel},
  journal={arXiv preprint},
  eprint={2404.08144},
  archivePrefix={arXiv},
  year={2024}
}

@article{bib198,
  title={Artificial Intelligence as the New Hacker: Developing Agents for Offensive Security},
  author={Valencia, Leroy Jacob},
  journal={arXiv preprint},
  eprint={2406.07561},
  archivePrefix={arXiv},
  year={2024}
}

@misc{bib199,
  title={2025 Verizon data breach investigations report}, 
  author={Keepnet Labs},
  year={2025},
  url="https://keepnetlabs.com/blog/2025-verizon-data-breach-investigations-report"
}

@misc{bib200,
  title={Cost of a Data Breach Report 2024}, 
  author={IBM},
  year={2024},
  url="https://www.ibm.com/downloads/documents/us-en/107a02e94948f4ec.https://www.ibm.com/security/digital-assets/cost-data-breach-report/1Cost of a Data Breach Report 2020.pdf"
}

@misc{bib201,
  title={Sony Hackers Used Phishing Emails to Breach Company Networks},
  author={David Bisson},
  year={2015},
  url="https://www.tripwire.com/state-of-security/sony-hackers-used-phishing-emails-to-breach-company-networks"
}

@article{bib202,
  title={The perfect weapon: How Russian cyberpower invaded the US},
  author={Lipton, Eric and Sanger, David E and Shane, Scott},
  journal={The New York Times},
  year={2016}
}

@inproceedings{bib203,
    title = "Defending Against Social Engineering Attacks in the Age of {LLM}s",
    author = "Ai, Lin  and
      Kumarage, Tharindu Sandaruwan  and
      Bhattacharjee, Amrita  and
      Liu, Zizhou  and
      Hui, Zheng  and
      Davinroy, Michael S.  and
      Cook, James  and
      Cassani, Laura  and
      Trapeznikov, Kirill  and
      Kirchner, Matthias  and
      Basharat, Arslan  and
      Hoogs, Anthony  and
      Garland, Joshua  and
      Liu, Huan  and
      Hirschberg, Julia",
    booktitle = "Proceedings of the 2024 Conference on Empirical Methods in Natural Language Processing",
    month = nov,
    year = "2024",
    doi = "10.18653/v1/2024.emnlp-main.716",
    pages = "12880--12902"
}

@article{bib204,
  title={From Chatbots to PhishBots?--Preventing Phishing scams created using ChatGPT, Google Bard and Claude},
  author={Roy, Sayak Saha and Thota, Poojitha and Naragam, Krishna Vamsi and Nilizadeh, Shirin},
  journal={arXiv preprint},
  eprint={2310.19181},
  archivePrefix={arXiv},
  year={2023}
}

@article{bib205,
  title={Spear phishing with large language models},
  author={Hazell, Julian},
  journal={arXiv preprint},
  eprint={2305.06972},
  archivePrefix={arXiv},
  year={2023}
}

@article{bib207,
  title={Spear phishing with large language models},
  author={Hazell, Julian},
  journal={arXiv preprint},
  eprint={2305.06972},
  archivePrefix={arXiv},
  year={2023}
}

@article{bib208,
  title={Neural codec language models are zero-shot text to speech synthesizers},
  author={Wang, Chengyi and Chen, Sanyuan and Wu, Yu and Zhang, Ziqiang and Zhou, Long and Liu, Shujie and Chen, Zhuo and Liu, Yanqing and Wang, Huaming and Li, Jinyu and others},
  journal={arXiv preprint},
  eprint={2301.02111},
  archivePrefix={arXiv},
  year={2023}
}

@article{bib210,
title = {SpearBot: Leveraging large language models in a generative-critique framework for spear-phishing email generation},
journal = {Information Fusion},
volume = {122},
pages = {103176},
year = {2025},
issn = {1566-2535},
doi = {https://doi.org/10.1016/j.inffus.2025.103176},
author = {Qinglin Qi and Yun Luo and Yijia Xu and Wenbo Guo and Yong Fang},
}

@article{bib211,
  title={Llms-as-judges: a comprehensive survey on llm-based evaluation methods},
  author={Li, Haitao and Dong, Qian and Chen, Junjie and Su, Huixue and Zhou, Yujia and Ai, Qingyao and Ye, Ziyi and Liu, Yiqun},
  journal={arXiv preprint},
  eprint={2412.05579},
  archivePrefix={arXiv},
  year={2024}
}

@article{bib212,
  title={Agent-as-a-judge: Evaluate agents with agents},
  author={Zhuge, Mingchen and Zhao, Changsheng and Ashley, Dylan and Wang, Wenyi and Khizbullin, Dmitrii and Xiong, Yunyang and Liu, Zechun and Chang, Ernie and Krishnamoorthi, Raghuraman and Tian, Yuandong and others},
  journal={arXiv preprint},
  eprint={2410.10934},
  archivePrefix={arXiv},
  year={2024}
}

@article{bib213,
  title={Evaluating Large Language Models' Capability to Launch Fully Automated Spear Phishing Campaigns: Validated on Human Subjects},
  author={Heiding, Fred and Lermen, Simon and Kao, Andrew and Schneier, Bruce and Vishwanath, Arun},
  journal={arXiv preprint},
  eprint={2412.00586},
  archivePrefix={arXiv},
  year={2024}
}

@article{bib214,
  title={Voice-Enabled AI Agents can Perform Common Scams},
  author={Fang, Richard and Bowman, Dylan and Kang, Daniel},
  journal={arXiv preprint},
  eprint={2410.15650},
  archivePrefix={arXiv},
  year={2024}
}

@inproceedings {bib216,
    author = {Ruofan Liu and Yun Lin and Yifan Zhang and Penn Han Lee and Jin Song Dong},
    title = {Knowledge Expansion and Counterfactual Interaction for {Reference-Based} Phishing Detection},
    booktitle = {32nd USENIX Security Symposium (USENIX Security 23)},
    year = {2023},
    isbn = {978-1-939133-37-3},
    pages = {4139--4156},
    url = {https://www.usenix.org/conference/usenixsecurity23/presentation/liu-ruofan},
    month = aug
}

@inproceedings{bib217,
  title={Phishagent: a robust multimodal agent for phishing webpage detection},
  author={Cao, Tri and Huang, Chengyu and Li, Yuexin and Huilin, Wang and He, Amy and Oo, Nay and Hooi, Bryan},
  booktitle={Proceedings of the AAAI Conference on Artificial Intelligence},
  volume={39},
  number={27},
  pages={27869--27877},
  doi="10.1609/aaai.v39i27.35003",
  year={2025}
}

@inproceedings{bib218,
  title={Large Multimodal Agents for Accurate Phishing Detection with Enhanced Token Optimization and Cost Reduction},
  author={Trad, Fouad and Chehab, Ali},
  booktitle={2024 2nd International Conference on Foundation and Large Language Models (FLLM)},
  pages={229--237},
  year={2024},
}

@article{bib239,
  title={Autodefense: Multi-agent llm defense against jailbreak attacks},
  author={Zeng, Yifan and Wu, Yiran and Zhang, Xiao and Wang, Huazheng and Wu, Qingyun},
  journal={arXiv preprint},
  eprint={2403.04783},
  archivePrefix={arXiv},
  year={2024}
}

@inproceedings{bib243,
  title={A Perfect Collusion Benchmark: How can AI agents be prevented from colluding with information-theoretic undetectability?},
  author={Motwani, Sumeet Ramesh and Baranchuk, Mikhail and Hammond, Lewis and de Witt, Christian Schroeder},
  booktitle={Multi-Agent Security Workshop@ NeurIPS'23},
  year={2023}
}

@article{bib244,
title = {Autonomous agents modelling other agents: A comprehensive survey and open problems},
journal = {Artificial Intelligence},
volume = {258},
pages = {66-95},
year = {2018},
issn = {0004-3702},
doi = {https://doi.org/10.1016/j.artint.2018.01.002},
author = {Stefano V. Albrecht and Peter Stone},
}

@ARTICLE{bib245,
  author={Dorri, Ali and Kanhere, Salil S. and Jurdak, Raja},
  journal={IEEE Access}, 
  title={Multi-Agent Systems: A Survey}, 
  year={2018},
  volume={6},
  number={},
  pages={28573-28593},
  doi={10.1109/ACCESS.2018.2831228}
}

@article{bib246,
author = {Hernandez-Leal, Pablo and Kartal, Bilal and Taylor, Matthew E.},
title = {A survey and critique of multiagent deep reinforcement learning},
year = {2019},
issue_date = {Nov 2019},
address = {USA},
volume = {33},
number = {6},
issn = {1387-2532},
doi = {10.1007/s10458-019-09421-1},
journal = {Autonomous Agents and Multi-Agent Systems},
month = nov,
pages = {750–797},
numpages = {48},
}

@article{bib247,
  title={Imprompter: Tricking LLM Agents into Improper Tool Use},
  author={Fu, Xiaohan and Li, Shuheng and Wang, Zihan and Liu, Yihao and Gupta, Rajesh K and Berg-Kirkpatrick, Taylor and Fernandes, Earlence},
  journal={arXiv preprint},
  eprint={2410.14923},
  archivePrefix={arXiv},
  year={2024}
}

@article{bib249,
author = {Deng, Zehang and Guo, Yongjian and Han, Changzhou and Ma, Wanlun and Xiong, Junwu and Wen, Sheng and Xiang, Yang},
title = {AI Agents Under Threat: A Survey of Key Security Challenges and Future Pathways},
year = {2025},
issue_date = {July 2025},
volume = {57},
number = {7},
issn = {0360-0300},
doi = {10.1145/3716628},
journal = {ACM Comput. Surv.},
month = feb,
articleno = {182},
numpages = {36},
}

@article{bib250,
  title={The emerged security and privacy of llm agent: A survey with case studies},
  author={He, Feng and Zhu, Tianqing and Ye, Dayong and Liu, Bo and Zhou, Wanlei and Yu, Philip S},
  journal={arXiv preprint},
  eprint={2407.19354},
  archivePrefix={arXiv},
  year={2024}
}

@misc{bib252,
  title={AutoGPT},
  author={AutoGPT},
  year={2023},
  url="https://github.com/Significant-Gravitas/AutoGPT"
}

@InProceedings{bib258,
  title = 	 {Language Agents with Reinforcement Learning for Strategic Play in the Werewolf Game},
  author =       {Xu, Zelai and Yu, Chao and Fang, Fei and Wang, Yu and Wu, Yi},
  booktitle = 	 {Proceedings of the 41st International Conference on Machine Learning},
  pages = 	 {55434--55464},
  year = 	 {2024},
  volume = 	 {235},
  series = 	 {Proceedings of Machine Learning Research},
  month = 	 {21--27 Jul},
  url = 	 {https://proceedings.mlr.press/v235/xu24ad.html},
}

@inproceedings{bib259,
  title={MASTERKEY: Automated Jailbreaking of Large Language Model Chatbots},
  author={Deng, Gelei and Liu, Yi and Li, Yuekang and Wang, Kailong and Zhang, Ying and Li, Zefeng and Wang, Haoyu and Zhang, Tianwei and Liu, Yang},
  booktitle={NDSS},
  year={2024}
}

@article{bib260,
  title={Towards Action Hijacking of Large Language Model-based Agent},
  author={Zhang, Yuyang and Chen, Kangjie and Jiang, Xudong and Sun, Yuxiang and Wang, Run and Wang, Lina},
  journal={arXiv preprint},
  eprint={2412.10807},
  archivePrefix={arXiv},
  year={2024}
}

@article{bib261,
  title={Evil geniuses: Delving into the safety of llm-based agents},
  author={Tian, Yu and Yang, Xiao and Zhang, Jingyuan and Dong, Yinpeng and Su, Hang},
  journal={arXiv preprint},
  eprint={2311.11855},
  archivePrefix={arXiv},
  year={2023}
}

@article{bib263,
  title={The philosopher's stone: Trojaning plugins of large language models},
  author={Dong, Tian and Xue, Minhui and Chen, Guoxing and Holland, Rayne and Meng, Yan and Li, Shaofeng and Liu, Zhen and Zhu, Haojin},
  journal={arXiv preprint},
  eprint={2312.00374},
  archivePrefix={arXiv},
  year={2023}
}

@article{bib264,
  title={Hacking Back the AI-Hacker: Prompt Injection as a Defense Against LLM-driven Cyberattacks},
  author={Pasquini, Dario and Kornaropoulos, Evgenios M and Ateniese, Giuseppe},
  journal={arXiv preprint},
  eprint={2410.20911},
  archivePrefix={arXiv},
  year={2024}
}

@article{bib265,
  title={Breaking agents: Compromising autonomous llm agents through malfunction amplification},
  author={Zhang, Boyang and Tan, Yicong and Shen, Yun and Salem, Ahmed and Backes, Michael and Zannettou, Savvas and Zhang, Yang},
  journal={arXiv preprint},
  eprint={2407.20859},
  archivePrefix={arXiv},
  year={2024}
}

@article{bib266,
  title={Large language model sentinel: Llm agent for adversarial purification},
  author={Lin, Guang and Tanaka, Toshihisa and Zhao, Qibin},
  journal={arXiv preprint},
  eprint={2405.20770},
  archivePrefix={arXiv},
  year={2024}
}

@article{bib267,
  title={Aegis: Online adaptive ai content safety moderation with ensemble of llm experts},
  author={Ghosh, Shaona and Varshney, Prasoon and Galinkin, Erick and Parisien, Christopher},
  journal={arXiv preprint},
  eprint={2404.05993},
  archivePrefix={arXiv},
  year={2024}
}

@article{bib268,
  title={Breaking Event Rumor Detection via Stance-Separated Multi-Agent Debate},
  author={Zhang, Mingqing and Gong, Haisong and Liu, Qiang and Wu, Shu and Wang, Liang},
  journal={arXiv preprint},
  eprint={2412.04859},
  archivePrefix={arXiv},
  year={2024}
}

@article{bib269,
  title={Chain of natural language inference for reducing large language model ungrounded hallucinations},
  author={Lei, Deren and Li, Yaxi and Hu, Mengya and Wang, Mingyu and Yun, Vincent and Ching, Emily and Kamal, Eslam},
  journal={arXiv preprint},
  eprint={2310.03951},
  archivePrefix={arXiv},
  year={2023}
}

@inproceedings{bib270,
  title={Towards Safe and Honest AI Agents with Neural Self-Other Overlap},
  author={Carauleanu, Marc and Vaiana, Michael and Rosenblatt, Judd and Berg, Cameron and de Lucena, Diogo S},
  booktitle={Neurips Safe Generative AI Workshop 2024},
  year={2024}
}

@article{bib271,
  title={Agentmonitor: A plug-and-play framework for predictive and secure multi-agent systems},
  author={Chan, Chi-Min and Yu, Jianxuan and Chen, Weize and Jiang, Chunyang and Liu, Xinyu and Shi, Weijie and Liu, Zhiyuan and Xue, Wei and Guo, Yike},
  journal={arXiv preprint},
  eprint={2408.14972},
  archivePrefix={arXiv},
  year={2024}
}

@inproceedings{bib272,
  title={Concept-guided llm agents for human-ai safety codesign},
  author={Geissler, Florian and Roscher, Karsten and Trapp, Mario},
  booktitle={Proceedings of the AAAI Symposium Series},
  volume={3},
  number={1},
  pages={100--104},
  year={2024},
  doi="10.1609/aaaiss.v3i1.31188"
}

@inproceedings{bib273,
    title = "{MA}g{IC}: Investigation of Large Language Model Powered Multi-Agent in Cognition, Adaptability, Rationality and Collaboration",
    author = "Xu, Lin  and
      Hu, Zhiyuan  and
      Zhou, Daquan  and
      Ren, Hongyu  and
      Dong, Zhen  and
      Keutzer, Kurt  and
      Ng, See-Kiong  and
      Feng, Jiashi",
    booktitle = "Proceedings of the 2024 Conference on Empirical Methods in Natural Language Processing",
    month = nov,
    year = "2024",
    doi = "10.18653/v1/2024.emnlp-main.416",
    pages = "7315--7332",
}

@inproceedings{bib274,
  title={MetaGPT: Meta Programming for A Multi-Agent Collaborative Framework},
  author={Hong, Sirui and Zhuge, Mingchen and Chen, Jonathan and Zheng, Xiawu and Cheng, Yuheng and Wang, Jinlin and Zhang, Ceyao and Wang, Zili and Yau, Steven Ka Shing and Lin, Zijuan and others},
  booktitle={The Twelfth International Conference on Learning Representations},
  year={2024}
}

@article{bib277,
  title={iAgent: LLM Agent as a Shield between User and Recommender Systems},
  author={Xu, Wujiang and Shi, Yunxiao and Liang, Zujie and Ning, Xuying and Mei, Kai and Wang, Kun and Zhu, Xi and Xu, Min and Zhang, Yongfeng},
  journal={arXiv preprint},
  eprint={2502.14662},
  archivePrefix={arXiv},
  year={2025}
}

@article{bib278,
  title={Mrj-agent: An effective jailbreak agent for multi-round dialogue},
  author={Wang, Fengxiang and Duan, Ranjie and Xiao, Peng and Jia, Xiaojun and Zhao, Shiji and Wei, Cheng and Chen, YueFeng and Wang, Chongwen and Tao, Jialing and Su, Hang and others},
  journal={arXiv preprint},
  eprint={2411.03814},
  archivePrefix={arXiv},
  year={2024}
}

@INPROCEEDINGS{bib279,
  author={Chao, Patrick and Robey, Alexander and Dobriban, Edgar and Hassani, Hamed and Pappas, George J. and Wong, Eric},
  booktitle={2025 IEEE Conference on Secure and Trustworthy Machine Learning (SaTML)}, 
  title={Jailbreaking Black Box Large Language Models in Twenty Queries}, 
  year={2025},
  volume={},
  number={},
  pages={23-42},
  doi={10.1109/SaTML64287.2025.00010}
}

@inproceedings{bib280,
 author = {Mehrotra, Anay and Zampetakis, Manolis and Kassianik, Paul and Nelson, Blaine and Anderson, Hyrum and Singer, Yaron and Karbasi, Amin},
 booktitle = {Advances in Neural Information Processing Systems},
 pages = {61065--61105},
 title = {Tree of Attacks: Jailbreaking Black-Box LLMs Automatically},
 url = {https://proceedings.neurips.cc/paper_files/paper/2024/file/70702e8cbb4890b4a467b984ae59828a-Paper-Conference.pdf},
 volume = {37},
 year = {2024}
}

@article{bib291,
  title={Agentsafe: Safeguarding large language model-based multi-agent systems via hierarchical data management},
  author={Mao, Junyuan and Meng, Fanci and Duan, Yifan and Yu, Miao and Jia, Xiaojun and Fang, Junfeng and Liang, Yuxuan and Wang, Kun and Wen, Qingsong},
  journal={arXiv preprint},
  eprint={2503.04392},
  archivePrefix={arXiv},
  year={2025}
}

@article{bib292,
  title={Agrail: A lifelong agent guardrail with effective and adaptive safety detection},
  author={Luo, Weidi and Dai, Shenghong and Liu, Xiaogeng and Banerjee, Suman and Sun, Huan and Chen, Muhao and Xiao, Chaowei},
  journal={arXiv preprint},
  eprint={2502.11448},
  archivePrefix={arXiv},
  year={2025}
}

@article{bib293,
  title={Evaluating the Robustness of Multimodal Agents Against Active Environmental Injection Attacks},
  author={Chen, Yurun and Hu, Xavier and Yin, Keting and Li, Juncheng and Zhang, Shengyu},
  journal={arXiv preprint},
  eprint={2502.13053},
  archivePrefix={arXiv},
  year={2025}
}

@inproceedings{bib294,
    title = "The Task Shield: Enforcing Task Alignment to Defend Against Indirect Prompt Injection in {LLM} Agents",
    author = "Jia, Feiran  and
      Wu, Tong  and
      Qin, Xin  and
      Squicciarini, Anna",
    editor = "Che, Wanxiang  and
      Nabende, Joyce  and
      Shutova, Ekaterina  and
      Pilehvar, Mohammad Taher",
    booktitle = "Proceedings of the 63rd Annual Meeting of the Association for Computational Linguistics (Volume 1: Long Papers)",
    month = jul,
    year = "2025",
    address = "Vienna, Austria",
    publisher = "Association for Computational Linguistics",
    url = "https://aclanthology.org/2025.acl-long.1435/",
    doi = "10.18653/v1/2025.acl-long.1435",
    pages = "29680--29697",
    ISBN = "979-8-89176-251-0",
}

@article{bib295,
  title={Rtbas: Defending llm agents against prompt injection and privacy leakage},
  author={Zhong, Peter Yong and Chen, Siyuan and Wang, Ruiqi and McCall, McKenna and Titzer, Ben L and Miller, Heather and Gibbons, Phillip B},
  journal={arXiv preprint},
  eprint={2502.08966},
  archivePrefix={arXiv},
  year={2025}
}

@article{bib296,
  title={Progent: Programmable privilege control for LLM agents},
  author={Shi, Tianneng and He, Jingxuan and Wang, Zhun and Wu, Linyu and Li, Hongwei and Guo, Wenbo and Song, Dawn},
  journal={arXiv preprint arXiv:2504.11703},
  eprint={2504.11703},
  archivePrefix={arXiv},
  year={2025}
}

@article{bib297,
  title={Prompt Injection Attack to Tool Selection in LLM Agents},
  author={Shi, Jiawen and Yuan, Zenghui and Tie, Guiyao and Zhou, Pan and Gong, Neil Zhenqiang and Sun, Lichao},
  journal={arXiv preprint},
  eprint={2504.19793},
  archivePrefix={arXiv},
  year={2025}
}

@inproceedings{bib298,
    title = "The Dark Side of Function Calling: Pathways to Jailbreaking Large Language Models",
    author = "Wu, Zihui  and
      Gao, Haichang  and
      He, Jianping  and
      Wang, Ping",
    booktitle = "Proceedings of the 31st International Conference on Computational Linguistics",
    month = jan,
    year = "2025",
    url = "https://aclanthology.org/2025.coling-main.39/",
    pages = "584--592",
}

@article{bib299,
  title={Model context protocol (mcp): Landscape, security threats, and future research directions},
  author={Hou, Xinyi and Zhao, Yanjie and Wang, Shenao and Wang, Haoyu},
  journal={arXiv preprint},
  eprint={2503.23278},
  archivePrefix={arXiv},
  year={2025}
}

@article{bib300,
  title={Enterprise-grade security for the model context protocol (mcp): Frameworks and mitigation strategies},
  author={Narajala, Vineeth Sai and Habler, Idan},
  journal={arXiv preprint},
  eprint={2504.08623},
  archivePrefix={arXiv},
  year={2025}
}

@misc{bib301,
  title={LangChain: Framework for developing applications powered by language models},
  author={LangChain},
  year={2022},
  url="https://github. com/langchain-ai/langchain"
}

@misc{bib302,
  title={Funcation Calling},
  author={OpenAI},
  year={2023},
  url="https://platform.openai.com/docs/guides/function-calling?api-mode=responses"
}

@misc{bib303,
  title={Model Context Protocol(MCP)},
  author={Anthropic},
  year={2024},
  url="https://docs.anthropic.com/en/docs/agents-and-tools/mcp"
}

@misc{bib304,
  title={Perspective API},
  author={Perspective},
  url="https://perspectiveapi.com/#/home",
  year={n.d.}
}

@misc{bib305,
  title={Azure Content Moderator},
  author={Microsoft},
  url="https://learn.microsoft.com/en-us/azure/ai-services/content-moderator",
  year={n.d.}
}

@misc{bib306,
  title={Moderation},
  author={OpenAI},
  url="https://platform.openai.com/docs/guides/moderation",
  year={n.d.}
}

@article{bib307,
  title={Mcp safety audit: Llms with the model context protocol allow major security exploits},
  author={Radosevich, Brandon and Halloran, John},
  journal={arXiv preprint},
  eprint={2504.03767},
  archivePrefix={arXiv},
  year={2025}
}

@article{bib308,
  title={Securing genai multi-agent systems against tool squatting: A zero trust registry-based approach},
  author={Narajala, Vineeth Sai and Huang, Ken and Habler, Idan},
  journal={arXiv preprint},
  eprint={2504.19951},
  archivePrefix={arXiv},
  year={2025}
}

@misc{bib309,
  title={Baichuan2-13B-Chat},
  author={Baichuan Intelligent Technology},
  year={2023},
  url="https://huggingface.co/baichuan-inc/Baichuan2-13B-Chat"
}

@misc{bib310,
  title={Baichuan2-13B-Chat},
  author={Baichuan Intelligent Technology},
  year={2023},
  url="https://huggingface.co/baichuan-inc/Baichuan2-7B-Chat"
}

@article{bib311,
  title={The Traitors: Deception and Trust in Multi-Agent Language Model Simulations},
  author={Curvo, Pedro MP},
  journal={arXiv preprint},
  eprint={2505.12923},
  archivePrefix={arXiv},
  year={2025}
}

@article{bib312,
  title={Attention Mechanism for LLM-based Agents Dynamic Diffusion under Information Asymmetry},
  author={Zhang, Yiwen and Wu, Yifu and Hua, Wenyue and Lu, Xiang and Hu, Xuming},
  journal={arXiv preprint},
  eprint={2502.13160},
  archivePrefix={arXiv},
  year={2025}
}

@article{bib313,
  title={The rise and potential of large language model based agents: A survey},
  author={Xi, Zhiheng and Chen, Wenxiang and Guo, Xin and He, Wei and Ding, Yiwen and Hong, Boyang and Zhang, Ming and Wang, Junzhe and Jin, Senjie and Zhou, Enyu and others},
  journal={Science China Information Sciences},
  volume={68},
  number={2},
  pages={121101},
  year={2025},
  doi="10.1007/s11432-024-4222-0",
}

@article{bib314,
  title={Navigating the risks: A survey of security, privacy, and ethics threats in llm-based agents},
  author={Gan, Yuyou and Yang, Yong and Ma, Zhe and He, Ping and Zeng, Rui and Wang, Yiming and Li, Qingming and Zhou, Chunyi and Li, Songze and Wang, Ting and others},
  journal={arXiv preprint},
  eprint={2411.09523},
  archivePrefix={arXiv},
  year={2024}
}

@article{bib315,
  title={A survey on trustworthy llm agents: Threats and countermeasures},
  author={Yu, Miao and Meng, Fanci and Zhou, Xinyun and Wang, Shilong and Mao, Junyuan and Pang, Linsey and Chen, Tianlong and Wang, Kun and Li, Xinfeng and Zhang, Yongfeng and others},
  journal={arXiv preprint},
  eprint={2503.09648},
  archivePrefix={arXiv},
  year={2025}
}

@article{bib316,
  title={Large language models surpass human experts in predicting neuroscience results},
  author={Luo, Xiaoliang and Rechardt, Akilles and Sun, Guangzhi and Nejad, Kevin K and Y{\'a}{\~n}ez, Felipe and Yilmaz, Bati and Lee, Kangjoo and Cohen, Alexandra O and Borghesani, Valentina and Pashkov, Anton and others},
  journal={Nature human behaviour},
  volume={9},
  number={2},
  pages={305--315},
  year={2025},
  doi="10.1038/s41562-024-02046-9",
}

@misc{bib317,
  title={Gpt-4o},
  author={OpenAI},
  year={2024},
  url="https://openai.com/index/hello-gpt-4o/"
}

@misc{bib318,
  title={Gpt-4},
  author={OpenAI},
  year={2023},
  url="https://openai.com/index/gpt-4-research/"
}

@article{bib319,
  title={Osworld: Benchmarking multimodal agents for open-ended tasks in real computer environments},
  author={Xie, Tianbao and Zhang, Danyang and Chen, Jixuan and Li, Xiaochuan and Zhao, Siheng and Cao, Ruisheng and Hua, Toh J and Cheng, Zhoujun and Shin, Dongchan and Lei, Fangyu and others},
  journal={Advances in Neural Information Processing Systems},
  volume={37},
  pages={52040--52094},
  year={2024}
}

@article{bib320,
  title={Visualwebarena: Evaluating multimodal agents on realistic visual web tasks},
  author={Koh, Jing Yu and Lo, Robert and Jang, Lawrence and Duvvur, Vikram and Lim, Ming Chong and Huang, Po-Yu and Neubig, Graham and Zhou, Shuyan and Salakhutdinov, Ruslan and Fried, Daniel},
  journal={arXiv preprint},
  eprint={2401.13649},
  archivePrefix={arXiv},
  year={2024}
}

@misc{bib321,
  title={MCP Market},
  author={MCPMarket},
  year={n.d.},
  url="https://mcpmarket.com"
}

@misc{bib322,
  title={MCP Marketplace},
  author={Cline},
  year={n.d.},
  url="https://cline.bot/mcp-marketplace"
}

@article{bib323,
  title={Abusing images and sounds for indirect instruction injection in multi-modal LLMs},
  author={Bagdasaryan, Eugene and Hsieh, Tsung-Yin and Nassi, Ben and Shmatikov, Vitaly},
  journal={arXiv preprint},
  eprint={2307.10490},
  archivePrefix={arXiv},
  year={2023}
}

@article{VulnBot,
  title={Vulnbot: Autonomous penetration testing for a multi-agent collaborative framework},
  author={Kong, He and Hu, Die and Ge, Jingguo and Li, Liangxiong and Li, Tong and Wu, Bingzhen},
  journal={arXiv preprint},
  eprint={2501.13411},
  archivePrefix={arXiv},
  year={2025}
}

@INPROCEEDINGS{PENTEST-AI,
  author={Bianou, Stanislas G. and Batogna, Rodrigue G.},
  booktitle={2024 IEEE International Conference on Cyber Security and Resilience (CSR)}, 
  title={PENTEST-AI, an LLM-Powered Multi-Agents Framework for Penetration Testing Automation Leveraging Mitre Attack}, 
  year={2024},
  volume={},
  number={},
  pages={763-770},
  doi={10.1109/CSR61664.2024.10679480}}

@inproceedings{FOFA,
author = {Liu, Ye and Yuan, Yulei and Zhu, Yuntian and Hu, Luolin and Wang, Lei},
title = {Research on Design and Implementation of an Intelligent Network Asset Search System Based on LLM Agent and FOFA},
year = {2025},
isbn = {9798400712715},
publisher = {Association for Computing Machinery},
address = {New York, NY, USA},
url = {https://doi.org/10.1145/3729706.3729783},
doi = {10.1145/3729706.3729783},
booktitle = {Proceedings of the 2025 4th International Conference on Cyber Security, Artificial Intelligence and the Digital Economy},
pages = {483–488},
numpages = {6},
location = {
},
series = {CSAIDE '25}
}

@article{RapidPen,
  title={RapidPen: Fully automated IP-to-shell penetration testing with LLM-based agents},
  author={Nakatani, Sho},
  journal={arXiv preprint},
  eprint={2502.16730},
  archivePrefix={arXiv},
  year={2025}
}

@article{AutoPentest,
  title={AutoPentest: Enhancing Vulnerability Management With Autonomous LLM Agents},
  author={Henke, Julius},
  journal={arXiv preprint},
  eprint={2505.10321},
  archivePrefix={arXiv},
  year={2025}
}

@article{BLMProbe,
  author={Tian, Zhenhao and He, Yi and Zhang, Nuo and Lin, Qixiao and Shi, Hetian and Zhuge, Jianwei and Mao, Jian and Chang, Deliang},
  journal={IEEE Transactions on Information Forensics and Security}, 
  title={BLMProbe: Enhancing Internet-Connected Device Discovery by Automated Device Labeling and Label Migration}, 
  year={2025},
  volume={20},
  number={},
  pages={7227-7242},
  doi={10.1109/TIFS.2025.3587211}}

@INPROCEEDINGS{c5-3-1,
  author={Enoch, Simon Yusuf and Kim, Dan Dongseong},
  booktitle={2024 International Conference on Ubiquitous Computing and Communications (IUCC)}, 
  title={Model-Based Attack Planning Strategies for Automated Penetration Testing Exercises}, 
  year={2024},
  volume={},
  number={},
  pages={115-122},
  doi={10.1109/IUCC65928.2024.00033}}

@Article{c5-3-2,
AUTHOR = {Chen, Ziyang and Kang, Fei and Xiong, Xiaobing and Shu, Hui},
TITLE = {A Survey on Penetration Path Planning in Automated Penetration Testing},
JOURNAL = {Applied Sciences},
VOLUME = {14},
YEAR = {2024},
NUMBER = {18},
ARTICLE-NUMBER = {8355},
URL = {https://www.mdpi.com/2076-3417/14/18/8355},
ISSN = {2076-3417},
DOI = {10.3390/app14188355}
}

@article{ATAG,
  title={ATAG: AI-Agent Application Threat Assessment with Attack Graphs},
  author={Gandhi, Parth Atulbhai and Shukla, Akansha and Tayouri, David and Ifland, Beni and Elovici, Yuval and Puzis, Rami and Shabtai, Asaf},
  journal={arXiv preprint},
  eprint={2506.02859},
  archivePrefix={arXiv},
  year={2025}
}

@inproceedings{MulVAL,
  title={MulVAL: A logic-based network security analyzer.},
  author={Ou, Xinming and Govindavajhala, Sudhakar and Appel, Andrew W and others},
  booktitle={USENIX security symposium},
  volume={8},
  pages={113--128},
  year={2005},
  organization={Baltimore, MD}
}

@article{Aurora,
  title={From Sands to Mansions: Towards Automated Cyberattack Emulation with Classical Planning and Large Language Models},
  author={Wang, Lingzhi and Li, Zhenyuan and Jiang, Yi and Wang, Zhengkai and Guo, Zonghan and Wang, Jiahui and Wei, Yangyang and Shen, Xiangmin and Ruan, Wei and Chen, Yan},
  journal={arXiv preprint},
  eprint={2407.16928},
  archivePrefix={arXiv},
  year={2024}
}

@article{MM-AttacKG,
  title={MM-AttacKG: A Multimodal Approach to Attack Graph Construction with Large Language Models},
  author={Zhang, Yongheng and Zhao, Xinyun and Ma, Yunshan and Ma, Haokai and Guan, Yingxiao and Yang, Guozheng and Lu, Yuliang and Wang, Xiang},
  journal={arXiv preprint},
  eprint={2506.16968},
  archivePrefix={arXiv},
  year={2025}
}

@ARTICLE{phishingsurvey,
  author={Nowakowski, Wojciech},
  journal={IEEE Access}, 
  title={Social Engineering Analysis Framework: A Comprehensive Playbook for Human Hacking}, 
  year={2025},
  volume={13},
  number={},
  pages={18827-18849},
  doi={10.1109/ACCESS.2025.3532999}}

@ARTICLE{phishing2,
AUTHOR={Rajivan, Prashanth  and Gonzalez, Cleotilde }, 
TITLE={Creative Persuasion: A Study on Adversarial Behaviors and Strategies in Phishing Attacks},
JOURNAL={Frontiers in Psychology},
VOLUME={Volume 9 - 2018},
YEAR={2018},
URL={https://www.frontiersin.org/journals/psychology/articles/10.3389/fpsyg.2018.00135},
DOI={10.3389/fpsyg.2018.00135},
ISSN={1664-1078}}

@article{phishing3,
  title={Evaluating Large Language Models' Capability to Launch Fully Automated Spear Phishing Campaigns: Validated on Human Subjects},
  author={Heiding, Fred and Lermen, Simon and Kao, Andrew and Schneier, Bruce and Vishwanath, Arun},
  journal={arXiv preprint},
  eprint={2412.00586},
  archivePrefix={arXiv},
  year={2024}
}

@inproceedings{wLLMgonline,
author = {Kim, Hanna and Song, Minkyoo and Na, Seung Ho and Shin, Seungwon and Lee, Kimin},
title = {When LLMs go online: the emerging threat of web-enabled LLMs},
year = {2025},
isbn = {978-1-939133-52-6},
publisher = {USENIX Association},
address = {USA},
booktitle = {Proceedings of the 34th USENIX Conference on Security Symposium},
articleno = {90},
numpages = {20},
location = {Seattle, WA, USA},
series = {SEC '25}
}

@article{hutchins2011intelligence,
  title={Intelligence-driven computer network defense informed by analysis of adversary campaigns and intrusion kill chains},
  author={Hutchins, Eric M and Cloppert, Michael J and Amin, Rohan M and others},
  journal={Leading Issues in Information Warfare \& Security Research},
  volume={1},
  number={1},
  pages={80},
  year={2011}
}

@INPROCEEDINGS{autoCTI,
  author={Loevenich, Johannes F. and Adler, Erik and Hürten, Tobias and Spelter, Florian and Roncevic, Damian and Lopes, Roberto Rigolin F.},
  booktitle={2025 International Conference on Military Communication and Information Systems (ICMCIS)}, 
  title={Automating Cyber Threat Intelligence and Attack Chain Generation using Cyber Security Knowledge Graphs and Large Language Models}, 
  year={2025},
  volume={},
  number={},
  pages={1-10},
  doi={10.1109/ICMCIS64378.2025.11047951}}

@inproceedings{yao2023react,
  title={React: Synergizing reasoning and acting in language models},
  author={Yao, Shunyu and Zhao, Jeffrey and Yu, Dian and Du, Nan and Shafran, Izhak and Narasimhan, Karthik and Cao, Yuan},
  booktitle={International Conference on Learning Representations (ICLR)},
  year={2023}
}

@article{CAPTCHAsurvey,
author = {Guerar, Meriem and Verderame, Luca and Migliardi, Mauro and Palmieri, Francesco and Merlo, Alessio},
title = {Gotta CAPTCHA ’Em All: A Survey of 20 Years of the Human-or-computer Dilemma},
year = {2021},
issue_date = {December 2022},
publisher = {Association for Computing Machinery},
address = {New York, NY, USA},
volume = {54},
number = {9},
issn = {0360-0300},
url = {https://doi.org/10.1145/3477142},
doi = {10.1145/3477142},
journal = {ACM Comput. Surv.},
month = oct,
articleno = {192},
numpages = {33}
}

@article{CAPTCHAsok,
title = {SoK: Machine vs. machine – A systematic classification of automated machine learning-based CAPTCHA solvers},
journal = {Computers \& Security},
volume = {97},
pages = {101947},
year = {2020},
issn = {0167-4048},
doi = {https://doi.org/10.1016/j.cose.2020.101947},
url = {https://www.sciencedirect.com/science/article/pii/S0167404820302236},
author = {Antreas Dionysiou and Elias Athanasopoulos},
}

@inproceedings{teoh2025captchas,
  title={Are $\{$CAPTCHAs$\}$ Still Bot-hard? Generalized Visual $\{$CAPTCHA$\}$ Solving with Agentic Vision Language Model},
  author={Teoh, Xiwen and Lin, Yun and Li, Siqi and Liu, Ruofan and Sollomoni, Avi and Harel, Yaniv and Dong, Jin Song},
  booktitle={34th USENIX Security Symposium (USENIX Security 25)},
  pages={3747--3766},
  year={2025}
}

@article{deng2024oedipus,
  title={Oedipus: Llm-enchanced reasoning captcha solver},
  author={Deng, Gelei and Ou, Haoran and Liu, Yi and Zhang, Jie and Zhang, Tianwei and Liu, Yang},
  journal={arXiv preprint},
  eprint={2405.07496},
  archivePrefix={arXiv},
  year={2024}
}

@article{zhang2024timing,
  title={Timing side-channel attacks and countermeasures in CPU microarchitectures},
  author={Zhang, Jiliang and Chen, Congcong and Cui, Jinhua and Li, Keqin},
  journal={ACM Computing Surveys},
  volume={56},
  number={7},
  pages={1--40},
  year={2024},
  publisher={ACM New York, NY}
}

@book{yaman2023agent,
  title={Agent SCA: Advanced physical side channel analysis agent with LLMs},
  author={Yaman, Ferhat},
  year={2023},
  publisher={North Carolina State University}
}

@article{skopik2020under,
  title={Under false flag: using technical artifacts for cyber attack attribution},
  author={Skopik, Florian and Pahi, Timea},
  journal={Cybersecurity},
  volume={3},
  number={1},
  pages={8},
  year={2020},
  publisher={Springer},
  doi={https://doi.org/10.1186/s42400-020-00048-4}
}

@article{happe2023llms,
  title={Llms as hackers: Autonomous linux privilege escalation attacks},
  author={Happe, Andreas and Kaplan, Aaron and Cito, Juergen},
  journal={arXiv preprint},
  eprint={2310.11409},
  archivePrefix={arXiv},
  year={2023}
}

@article{saha2025malgen,
  title={MalGEN: A Generative Agent Framework for Modeling Malicious Software in Cybersecurity},
  author={Saha, Bikash and Shukla, Sandeep Kumar},
  journal={arXiv preprint},
  eprint={2506.07586},
  archivePrefix={arXiv},
  year={2025}
}

@article{lupinacci2025dark,
  title={The Dark Side of LLMs: Agent-based Attacks for Complete Computer Takeover},
  author={Lupinacci, Matteo and Pironti, Francesco Aurelio and Blefari, Francesco and Romeo, Francesco and Arena, Luigi and Furfaro, Angelo},
  journal={arXiv preprint},
  eprint={2507.06850},
  archivePrefix={arXiv},
  year={2025}
}

@article{ITURBE2024104077,
title = {Unleashing offensive artificial intelligence: Automated attack technique code generation},
journal = {Computers \& Security},
volume = {147},
pages = {104077},
year = {2024},
issn = {0167-4048},
doi = {https://doi.org/10.1016/j.cose.2024.104077},
url = {https://www.sciencedirect.com/science/article/pii/S0167404824003821},
author = {Eider Iturbe and Oscar Llorente-Vazquez and Angel Rego and Erkuden Rios and Nerea Toledo},
}

@Inbook{Conti2018,
author="Conti, Mauro
and Dargahi, Tooska
and Dehghantanha, Ali",
editor="Dehghantanha, Ali
and Conti, Mauro
and Dargahi, Tooska",
title="Cyber Threat Intelligence: Challenges and Opportunities",
bookTitle="Cyber Threat Intelligence",
year="2018",
publisher="Springer International Publishing",
address="Cham",
pages="1--6",
isbn="978-3-319-73951-9",
doi="10.1007/978-3-319-73951-9_1",
url="https://doi.org/10.1007/978-3-319-73951-9_1"
}

@INPROCEEDINGS{10628558,
  author={Fieblinger, Romy and Alam, Md Tanvirul and Rastogi, Nidhi},
  booktitle={2024 IEEE European Symposium on Security and Privacy Workshops (EuroS\&PW)}, 
  title={Actionable Cyber Threat Intelligence Using Knowledge Graphs and Large Language Models}, 
  year={2024},
  volume={},
  number={},
  pages={100-111},
  doi={10.1109/EuroSPW61312.2024.00018}}

@article{HU2024103999,
title = {LLM-TIKG: Threat intelligence knowledge graph construction utilizing large language model},
journal = {Computers \& Security},
volume = {145},
pages = {103999},
year = {2024},
issn = {0167-4048},
doi = {https://doi.org/10.1016/j.cose.2024.103999},
url = {https://www.sciencedirect.com/science/article/pii/S0167404824003043},
author = {Yuelin Hu and Futai Zou and Jiajia Han and Xin Sun and Yilei Wang},
}

@inproceedings{huang2024ctikg,
  title={Ctikg: Llm-powered knowledge graph construction from cyber threat intelligence},
  author={Huang, Liangyi and Xiao, Xusheng},
  booktitle={First Conference on Language Modeling},
  year={2024}
}

@article{balasubramanian2025generative,
  title={Generative AI for cyber threat intelligence: applications, challenges, and analysis of real-world case studies},
  author={Balasubramanian, Prasasthy and Liyana, Sonali and Sankaran, Hamsini and Sivaramakrishnan, Shambavi and Pusuluri, Sruthi and Pirttikangas, Susanna and Peltonen, Ella},
  journal={Artificial Intelligence Review},
  volume={58},
  number={11},
  pages={1--134},
  year={2025},
  doi={https://doi.org/10.1007/s10462-025-11338-z},
  publisher={Springer}
}

@InProceedings{10.1007/978-3-031-87496-3_5,
author="Mitra, Shaswata
and Neupane, Subash
and Chakraborty, Trisha
and Mittal, Sudip
and Piplai, Aritran
and Gaur, Manas
and Rahimi, Shahram",
editor="Adi, Kamel
and Bourdeau, Simon
and Durand, Christel
and Viet Triem Tong, Val{\'e}rie
and Dulipovici, Alina
and Kermarrec, Yvon
and Garcia-Alfaro, Joaquin",
title="LocalIntel: Generating Organizational Threat Intelligence from Global and Local Cyber Knowledge",
booktitle="Foundations and Practice of Security",
year="2025",
publisher="Springer Nature Switzerland",
address="Cham",
pages="63--78",
isbn="978-3-031-87496-3"
}

@article{cichonski2012computer,
  title={Computer security incident handling guide},
  author={Cichonski, Paul and Millar, Tom and Grance, Tim and Scarfone, Karen and others},
  journal={NIST Special Publication},
  volume={800},
  number={61},
  pages={1--147},
  year={2012}
}

@article{beretas2024information,
  title={Information systems security, detection and recovery from cyber attacks},
  author={Beretas, Christos},
  journal={Universal Library of Engineering Technology},
  volume={1},
  number={1},
  year={2024}
}

@article{SUBSTANTIATION,
author = {Golovan, V. and Tarasenko, S. and Budur, O. and Dekhtyarenko, K. and Bubenshchikov, R.},
year = {2021},
month = {01},
pages = {112-122},
title = {SUBSTANTIATION OF PRINCIPLES OF STRUCTURE AND EVALUATION OF EFFICIENCY OF PERIMETER SYSTEMS OF PROTECTION OF ARSENALS, BASES AND WAREHOUSES},
volume = {1},
journal = {Collection of scientific works of Odesa Military Academy},
doi = {10.37129/2313-7509.2020.14.1.112-122}
}

@article{ahmad2015case,
  title={A case analysis of information systems and security incident responses},
  author={Ahmad, Atif and Maynard, Sean B and Shanks, Graeme},
  journal={International Journal of Information Management},
  volume={35},
  number={6},
  pages={717--723},
  year={2015},
  publisher={Elsevier}
}

@inproceedings{lin2024optimization,
  title={Optimization and Implementation of Network Security Incident Handling Process Based on Secure Arrangement},
  author={Lin, Ni and Yang, Qian and Shuai, Zhang and Sicheng, Tong and Chakun, Bao},
  booktitle={2024 IEEE 2nd International Conference on Sensors, Electronics and Computer Engineering (ICSECE)},
  pages={472--478},
  year={2024},
  organization={IEEE}
}

@misc{hillstone,
  title={Detecting Post-Breach Threats Using the Cyber Kill Chain},
  author={Hillstone Network},
  year={n.d.},
  url="https://www.hillstonenet.com/wp-content/uploads/Detecting-post-breach-threats-using-the-Cyber-Kill-Chain.pdf"
}

@article{portnoy2024towards,
  title={Towards Automatic Hands-on-Keyboard Attack Detection Using LLMs in EDR Solutions},
  author={Portnoy, Amit and Azikri, Ehud and Kels, Shay},
  journal={arXiv preprint},
  eprint={2408.01993},
  archivePrefix={arXiv},
  year={2024}
}

@article{narajala2025securing,
  title={Securing agentic ai: A comprehensive threat model and mitigation framework for generative ai agents},
  author={Narajala, Vineeth Sai and Narayan, Om},
  journal={arXiv preprint},
  eprint={2504.19956},
  archivePrefix={arXiv},
  year={2025}
}

@Article{app15137237,
AUTHOR = {Lee, Juyoung and Jeong, Yeonsu and Han, Taehyun and Lee, Taejin},
TITLE = {LogRESP-Agent: A Recursive AI Framework for Context-Aware Log Anomaly Detection and TTP Analysis},
JOURNAL = {Applied Sciences},
VOLUME = {15},
YEAR = {2025},
NUMBER = {13},
ARTICLE-NUMBER = {7237},
URL = {https://www.mdpi.com/2076-3417/15/13/7237},
ISSN = {2076-3417},
DOI = {10.3390/app15137237}
}

@article{de2025open,
  title={Open challenges in multi-agent security: Towards secure systems of interacting ai agents},
  author={de Witt, Christian Schroeder},
  journal={arXiv preprint},
  eprint={2505.02077},
  archivePrefix={arXiv},
  year={2025}
}

@article{yang2025large,
  title={Large Language Models for Network Intrusion Detection Systems: Foundations, Implementations, and Future Directions},
  author={Yang, Shuo and Zheng, Xinran and Zhang, Xinchen and Xu, Jinfeng and Li, Jinze and Xie, Donglin and Long, Weicai and Ngai, Edith CH},
  journal={arXiv preprint},
  eprint={2507.04752},
  archivePrefix={arXiv},
  year={2025}
}

@article{song2024audit,
  title={Audit-llm: Multi-agent collaboration for log-based insider threat detection},
  author={Song, Chengyu and Ma, Linru and Zheng, Jianming and Liao, Jinzhi and Kuang, Hongyu and Yang, Lin},
  journal={arXiv preprint},
  eprint={2408.08902},
  archivePrefix={arXiv},
  year={2024}
}

@ARTICLE{11141466,
  author={Hmimou, Yasser and Tabaa, Mohamed and Khiat, Azeddine and Hidila, Zineb},
  journal={IEEE Access}, 
  title={A Multi-Agent System for Cybersecurity Threat Detection and Correlation Using Large Language Models}, 
  year={2025},
  volume={13},
  number={},
  pages={150199-150215},
  doi={10.1109/ACCESS.2025.3602681}}

@article{kim2025prompt,
  title={Prompt flow integrity to prevent privilege escalation in llm agents},
  author={Kim, Juhee and Choi, Woohyuk and Lee, Byoungyoung},
  journal={arXiv preprint},
  eprint={2503.15547},
  archivePrefix={arXiv},
  year={2025}
}

@article{HE2022102435,
title = {Agile incident response (AIR): Improving the incident response process in healthcare},
journal = {International Journal of Information Management},
volume = {62},
pages = {102435},
year = {2022},
issn = {0268-4012},
doi = {https://doi.org/10.1016/j.ijinfomgt.2021.102435},
url = {https://www.sciencedirect.com/science/article/pii/S0268401221001286},
author = {Ying He and Efpraxia D. Zamani and Stefan Lloyd and Cunjin Luo},
}

@article{hays2024employing,
  title={Employing llms for incident response planning and review},
  author={Hays, Sam and White, Jules},
  journal={arXiv preprint},
  eprint={2403.01271},
  archivePrefix={arXiv},
  year={2024}
}

@inproceedings{kramer2025integrating,
  title={Integrating Large Language Models into Security Incident Response},
  author={Kramer, Diana and Rosique, Lambert and Narotam, Ajay and Bursztein, Elie and Kelley, Patrick Gage and Thomas, Kurt and Woodruff, Allison},
  booktitle={Twenty-First Symposium on Usable Privacy and Security (SOUPS 2025)},
  pages={133--148},
  year={2025}
}

@article{fumero2025cybersleuth,
  title={CyberSleuth: Autonomous Blue-Team LLM Agent for Web Attack Forensics},
  author={Fumero, Stefano and Huang, Kai and Boffa, Matteo and Giordano, Danilo and Mellia, Marco and Houidi, Zied Ben and Rossi, Dario},
  journal={arXiv preprint},
  eprint={2508.20643},
  archivePrefix={arXiv},
  year={2025}
}

@article{molleti2024automated,
  title={Automated threat detection and response using LLM agents},
  author={Molleti, R and Goje, V and Luthra, P and Raghavan, P},
  journal={World J. Adv. Res. Rev},
  year={2024}
}

@article{lin2025ircopilot,
  title={IRCopilot: Automated Incident Response with Large Language Models},
  author={Lin, Xihuan and Zhang, Jie and Deng, Gelei and Liu, Tianzhe and Liu, Xiaolong and Yang, Changcai and Zhang, Tianwei and Guo, Qing and Chen, Riqing},
  journal={arXiv preprint},
  eprint={2505.20945},
  archivePrefix={arXiv},
  year={2025}
}

@INPROCEEDINGS{11012055,
  author={Liu, Zefang},
  booktitle={2025 13th International Symposium on Digital Forensics and Security (ISDFS)}, 
  title={AutoBnB: Multi-Agent Incident Response with Large Language Models}, 
  year={2025},
  volume={},
  number={},
  pages={1-6},
  doi={10.1109/ISDFS65363.2025.11012055}}

\end{document}